\documentclass[a4paper,12pt,oneside]{smartthesis}
\usepackage[numbers]{natbib}
\pdfoutput=1
\usepackage{times}
\usepackage{here}
\usepackage{epsfig}
\usepackage{rotating}
\usepackage{longtable}
\usepackage{makeidx}
\usepackage{setspace}
\usepackage{amssymb,amsmath,euscript}
\usepackage{floatflt}
\usepackage{mathrsfs}

\usepackage{epsf,latexsym}
\usepackage{color,graphicx,graphics}
\usepackage[sf,small,hang]{caption}
\usepackage{pstcol}
\newcommand{\avg}[1]{\left< #1 \right>} 
\newcommand{\pd}[2]{\frac{\partial #1}{\partial #2}}

\usepackage{ctable}
 \usepackage{caption}
\usepackage{enumerate}
\usepackage{ragged2e}

\begin{document}
\frontmatter
\thispagestyle{empty}


\LARGE

\begin{center}

{\bf Broadband spectral modelling of bent jets of Active Galactic Nuclei}\\

\vspace{1.5 cm}

\Large

{A Thesis submitted\\
to the}\\

\vspace{0.5 cm}

{\bf University of Mumbai \\
for the \\
Ph. D. (SCIENCE) Degree\\
in PHYSICS} \\

\vspace{1.0 cm}

{Submitted By} \\

\vspace{0.5 cm}

{\bf B. Sunder Sahayanathan} \\
\vspace{1.0 cm}
	
{Under the Guidance of} \\

{\bf Dr. A. K. Mitra} \\
\vspace{1.5 cm}
 
{\bf Astrophysical Sciences Division} \\
{\bf Bhabha Atomic Research Centre} \\
{\bf Trombay, Mumbai - 400085} \\
{\bf October 2010} \\

\end{center}
\normalsize

\newpage
\newpage
\thispagestyle{empty}
\vspace*{3in}
\begin{center}
{\large {\it In memory of \\ my beloved grandma \\ Josephine...\\}}
\end{center}

\newpage
\thispagestyle{empty}
\newpage
\renewcommand{\thepage}{\roman{page}}
\setcounter{page}{1}
\setlength{\parindent}{0.0in}
\setlength{\parskip}{0.1in}
\chapter{Abstract}
The understanding of the physics of relativistic jets from active galactic 
nuclei (AGN) is still incomplete. A way to understand the different features 
of the AGN jets is to study it broadband spectra. In general, within the 
limits of present observations, AGN jets are observed in radio-to-X-ray energy 
band and they exhibit various intrinsic features such as \emph{knots}.
Particularly, the blazar jets which are pointed towards the observer, are 
observed in radio-to-$\gamma$-ray and their radio maps exhibit internal jet 
structures. Moreover, the high energy emission from blazars show rapid variability.

In this thesis, models have been developed to study the radiation 
emission processes from the knots of AGN jets as well as for blazar jets. 
A continuous injection plasma model is developed to study the X-ray emission 
from the knots of sources 1136-135, 1150+497, 1354+195 and 3C 371. The knot 
dynamics is then studied within the framework of internal shock model. In such 
a scenario, knots are formed due to the collision of two successive matter 
blobs emitted sporadically from the central engine of AGN. Shocks, generated 
in such collisions, accelerate electrons to relativistic energies. These
electrons subsequently emit radiation via synchrotron and/or inverse Compton 
processes in the radio-to-X-ray energy range. The study of M87 knots involves 
a two zone model where the electrons with a power-law distribution are further
accelerated. The synchrotron emission from these energetic electrons is then 
used to explain the observed spectrum. 

Regarding the blazar jets, the limb-brightening feature observed in
the radio maps of the BL Lac object MKN501 is studied considering shear 
acceleration of electrons at the boundary 
of the jet. This interpretation does not require a large viewing angle of the jet
as demanded by the earlier models and is consistent with the constraints obtained
from very high energy studies. In case of MKN421, the dependence of temporal 
behaviour of radiation emission on the particle acceleration mechanism has
been studied within the framework of two zone model.

\newpage

\chapter{List of Publications relevant to this Thesis}
\begin{enumerate}[(1)]
\item
{\bf A Continuous Injection Plasma Model for the X-Ray/Radio Knots in 
Kiloparsec-Scale Jets of Active Galactic Nuclei}\\
Sahayanathan, S.; Misra, R.; Kembhavi, A. K.; 
Kaul, C. L. \\
{\it Astrophysical Journal Letters} (2003),\; {\bf 588},\; L77-L80\\
\item
{\bf Interpretation of the Radio/X-Ray Knots of AGN Jets within the 
Internal Shock Model Framework}\\
Sahayanathan, S.; Misra, R.\\
{\it Astrophysical Journal} (2005), \;{\bf 628}, \;611-616\\

\item
{\bf Particle acceleration process and temporal behaviour of 
non-thermal emission from blazar}\\
Bhattacharyya, S.; {Sahayanathan, S.}; Bhatt, N.\\
{\it New Astronomy} (2005),\; {\bf 11},\; 17-26\\

\item
{\bf A two-zone synchrotron model for the knots in the M87 jet}\\
{Sahayanathan, S.}\\
{\it Monthly Notices of the Royal Astronomical Society Letters} (2008), 
\;{\bf 388},\; L49-L53\\

\item
{\bf Boundary shear acceleration in the jet of MKN501}\\
{Sahayanathan, S.}\\
{\it Monthly Notices of the Royal Astronomical Society Letters} (2009), 
\;{\bf 398},\; L49-L53\\

\end{enumerate}

\newpage
\chapter{Acknowledgements}
I thank my advisor Dr. A. K. Mitra for his kind cooperation, encouragement
and suggestions rendered throughout the course of my Ph.D work as well as 
in preparing this thesis. 
I am grateful to Mr. R. Koul, Head, Astrophysical Sciences 
Division (ApSD), Bhabha Atomic Research Centre (BARC), for his 
continuous support and cooperation.

When I look back in time I see two persons who paved the way 
for my present position as a researcher. The first one is my friend
Subir Bhattacharyya who lifted me from the level of a post graduate
student to the state of a researcher. The discussions, suggestions
and the arguments we had in the initial phase of my career laid 
the foundation of my later research work. The second one is my
friend and collaborator Ranjeev Misra, IUCAA, Pune who showed me 
how to think
and pursue as an independent researcher. The long discussions we
had and his valuable advice (often at the dimmed light environment 
of SATHI, a restaurant cum bar) played an important role in moulding
me as a researcher. 
Besides these two, there is third one who is my close companion,
Nilay Bhatt. At the moments of joy, disturbances and frustrations 
he is the one who remains always at my side. He is a 
selfless person who is at help when you are in need and makes sure that 
your need is served. I don't have enough words to offer my thanks to 
these personalities.

I thank Prof. Ajit Khembhavi, IUCAA, Pune for his suggestions and 
support and also for providing me a pleasant stay during my visits 
at IUCAA. I thank my friends Manojendu Choudhary, Sagar Godambe and Mradul
Sharma for the useful discussions and support.  

A young boy with dreams from a small town located at the southern tip of  
Indian peninsula after spending decades is in a position to present 
the thesis of his
work to obtain a Ph.D degree, considered as a highest degree
in present education system. This would not have been possible 
without the efforts
of my loving parents who took care of me in every step of my life. 
I cannot just thank them for the pains they took in my upbringing 
but be grateful to
them throughout my life. My
grandma Josephine to whom I dedicate this thesis, spent a considerable
amount of her pension on me and there by sharing the burden. 
Though I lost her at the beginning of my career she lives in my 
heart until my last breath. I am also indebted to my uncle 
Fr. I. M. John who came as a saviour at every stage in my life when 
I was left helpless. Without his timely help and advice it would not
be possible for me to reach up to here in my career. I also
thank my elder brother Arockiaraj who took me along with him after
my under graduation and helped me in completing my post graduation. I
thank my sister-in-law Jacintha and my niece Sangeetha for providing 
a homely stay during the days of my post graduation. I thank my brother Joseph 
Prakash and sister Maria Selvi whose love and affection always kept
my morals up.

In the mid of my career a loving and caring heart entered my life, 
considered my dreams as her own and strived with me to achieve them. 
I thank my better half, Roseline, for her love and infinite tolerance
to listen all the blah-blah-blah of mine with an unbelievable curious face. 
She is
the one who supported me during the tough days of my research career. 
Our little naughty boy, Alvin, whose innocent face and smile often
explains me the meaning of life needs a special thanks for 
keeping me busy in repairing his broken toys.

I thank my parents-in-law for their wish and eagerness to see me obtain this 
degree. I feel sad that my father-in-law passed away recently 
and I will miss him in the celebration.

Last but not the least I thank all my colleagues at ApSD and my institute
BARC for the support and encouragement. 

\tableofcontents
\listoffigures
\listoftables
\chapter{Acronyms}
\begin{longtable}{>{\bfseries}p{2.5cm} l}
{3C}&{Third Cambridge Catalogue of Radio Sources}\\
{ACIS}&{Advanced CCD Imaging Spectrometer}\\
{ACT}&{Atmospheric Cherenkov Telescope}\\
{AGN}&{Active Galactic Nuclei} \\
{ASCA}&{Advanced Satellite for Cosmology and Astrophysics}\\
{BLRG}&{Broad Line Radio Galaxies}\\
{CANGAROO}&{Collaboration of Australia and Nippon (Japan) for a GAmma Ray}\\
{}& {Observatory in the Outback}\\
{CAT}&{Cherenkov Array at Themis}\\
{CCD}&{Charge Coupled Device}\\
{CELESTE}&{CErenkov Low Energy Sampling and Timing Experiment}\\
{CIAO}&{\emph{Chandra} Interactive Analysis of Observations software}\\
{CMB}&{Cosmic Microwave Background}\\
{CTA}&{Cherenkov Telescope Array}\\
{EGRET}&{Energetic Gamma-Ray Experiment Telescope}\\
{FR I}&{Fanaroff-Riley I}\\
{FR II}&{Fanaroff-Riley II}\\
{FSRQ}&{Flat spectrum Radio Quasars}\\
{HEGRA}&{High-Energy-Gamma-Ray Astronomy}\\
{HESS}&{High Energy Stereoscopic System}\\
{HST}&{Hubble Space Telescope}\\
{IACT}&{Imaging Atmospheric Cherenkov Telescope}\\
{IC/CMB}&{Inverse Compton scattering of Cosmic Microwave Background}\\
{M87}& {Messeir 87}\\
{MACE}& {Major Atmospheric Cherenkov Experiment}\\
{MAGIC}&{Major Atmospheric Gamma-ray Imaging Cherenkov Telescope}\\
{MERLIN}&{Multi-Element Radio Linked Interferometer Network}\\
{MKN}& {Sources from Markarian Catalogue }\\
{NASA}&{National Aeronautics and Space Administration}\\
{NGC}&{New General Catalogue}\\
{NLRG}&{Narrow Line Radio Galaxies}\\
{PACT}&{Pachmarhi Array of Cherenkov Telescopes}\\
{PKS}&{Parkes Catalogue of Radio Sources}\\
{QSO}&{Quasi Stellar Objects}\\
{SED}& {Spectral Energy Distribution}\\
{SSC}& {Synchrotron Self Compton}\\
{SSRQ}&{Steep Spectrum Radio Quasars}\\
{STACEE}&{Solar Tower Atmospheric Cherenkov Effect Experiment}\\
{TACTIC}&{TeV Atomospheric Cherenkov Telescope with Imaging Camera}\\
{VERITAS}&{The Very Energetic Radiation Imaging Telescope Array System}\\
{VLA}&{Very Large Array}\\
{VLBA}&{Very Long Baseline Array}\\
{VHE}&{Very High Energy}\\
{XJET}&{X-ray emission from Extragalactic Radio Jets}\\{}&{(\footnotesize{http://hea-www.harvard.edu/XJET/})}\\
{XSPEC}&{X-Ray Spectral Fitting Package}\\
\end{longtable}


\mainmatter
\chapter{Introduction}
\label{chap:agn}

\section[Active Galactic Nuclei]{Active Galactic Nuclei (AGN)}
The nucleus of a galaxy with luminosity $\gtrsim 10^{44}$ ergs s$^{-1}$ and exceeding 
the overall luminosity of the entire host galaxy is known as Active Galactic 
Nucleus. Galaxies which host an AGN are called as active galaxies.

\subsection{Historical Background}
AGN was first observed in optical by Fath \cite{1909v05p71} in 1908 at Lick 
Observatory. His aim was to test the claim of that time 
that the spectra of the spiral nebulae are continuous, consistent with a collection of stars.
The continuous spectrum and absorption lines he observed for most of the sources 
suggested the presence of 
an unresolved collection of stars. However, for the nebula NGC 1068, the spectrum 
was composite showing emission and absorption lines.
Later a higher quality and better resolution spectrum of NGC 1068 
obtained by Slipher \cite{1917v3p59} 
at Lowell Observatory confirmed the presence of emission and absorption lines. 
Slipher also observed the emission lines are broad and spread over a substantial
range of frequencies. Seyfert \cite{1943v97p28} was the first to do a systematic study of such 
galaxies with strong nuclear emission and broad lines. He
attributed the broadness of emission lines to Doppler shifts and obtained a
maximum velocity spread of $8500$ km s$^{-1}$ for the hydrogen lines of NGC 3516 and
NGC 7469.
This class of objects with broad emission lines and a luminous core was later 
called as Seyfert galaxies (the most numerous type of AGN known).

Though the observation of AGN began as early as the beginning of the twentieth century, 
a major study of these objects started only after the advent of radio astronomy 
initiated by Karl Jansky \cite{1933v132p66}, 
a radio engineer working at Bell Telephone Laboratories. Later 
Grote Reber \cite{1944v100p279}, a radio engineer working on radio astronomy during 
his spare time, published a map of the radio sky at $160$ MHz using his 31 feet 
reflector in his backyard. The map included the bright source located in the constellation
Cygnus which we know at present as an AGN and commonly called as Cygnus A. 
In 1951, Smith \cite{1951v168p555} used radio 
interferometry and obtained accurate positions of the radio sources Taurus A, 
Virgo A, Cygnus A and Cassiopeia A, which led Bade and Minkowski \cite{1954v119p206} 
to identify the latter two in the optical band. The observed emission lines of these 
sources were
very similar to that of Seyfert galaxies and radio sources with this characteristic 
feature in their spectrum are now identified as radio galaxies. 
There was also a considerable progress in the study of the radio source structures 
during this period. 
Jennison and Das Gupta \cite{1953v172p996} used 
radio interferometry to
study the structure of Cygnus A which showed two bright components separated by $\sim 1.5$
arcmin. This morphology was observed to be common for extragalactic radio sources.
Subsequent observations at GHz frequencies provided better angular resolution and 
revealed the presence of a compact nucleus
often associated with 
extended radio sources. 
In addition to these radio galaxies, some star-like radio sources were also detected.
Their spectra
were continuous with broad emission lines and no absorption lines. 
These peculiar stars were then called as ``radio stars''. Many attempts were made 
to understand them as an old novae or white dwarfs (\cite{1962v74p406} and 
references therein) until Schmidt \cite{1963v197p1040} 
identified the lines of 3C 273 as nebular emission lines with redshift 
$z=0.158$ and Greenstein and Matthews \cite{1963v197p1041} identified emission
lines in 3C 48 with redshift $z=0.367$ (the largest redshift known at that time).     
It was clear then that radio stars are highly luminous sources situated at large
distances. These radio stars are today called as ``quasi-stellar radio sources'' or simply 
``quasars'' (a term coined by Chinese-born U.S. astrophysicist Hong-Yee Chiu 
\cite{1964v17p21}). Seyfert galaxies, radio galaxies, quasars and objects of similar 
type are all collectively termed as \emph{Active Galactic Nuclei (AGN)}. 
Quasars are the distant AGN for which the host galaxy cannot be resolved.

Janksy \cite{1935v23p1158} suggested that the origin of the observed radio emission 
may be (a) from the stars or (b) secondary emission from the atmosphere due to the 
interaction of high energy particles emitted by the stars or (c) thermal emission 
from the interstellar dust. Former two options suggested maximum radio intensity 
from the direction of the sun since it being the closest star. However such an excess
from the direction of the sun was not observed and hence these options are not viable.
Meanwhile, Whipple and Greenstein \cite{1937v23p177} calculated the interstellar 
dust temperatures and it was found that they are too low to produce the radio 
intensity observed by Jansky.  
Reber \cite{1940v28p68} suggested the free-free emission by ionised gas in the 
interstellar medium as a plausible mechanism for the observed radio emission. However, 
Henyey and Keenan \cite{1940v91p625} and Townes \cite{1947v105p235} showed that 
even this cannot 
reconcile the required Jansky's brightness temperature. These studies thereby ruled out the 
possibility of the radio emission from the interstellar dust. 
Finally in 1950, Kiepenheuer 
\cite{1950v79p738} explained the galactic radio background in terms of synchrotron radiation
by cosmic rays in the galactic magnetic field. By the end of 1950's, the synchrotron 
theory got accepted for explaining the radio emission from the extragalactic radio 
sources. Later the power-law\footnote{We define a power-law flux as 
$F_\nu \propto \nu^{-\alpha}$ (ergs cm$^{-2}$ s$^{-1}$ Hz$^{-1}$)
where $\nu$ is the observed photon frequency and $\alpha$ the power-law spectral index.}
nature of the spectra and high degree of polarisation 
observed also supported this theory.

In 1959, Burbidge \cite{1959v129p849} used synchrotron theory and found that the 
minimum energy content 
of the radio sources is extremely large $\sim 10^{60}$ ergs. Burbidge attributed 
this large energy to a chain of supernovae explosions occurring in a tight packed 
set of stars at the nuclear region of the galaxy. Such a situation is plausible since 
the nuclear region is expected to be much denser than the average density of the galaxy. 
In 1962, Hoyle and Fowler \cite{1963v125p169} suggested that such close packed stars 
can as well form a single super massive object. Accordingly they proposed the 
existence of a super massive star at the nuclear region of the galaxy and conjectured 
that it would evolve into a `super-supernovae' providing the 
required energy. However the thermonuclear process is less efficient in converting mass 
into energy compared with the gravitational one for the masses of this order. Again, Hoyle 
and Fowler \cite{1963v197p533} proposed in their 
pioneering paper in Nature in early 1963, that the energy of the radio sources was of
gravitational origin derived from the slow contraction of a super massive object under its
own strong gravitational field. 
This theory of a collapsed super massive object powering
the AGN is widely accepted at present. We refer to this object as \emph{Central Engine}
hereafter.

Today AGN are detected up to $\gamma$-ray energies and their 
structures are studied even at X-ray energies (see Chapter 
\ref{chap:agn_knots}). AGN are first detected in $\gamma$-rays 
by satellite based experiments followed by ground based experiments 
using atmospheric Cherenkov techniques. 3C 273 was the first 
AGN detected in $\gamma$-ray (MeV) by \emph{Cos-B}
\footnote{\emph{Cos-B} was an European Space Research Organisation 
satellite mission to study $\gamma$-ray sources launched by \emph{NASA}.}
in 1978 \cite{1978v275p298S}. Later \emph{EGRET} 
\footnote{Energetic Gamma-Ray Experiment Telescope (\emph{EGRET}) 
is a satellite borne $\gamma$-ray telescope onboard Compton Gamma-Ray Observatory (\emph{CGRO})
launched by \emph{NASA}.} 
operating at MeV-GeV $\gamma$-ray energy range detected around 60 AGN
during 1991-2000 \cite{1999v123p79}. Around this period the ground based 
atmospheric Cherenkov telescopes (ACT) were operational with the detection of MKN 421 in 
TeV $\gamma$-ray by \emph{Whipple} 
\footnote{Whipple is located at the Fred 
Lawrence Whipple Observatory in Southern Arizona, USA.}
in 1992 \cite{1992v358p477}. Later imaging atmospheric Cherenkov telescopes (IACT) 
(e.g. \emph{HEGRA}
\footnote{High-Energy-Gamma-Ray Astronomy (\emph{HEGRA}) telescope is 
located at Roque de los Muchachos Observatory on La Palma.}, 
\emph{TACTIC}
\footnote{TeV Atomospheric Cherenkov Telescope with Imaging Camera (\emph{TACTIC}) is 
located at Mt. Abu, Rajasthan, INDIA.},
\emph{CAT}
\footnote{Cherenkov Array at Themis (\emph{CAT}) is located at Themis, France}, 
\emph{CANGAROO}
\footnote{Collaboration of Australia and Nippon (Japan) for a GAmma Ray 
Observatory in the Outback (\emph{CANGAROO}) is located at Woomera, Australia.} 
etc.) and wavefront sampling Cherenkov telescopes 
(e.g. 
\emph{CELESTE}
\footnote{CErenkov Low Energy Sampling and Timing Experiment (\emph{CELESTE}) is 
located at Themis, France}, 
\emph{PACT}
\footnote{Pachmarhi Array of Cherenkov Telescopes (\emph{PACT})  is 
located at Pachmarhi, Madhya Pradesh, INDIA}, 
\emph{STACEE} 
\footnote{Solar Tower Atmospheric Cherenkov Effect Experiment (\emph{STACEE}) is 
located near Albuquerque, New Mexico.}
etc.) detected 6 AGN at TeV $\gamma$-ray energies \cite{2002v384p56}. All 
AGN detected in $\gamma$-ray energies during this period are blazars (a class of AGN with 
a jet of matter flowing at relativistic speed towards the 
observer (\S \ref{sec1:unification} and \S \ref{sec1:classification})) except the 
radio galaxy Cen A which was detected by \emph{EGRET}. Presently the satellite based 
experiment \emph{Fermi}\footnote{Fermi Gamma-ray Space Telescope is a satellite 
based $\gamma$-ray telescope. The mission is a joint venture of NASA, the United 
States Department of Energy, and government agencies in France, Germany, Italy, 
Japan, and Sweden.}
operating at GeV-TeV $\gamma$-ray energies detected around 600 AGN 
till now which include blazars and radio galaxies \cite{2010v715p429}.
The second generation IACT namely 
\emph{MAGIC}\footnote{Major Atmospheric Gamma-ray Imaging Cherenkov Telescope 
(\emph{MAGIC}) is a system of two IACT situated at the Roque de los Muchachos 
Observatory, La Palma},
\emph{HESS} \footnote{High Energy Stereoscopic System (\emph{HESS}) is a 
stereoscopic IACT located at Namibia.}
and 
\emph{VERITAS} \footnote{Very Energetic Radiation Imaging Telescope Array System
(\emph{VERITAS}) is an array of four IACT located at Fred Lawrence Whipple 
Observatory in southern Arizona, USA.}
operating at GeV-TeV $\gamma$-ray 
energies detected around 40 AGN which again include blazars and radio 
galaxies\footnote{http://www.mppmu.mpg.de/~rwagner/sources/}. The number 
of AGN detected in $\gamma$-ray energy range are likely
to increase with the help of these experiments and the future experiments 
(\emph{MACE}\footnote{Major Atmospheric Cerenkov Telescope Experiment 
(\emph{MACE}) is an upcoming stereoscopic IACT at Hanle, India.}
and 
\emph{CTA}\footnote{Cherenkov Telescope Array (\emph{CTA}) is a proposed
open observatory and will consist of two arrays of IACT with one array 
at northern
hemisphere and a second array at southern hemisphere.}
).
Many of the AGN detected by \emph{Fermi} and second generation IACT along with 
simultaneous observation at radio, optical and X-ray energies cannot be 
explained with our present understanding and the problems 
regarding AGN physics are still open \cite{2009v703p1168}.

\subsection{Morphology} 
\begin{figure}[tp]
\begin{center}
\includegraphics[width=140mm,bb=0 0 192 129]{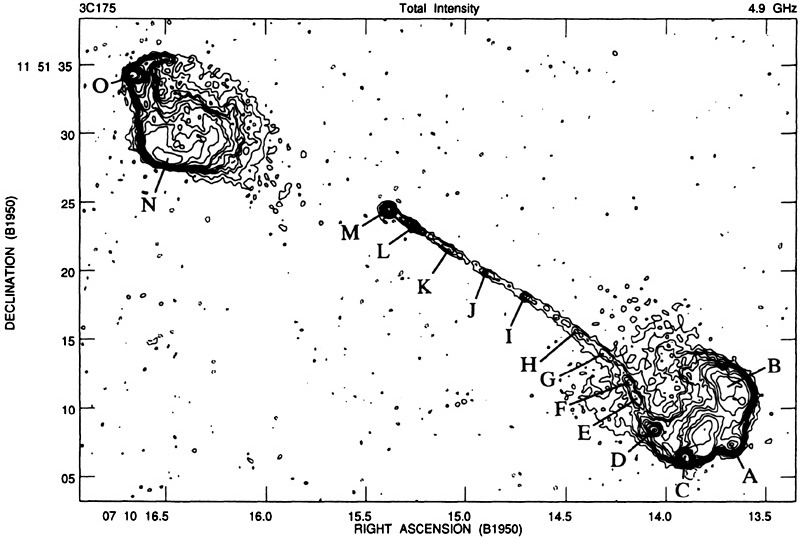} 
\caption[VLA image of the AGN 3C 175]{Very Large Array (VLA) map of the AGN 3C 175 at 
4.9 GHz. M is the core, C and O are the hotspots and D to L are the knots. Figure
reproduced from Bridle et al. \cite{1994v108p766}}
\label{fig1:3c175}
\end{center}
\end{figure}

AGN comes in a variety of morphological structures and sizes. At one extreme we find an
unresolved compact luminous core and in the other end we have, complex structures extending 
up to hundreds of kiloparsec (kpc, 1pc = $3.0857\times10^{18}$ cm). However one can develop 
a primary morphological feature
of an AGN based on the commonly observed characteristics (Figure \ref{fig1:3c175}).

\begin{itemize}
\item \emph{Core:} These are compact unresolved luminous radio components coinciding 
with the nucleus of the associated galaxy (if resolved). They mostly have a spectrum which
is a flat power-law with an index $\alpha<0.5$. Cores are found in almost all quasars 
and in nearly 80 per cent of all radio galaxies.
\item \emph{Lobes:} These are two extended regions of radio emission located on opposite 
sides of the galaxy or the nuclei. Lobes can be separated by several hundred 
kpc or in some extreme cases even up to a few megaparsec (Mpc) 
(e.g. 3C 236 has an overall 
size of $4$ Mpc). They have a steep power-law spectrum with index 
$\alpha > 0.5$. Within the lobes one often finds local intensity maxima commonly
called as \emph{hotspots}.
\item \emph{Jets:} These are narrow features that connect the compact core to the outer
regions commonly referred to \emph{radio jets}. They extend from pc scale to 
kpc scales. Radio jets often have bright regions along its length which are called 
as \emph{radio knots} or simply \emph{knots}. We shall consider the jets in detail in 
\S \ref{sec1:agnjet}.
\end{itemize}

\subsection{Classification}\label{sec1:classification}
AGN are broadly classified into two groups, namely \emph{radio loud} and
\emph{radio quiet} based on the ratio of their radio luminosity 
at 5 GHz to the optical luminosity at the B-band. 
Conventionally the demarcating value of this ratio is taken around 
$L_{5 \text{GHz}}/L_{B}\approx 10$ and roughly 15-20\% of AGN are radio loud.

\begin{figure}[tp]
\begin{center}
\includegraphics[width=135.46mm,bb=0 0 461 321]{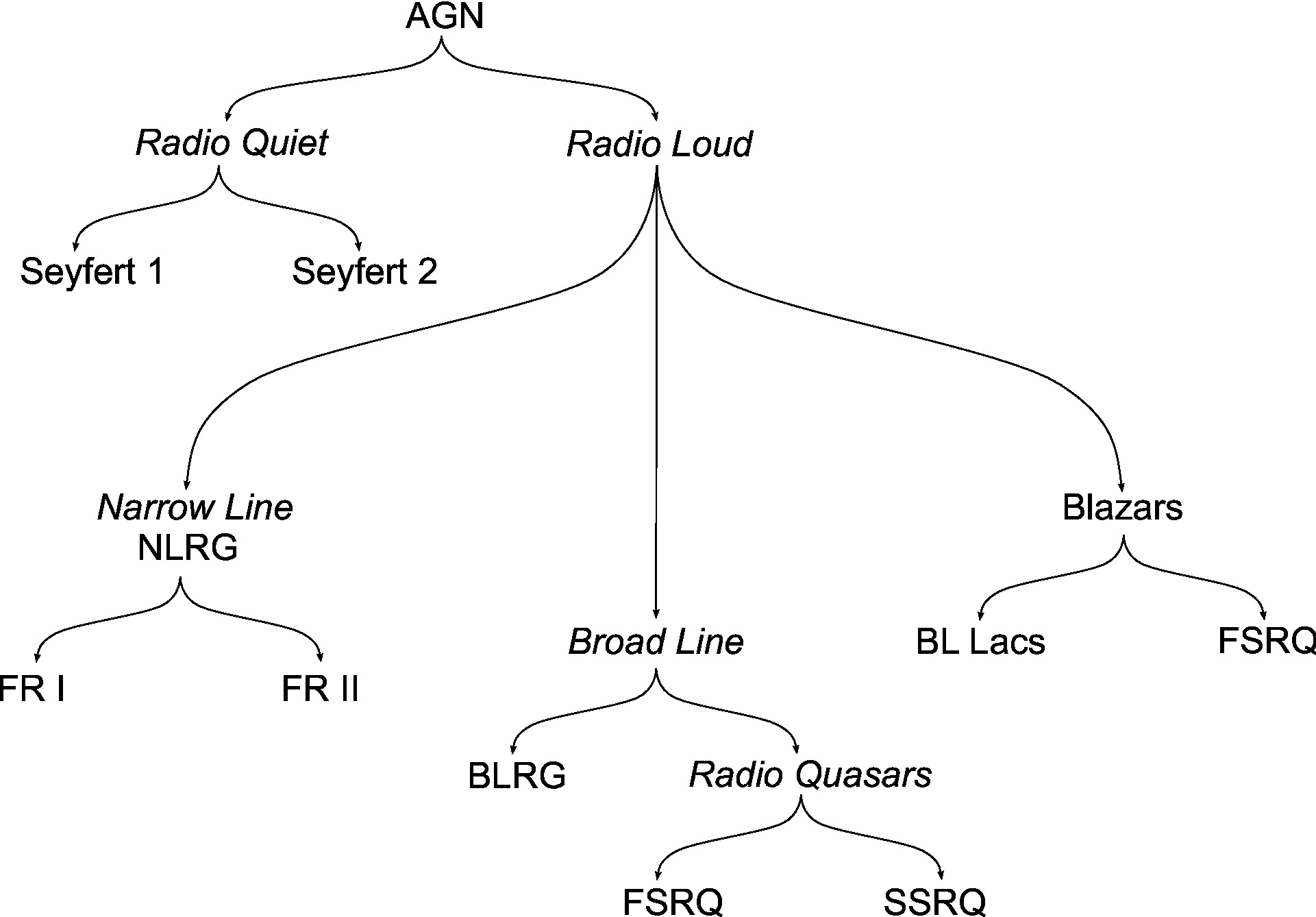} 
\caption{AGN Classification} 
\end{center}
\end{figure}

\subsubsection{Radio Quiet AGN}
The commonly observed radio quiet AGN are the \emph{Seyfert} galaxies which have a 
morphology of spiral galaxies. They are classified based on their optical/UV properties.
\begin{itemize}
\item \emph{Seyfert Type 1:} These are the Seyfert galaxies having a bright star-like
nucleus and emit a strong continuum from far infrared to the X-ray band. Their
spectrum contain broad emission lines with a width of the order of few thousand
kilometers per second.

\item \emph{Seyfert Type 2:} These are the Seyfert galaxies with weak continuum and 
narrow emission lines with a line width of the order of few hundred kilometers per
second.  
\end{itemize}

Seyfert galaxies with properties intermediate between type 1 and type 2 are 
classified as types 1.2, 1.5 and so on.  
Although Seyfert galaxies are categorized as radio quiet sources there also exists 
radio loud Seyfert galaxies detected even up to $\gamma$-ray energies \cite{2009v707p142}.

\subsubsection{Radio Loud AGN}
Radio loud AGN are classified based on their morphology and optical/UV  properties.
\begin{itemize}
\item \emph{Fanaroff-Riley I (FR I) radio galaxies:} These are extended 
sources often having morphology with core, jet  and lobes. Their jets are often 
symmetric and become 
fainter as one approaches the outer extreme. Due to this feature they are called as
\emph{edge-darkened} sources. Their spectrum contain narrow emission lines features.

\item \emph{Fanaroff-Riley II (FR II) radio galaxies:} These are more luminous than 
FR I radio galaxies with knotty jet and lobes with hot spots. In contrast to FR I 
type, their intensity falls off towards the nucleus and hence they are called as 
\emph{edge-brightened} sources. Jets of these types of sources are well collimated
and often one sided. Their spectrum contain narrow emission lines features.

\emph{FR I} and \emph{FR II} galaxies are often elliptical and together are referred to 
as \emph{Narrow Line Radio Galaxies (NLRG)}.

\item \emph{Broad Line Radio Galaxies (BLRG):} These sources have a continuum and emission
lines resembling those from Seyfert 1 galaxies.  

\item \emph{Radio Quasars:} These sources are distinguished from BLRG based on the 
luminosity and have a luminous nucleus which outshines the light from the
host galaxy. These sources are often observed to have one sided jets with 
superluminal knots (see \S \ref{sec1:relativisticjet}). They are further
classified into \emph{Steep Spectrum Radio Quasars (SSRQ)} and 
\emph{Flat Spectrum Radio Quasars (FSRQ)} depending on their radio spectral index is
either  $\alpha \gtrsim 0.5$ or $\alpha \lesssim 0.5$.

\item \emph{BL Lacs:} These sources have strong nuclear continuum with high polarisation and
rapid flux variability. Their continuum extends from radio to $\gamma$-ray energies with 
few of them detected at TeV energies \cite{2002v384p56}. Their emission lines are 
absent or weak.
\end{itemize}

BL Lacs and FSRQ are collectively called as \emph{Blazars} since they share many 
common properties like, rapid variability, high and variable polarisation, superluminal 
motion etc.

\subsection{Unification}\label{sec1:unification}
\begin{figure}[tp]
\begin{center}
\includegraphics[width=128.58mm, height=93.25mm,bb=0 0 546 397]{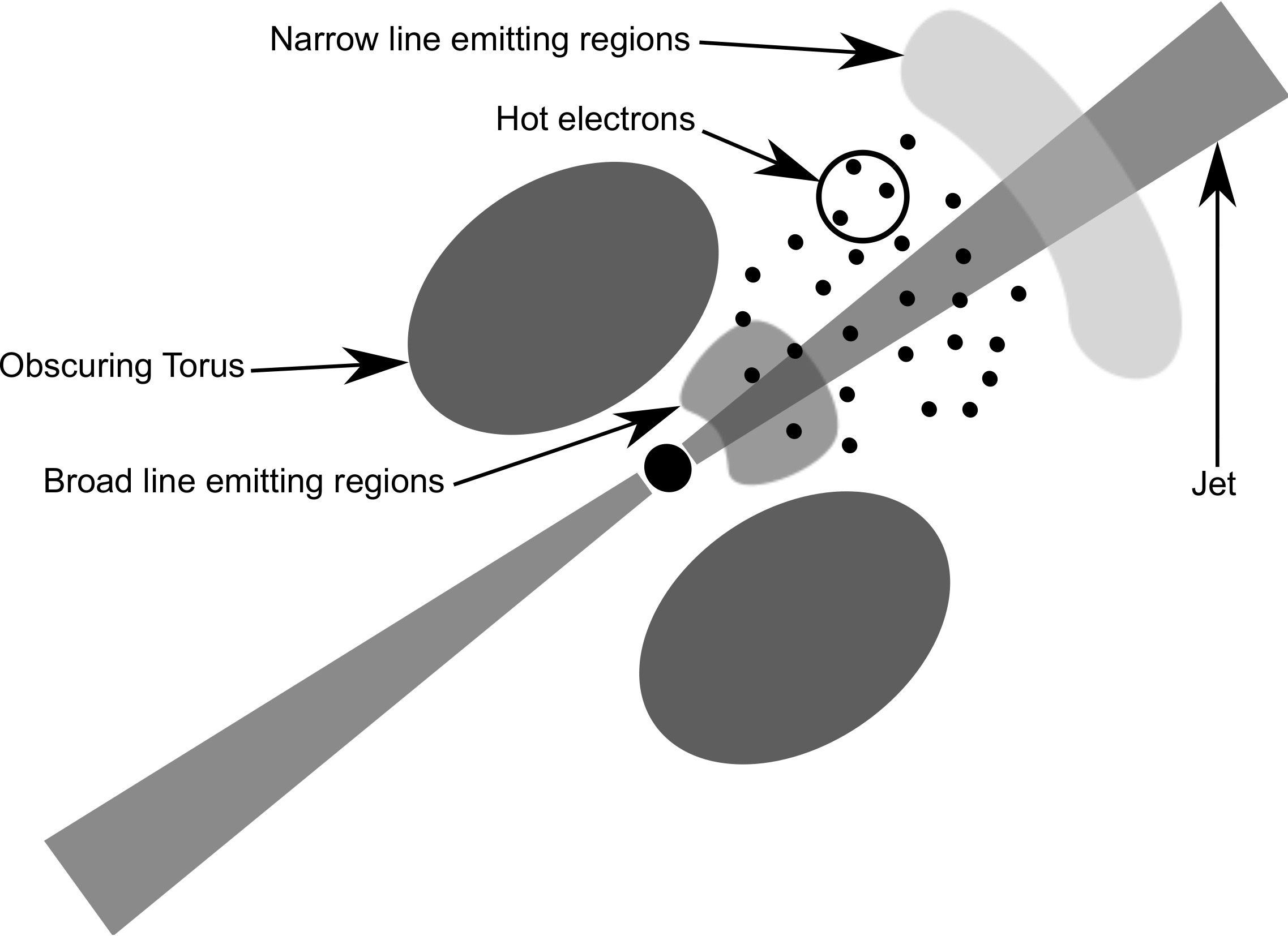} 
\caption[Schematic diagram of an AGN]{Schematic of an AGN based on Unification hypothesis (not to scale)}
\label{fig1:agnsche}
\end{center}
\end{figure}

The idea of unifying the different classes of AGN as a variant of objects
belonging to a single population emanated from the fact that they share many common
observational features. 
The most successful theory in AGN unification is based on the \emph{Orientation hypothesis},
which assumes that the observed differences between the different classes of AGN are due
to different orientations of similar objects with respect to the line of 
sight \cite{1993v31p473,1995v107p803} as shown in Figure \nolinebreak \ref{fig1:agnsche}. 

The unification of \emph{Seyfert 1} and \emph{Seyfert 2} is based on the fact that  
broad emission lines are present in the polarised spectra of the latter.
Hence the unified picture derived for these sources assume that the broad line emitting
regions are located closer to the central engine and a ``dusty torus'' is wrapped 
around the central region. Consequently the broad emission lines are hidden by the torus 
when the source is viewed edge-on (i.e. when the angle between the line of sight and the 
axis of 
the torus is large). 
On the other hand, narrow line emitting regions are farther from the central engine 
and hence are not obscured. Also, the unified picture assumes the presence of hot
electrons far from the central region which are believed to reflect the broad lines 
towards the observer 
when the source is viewed edge-on. These reflected lines will therefore be visible in the 
polarised spectrum. Considering the above mentioned picture of the source, the edge-on view 
will resemble as \emph{Seyfert 2} while the face-on 
view as \emph{Seyfert 1} \cite{1993v31p473}.
Also the unified picture of AGN contains twin relativistic jets emanating from the 
central engine, normal to the obscuring torus. 
High luminosity sources, \emph{Quasars} and \emph{FR II} 
are considered to be similar 
objects with the jets of the former aligned close to the line of sight (angle $<14^o$) and
thereby causing relativistic beaming. Similarly, \emph{BL Lacs} are the aligned jet 
version of \emph{FR I} (low luminosity sources) \cite{1995v107p803}.

\section{Jets of Active Galactic Nuclei}
\label{sec1:agnjet}

The bridge of radiation connecting the central compact source with the extended 
lobes of the AGN is called as a ``Jet''. 
They are the common feature observed in most of the AGN and are either
\emph{two-sided} or \emph{one-sided}. They may extend 
from pc to kpc scales and are visible in radio, optical
and X-rays. Jets are interpreted as conduits for the transport of
energetic particles from the nucleus to the extended radio structures
at relativistic velocities.

\subsection{Evidence for Relativistic flow} \label{sec1:relativistic}
\label{sec1:relativisticjet}

\subsubsection{Superluminal Motion}
\begin{figure}[tp]
\begin{center}
\includegraphics[width=101.98mm, height=25.71mm,bb=0 0 346 88]{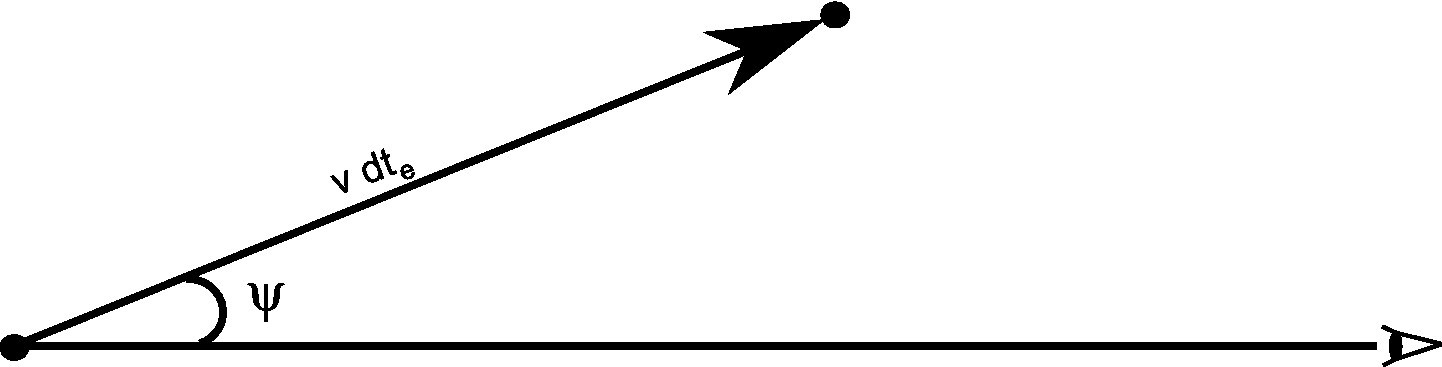} 
\caption[Superluminal motion - Representation]{A radiating blob moving at a 
relativistic speed $v$ at an angle $\psi$ with respect to 
the line of sight of the observer}
\label{fig1:suplum}
\end{center}
\end{figure}

\begin{figure}[tp]
\begin{center}
\includegraphics[width=120mm,bb=0 0 708 499]{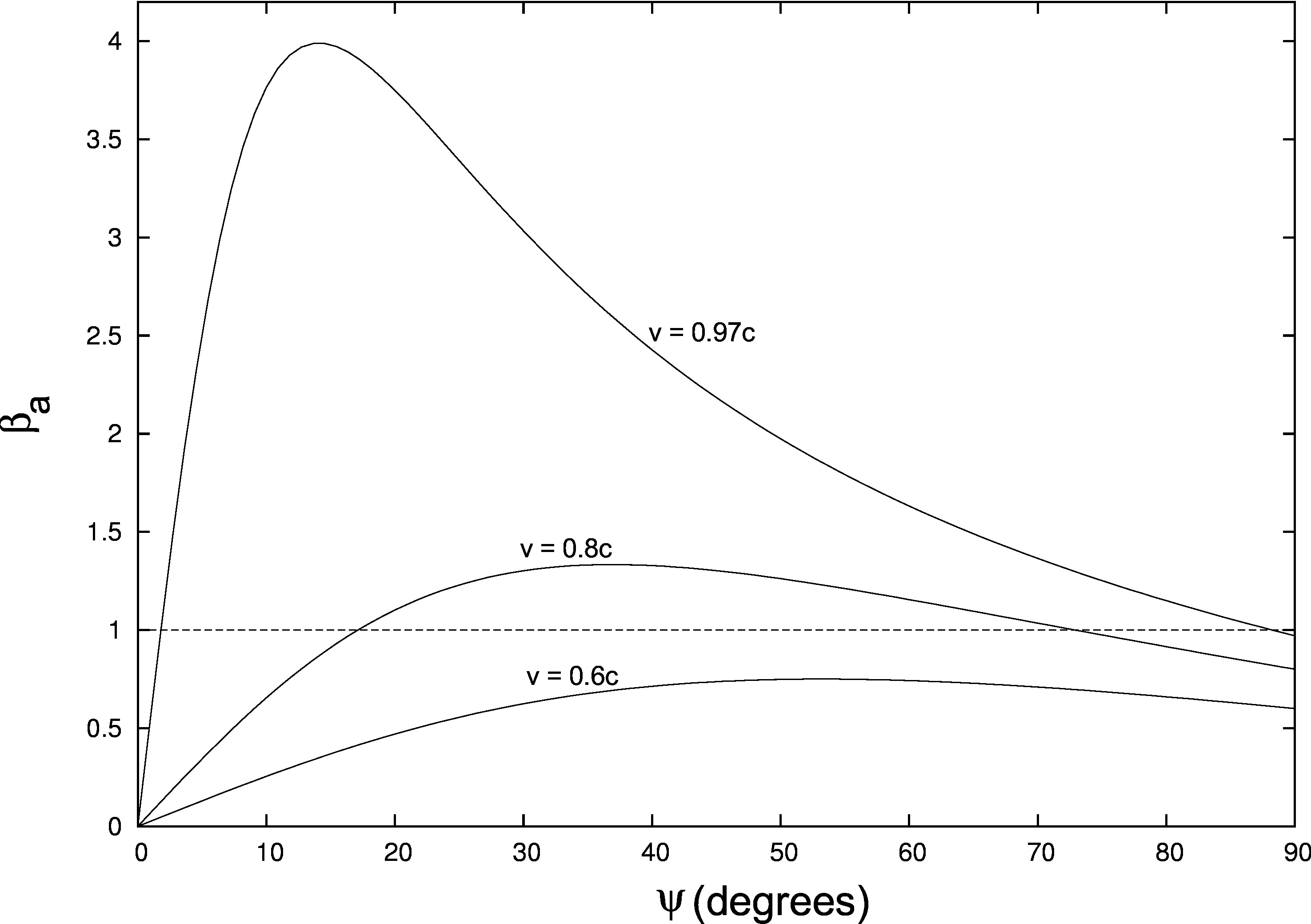} 
\caption[Superluminal motion - Plot of apparent velocity versus viewing angle]
{Apparent velocity of a relativistically moving blob (in units of $c$) with 
respect to the viewing angle of the observer. Each curve represents blob with
actual velocity $0.6c$, $0.8c$ and $0.97c$.}
\label{fig1:apparent}
\end{center}
\end{figure}

For many AGN, jet components are observed to move with velocities greater than
the velocity of light and this phenomenon is referred to \emph{superluminal} motion
\cite{1993v407p65}. Rees \cite{1966v211p468} interpreted this phenomenon as a result 
of bulk relativistic flow at an angle close to the line of sight.
To understand this, let us consider a blob ejected from a stationary source with a velocity $v$ 
at an angle $\psi$ with respect to the line of sight of the observer as shown in 
the Figure \ref{fig1:suplum}. Let the blob
emit signals for a duration $dt_e$, measured from the frame of the stationary 
source. The distance travelled by the blob during this time is $v\;dt_e$. 
Due to the inclined motion, the signal emitted by the blob at the end of the duration $dt_e$,
travel lesser distance compared to the one emitted at the beginning of $dt_e$ for 
a distant observer. Hence the observer will measure this interval as
\begin{align}
dt_o = dt_e\left(\;1-\frac{v}{c}\,\cos \psi\; \right)
\end{align}
where $c$ is the velocity of light. Since the projected distance travelled by the blob 
in the sky plane is $v\,dt_e\,\sin\psi$, the 
apparent velocity measured by the observer will be
\begin{align}
v_a &= \frac{v\;dt_e\;\sin\psi}{dt_o} \nonumber \\
 &= \frac{v\,\sin\psi}{1-\frac{v}{c}\,\cos\psi} 
\end{align}
or
\begin{align}
\label{eq1:suplum}
\beta_a = \frac{\beta\, \sin\psi}{1-\beta\, \cos\psi}
\end{align}
where $\beta_a=v_a/c$ and $\beta = v/c$ are dimensionless
velocities. In Figure \ref{fig1:apparent}, we plot the dependence of $\beta_a$
with the viewing angle $\psi$ for different relativistic motion of the blob.
It is evident from the plot that the apparent speed can 
exceed $c$ for the blobs with relativistic velocities and moving 
closer to the line of sight. 

\subsubsection{Gamma-ray transparency}\label{sec1:gamma_trans}
High energy $\gamma$-rays can interact with low energy photons to produce
electron-positron pairs. The cross-section for this process is maximum when
\begin{align}
\epsilon_\gamma \epsilon_x \sim 2\,(m_e\, c^2)^2
\end{align}
where $\epsilon_\gamma$ and $\epsilon_x$ are the energy of the high energy
and the low energy photons respectively and $m_e$ is the electron mass. For
example, a $100$ MeV photon can pair produce with $5$ keV X-ray photon. In general, 
the transparency of a medium to a radiation is described by its optical depth.
Optical depth of a medium corresponding to a radiation of frequency $\nu$ is 
defined as 
\begin{align}
\tau_\nu(s)= \int\limits_{s_0}^s \alpha_\nu(s^\prime)\;ds^\prime 
\end{align}
where $s_0 \rightarrow s$ is the path of light travel and $ \alpha_\nu$ is the 
absorption coefficient.
The medium is said to be \emph{optically thick} (or \emph{opaque}) when $\tau>1$
and \emph{optically thin} (or \emph{transparent}) when $\tau<1$. 
The optical depth
$\tau$ to pair production for a $\gamma$-ray of energy 
$\epsilon_\gamma$ in a homogeneous region of size $R$ can be approximated
as \cite{book:khembhavi_narlikar}
\begin{align}
\tau(\epsilon_\gamma)\simeq\frac{0.2\,\sigma_T\,L(2\epsilon_x)}{4\pi \,Rc}
\end{align}
where $L(2\epsilon_x)$ is the luminosity of the target photon at energy 
$2\epsilon_x$ and $\sigma_T$ is the Thompson cross-section. If the power-law
spectra of AGN and quasars in the X-ray region extend to much higher energies,
then the source may be opaque to $\gamma$-rays if $\tau(\epsilon_\gamma)>1$.
This situation can be conveniently stated in terms of \emph{compactness
parameter} ($l$) defined as \cite{1983v205p593}
\begin{align}
\label{eq1:compactness}
l = \frac{L}{R}\;\frac{\sigma_T}{m_e\,c^3} 
\end{align}
where $L$ is the total luminosity produced in the region of size $R$. 
The opacity condition, $\tau>1$, can now be translated in terms of compactness parameter
as $l>60$.
Estimation of compactness parameter requires the knowledge of 
X-ray luminosity and the size of the emission region. The latter can be 
estimated from the observed time variability of the flux as
\begin{align}
R\sim c\, t_{var}
\end{align}
where $t_{var}$ is the variability time-scale. The inferred values of $l$
for blazars from the observed X-ray luminosities are much larger 
than $60$. For 3C 279, $l\sim5000$ and for PKS 0528+134, $l\sim15000$.
However \emph{EGRET}
mission detected many blazars at energies greater 
than 100 MeV. In order to observe the $\gamma$-rays of these energies, the
actual luminosity $L$ of the source in its proper frame must be much smaller 
than the observed value and the actual size should be larger than the inferred 
value in order to fulfill the condition $l<60$. Indeed, this can happen if the 
source is moving at relativistic speed towards the observer. The relativistic
beaming effects will enhance the luminosity as \cite{1984v56p255}
\begin{align}
L_{obs}=\delta^4\, L_{int}
\end{align}
and the size of the emission region will be reduced as
\begin{align}
R\sim \frac{c\,t_{var}}{\delta}
\end{align}
where $L_{obs}$ and $L_{int}$ are the observed and the intrinsic luminosity
and the relativistic Doppler factor $\delta$ is given by
\begin{align}\label{eq1:dopplerfact}
\delta = \frac{1}{\Gamma(1-\beta\, \cos\psi)}
\end{align}
Here, $\Gamma= (1-\beta^2)^{-1/2}$ is the bulk Lorentz factor, 
$\beta=v/c$ is the dimensionless velocity and $\psi$ is the angle
between the line of sight of the observer and the direction of motion of the 
source. The compactness parameter $l$ in such a case will reduce to 
\begin{align}
l = \delta^{-5}\;\frac{L_{obs}}{t_{var}}\;\frac{\sigma_T}{m_e\,c^4}
\end{align}
Dondi and Ghisellini \cite{1995v273p583} estimated the minimum value 
of the Doppler factor, corresponding to the optical depth $\tau=1$, for 
the $\gamma$-ray bright blazars detected by \emph{EGRET}. Their inferred 
minimum values of $\delta$ were spread within $1.3$ to $11.3$.

\subsection{AGN Jet Features}
\subsubsection{Lobes}
Lobes are the regions where the jets terminate and release 
their energy and momentum into the ambient intergalactic medium. The radio
spectrum of the lobes are steep with index  $\alpha>0.5$ and the emission 
is usually polarised. This suggests the lobes to be an optically thin 
synchrotron sources driven by efficient cooling of relativistic particles. 
They contain enhanced emission regions known as \emph{hot spots}
which are often collinear with the central core (see Figure \ref{fig1:3c175}).
In low resolution radio maps of the lobes, the presence of hot spots will 
make it appear as an \emph{edge-brightened} source.
The radio spectrum of the hot spots are
flatter than the lobe spectrum with the index in the range 
$\alpha \sim 0.5-1$. This suggests the hot spots as the location where the
jet hit the ambient medium and the bulk kinetic energy of the beam is 
converted into the random energy through the shock formed at the 
collision (see \S \ref{sec3:shock_accln}). The shock accelerated energetic 
particles diffuse from hotspot to the lobes thereby providing a continuous supply
of energy.
  
\subsubsection{Knots}
\begin{figure}[tp]
\begin{center}
\includegraphics[width=141.1mm, height=122.81mm,bb=0 0 400 348]{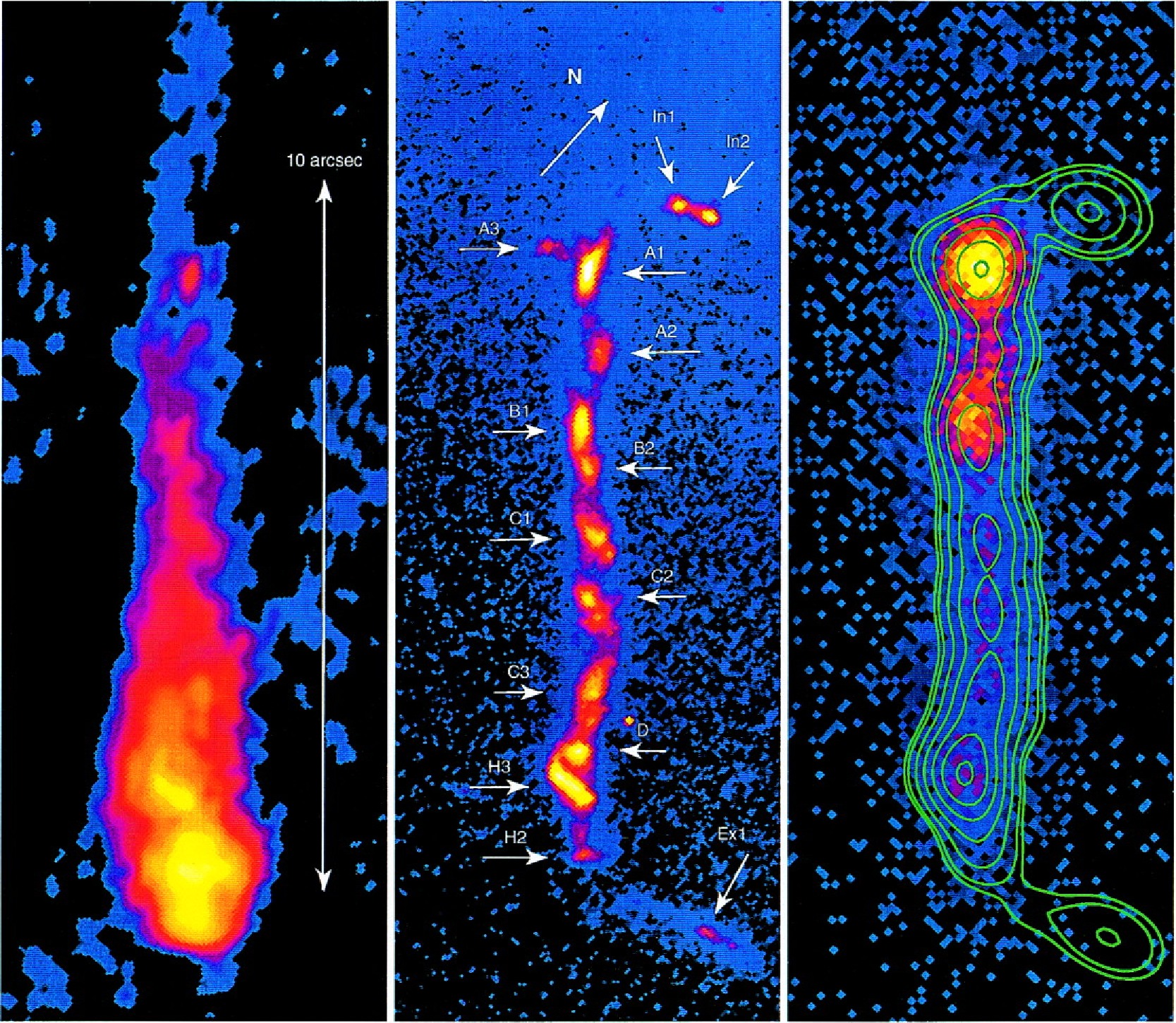} 
\caption[Multi wavelength image of 3C 273]{Image of 3C 273 with knots. Radio image by \emph{MERLIN} (left), 
optical image by \emph{Hubble Space Telescope} (middle) and X-ray image by \emph{Chandra}
with optical contour overlaid. Figure reproduced from Marshall et al. \cite{2001v549p167}}
\label{fig1:3c273}
\end{center}
\end{figure}

These are the bright regions observed along the length of the AGN jet
(Figure \ref{fig1:3c273}).
The emission from these regions are strongly polarised and the polarisation
angle is often perpendicular to the jet axis. 
Knots are seen in radio, optical and X-ray images of AGN jets and many 
show
\emph{superluminal} motion. The non-thermal (power-law) nature
of the knot spectrum along with the observed polarisation suggests that the 
low-energy(radio-to-optical) emission must be synchrotron radiation emitted
by relativistic distribution of charged particles cooling in a magnetic 
field. However, the X-ray emission from the knots can be an extension of 
synchrotron spectrum itself or the inverse Compton 
radiation emitted via scattering of soft target photons by relativistic 
electrons (see chapter \ref{chap:agn_knots}).
The target photons for the inverse Compton scattering can be either synchrotron 
photons itself or the photons external to the jet.
The enhanced brightness of the knots is often 
interpreted as a result of efficient 
acceleration of charged particles probably by a shock present in the jet.

\subsubsection{Jet Asymmetry}
\begin{figure}[tp]
\begin{center}
\includegraphics[width=144mm, height=59.3mm,bb=0 0 613 256]{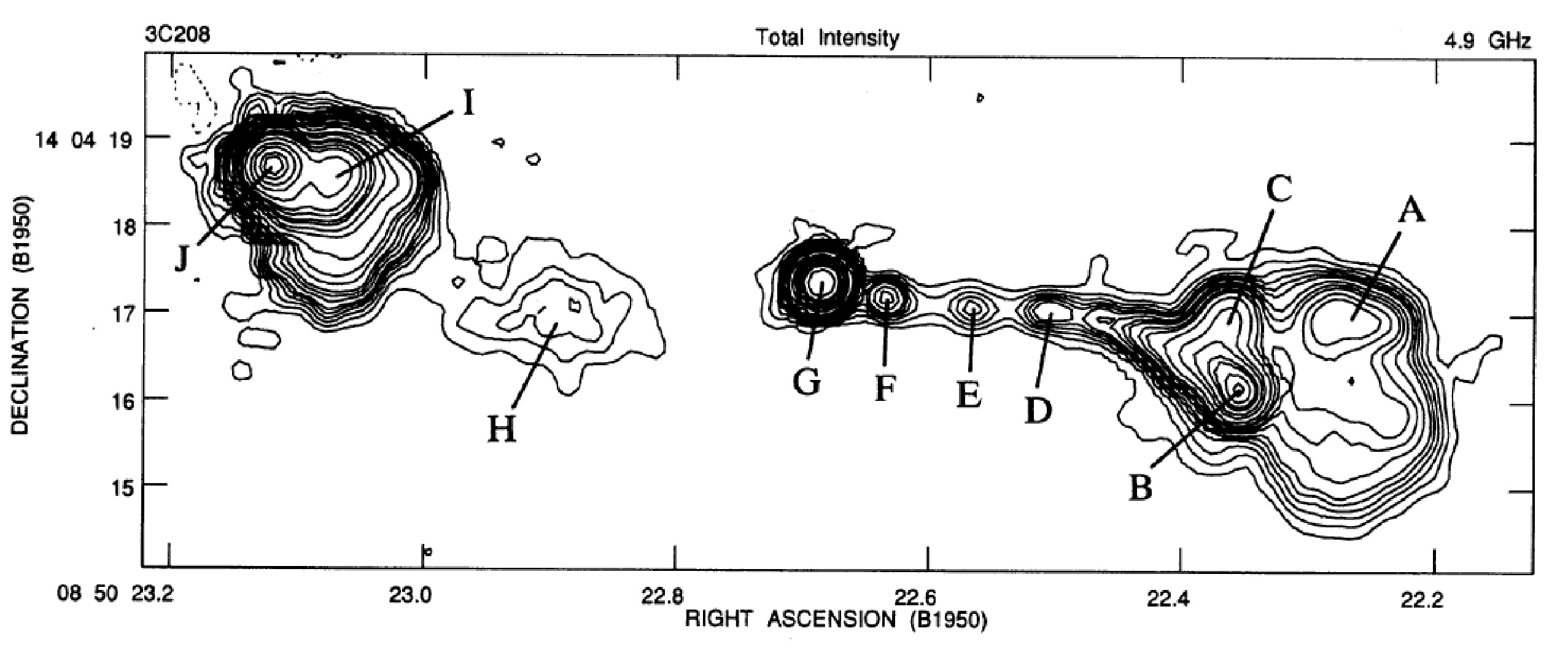} 
\caption[VLA image of 3C 308]{VLA image of 3C 308 at $4.9$ GHz. Figure reproduced from 
Bridle et al. \cite{1994v108p766}}
\label{fig1:3c308}
\end{center}
\end{figure}

One of the intriguing characteristics of the AGN jets is that they are often 
observed to be one sided (Figure \ref{fig1:3c308}). The straightforward way 
to interpret this asymmetry is to relate it to
relativistic beaming, since there are enough evidences that the jets are 
relativistic. If we assume that the central engine produces two similar 
jets ejected in 
opposite directions, then the brightness of the jet moving towards the 
observer will be enhanced due to relativistic beaming. Whereas the counter-jet 
will be dimmed (or invisible) since it moves away from the observer.
If $F_{adv}$ is the flux of the advancing jet and $F_{rec}$ is the flux of the 
receding one, then the jet/counter-jet flux ratio $J$ can be 
written as \cite{1993v407p65}
\begin{align}
J=\frac{F_{adv}}{F_{rec}}= 
\left(\frac{1+\beta \,\cos \psi}{1-\beta\, \cos \psi}\right)^p
\end{align}
where $\beta$ is the bulk velocity of the jet in units of $c$ and $\psi$ is 
the angle between the jet direction and the line of sight of the observer.
If $\alpha$ is the power-law spectral index of the intrinsic jet flux, then the
index $p$ can be either $2+\alpha$ in case of a continuous jet flow or $3+\alpha$ for a 
moving isotropic source. Using equation (\ref{eq1:suplum}), $J$ can be expressed in 
terms of  superluminal velocity ($\beta_a$) as 
\begin{align}
J=(\,\beta_a^2+\delta^2\,)^p
\end{align}
Here $\delta$ is the relativistic Doppler factor given by 
equation (\ref{eq1:dopplerfact}).
The fact that all quasars are detected with one sided jet suggests that their jets 
are directed close to the line of sight of the observer. For example, the predicted
value of $J$ for the quasar 3C 273 is $9\times10^6$ \cite{1993v407p65}.

The interpretation of the one sided jet as an outcome of relativistic beaming
effects can be tested by measuring the difference in polarisation between the 
lobe in jet and counter-jet side. The lobe situated at the counter-jet 
side should be less polarised than the lobe at
the jet side since the emission from the farther lobe face more 
depolarising medium along the line of sight. Indeed, such a difference was 
observed by Garrington and Conway \cite{1991v250p198} for 49 sources from a 
sample of 69, supporting the above interpretation.

\subsubsection{Bent Jets}
\begin{figure}[tp]
\begin{center}
\includegraphics[width=115.45mm, height=80.79mm,bb=0 0 1309 916]{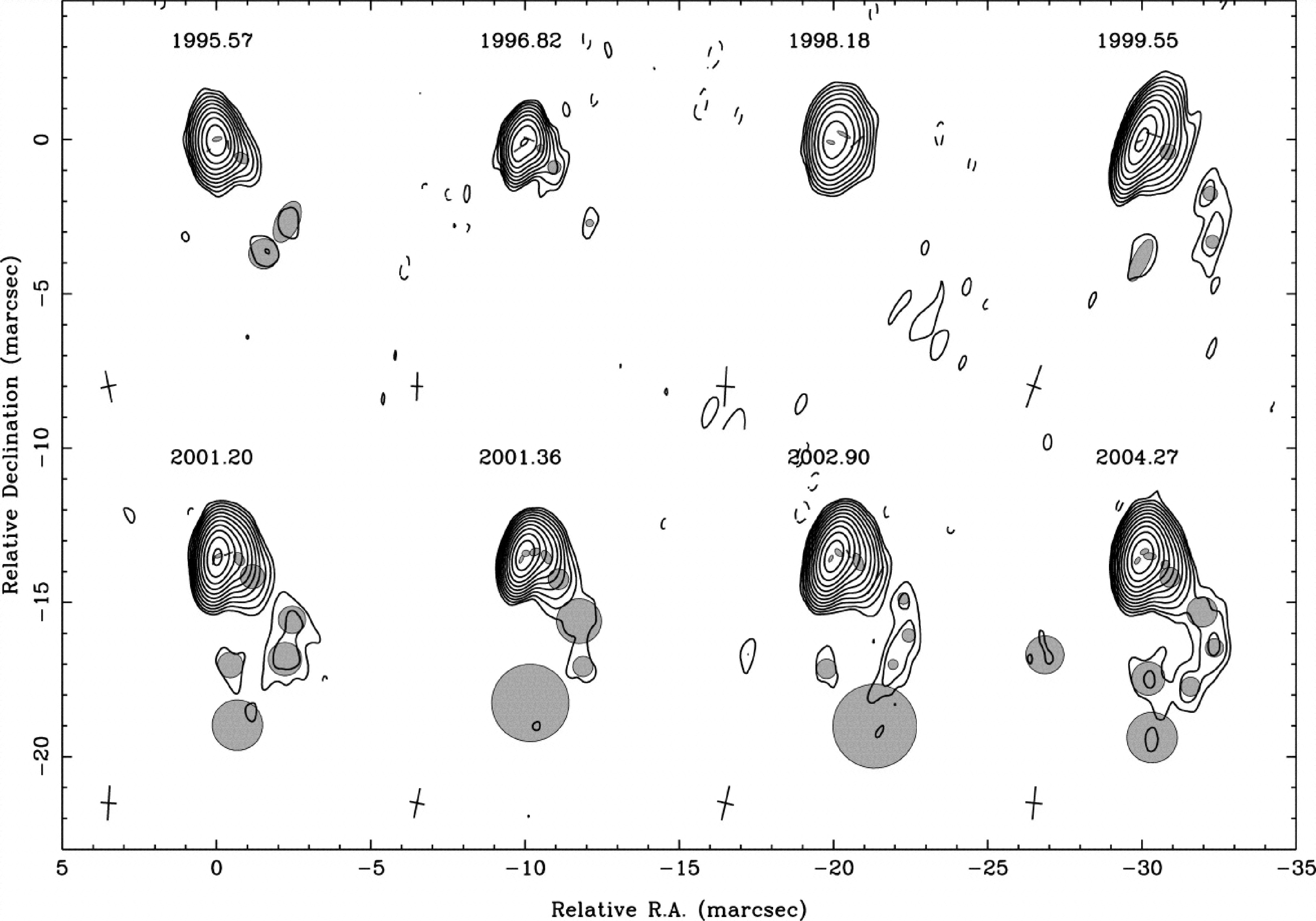} 
\caption[VLBA images of PKS 2136+141]{Very Large Baseline Array (VLBA) 
images of PKS 2136+141. Figure reproduced from 
Savolainen et al. \cite{2006v647p172}}
\label{fig1:pks2136}
\end{center}
\end{figure}

Sky maps of most AGN jets show curvature and in particular this feature 
is prominent 
in the \emph{radio trails} or \emph{head-tail}\footnote{These are sources 
with a radio morphology consisting of a bright nucleus and a single faint
jet streaming on one side.} sources. 
In high resolution radio maps of these
sources, one can see jets emerging from the nucleus and bent through a large
angle \cite{1986v301p841}. For example, the sky map of the radio loud quasar 
PKS 2136+141 show a jet which is bent by an angle $\sim 210^\circ$ 
and this is the largest bending angle seen till today \cite{2006v647p172}
(Figure \ref{fig1:pks2136}). The bending of the radio jet can be interpreted 
in many ways. A distortion in the jet shape can occur as a result of ram pressure 
associated with the motion of the host galaxy through a dense intracluster
medium. Bent jets with reflection symmetry can be explained if the 
host galaxy is associated with a companion. The jet curvature, in such a case 
is the result of the acceleration of the parent galaxy introduced by its 
companion \cite{1978v185p527}. Jets of the radio source can also appear as
an inversion symmetry if the jet precesses. In this case though the jet matter 
follows a linear path, it will appear curved in the sky plane due to
precession. Alternatively, the jet can also be deflected in spite of the 
kinetic motion of the host galaxy or by geometrical effects as explained 
above. As the jet moves through the surrounding medium, it gets disrupted
when the medium is an extremely dense intracluster gas. Pressure 
stratification of the intracluster gas can also cause the jet to 
curve \cite{1996v280p559,1996v280p567}.

\subsubsection{Limb-brightened Jets} 
\begin{figure}[tp]
\begin{center}
\includegraphics[width=115.45mm, height=80.79mm,bb=0 0 442 331]{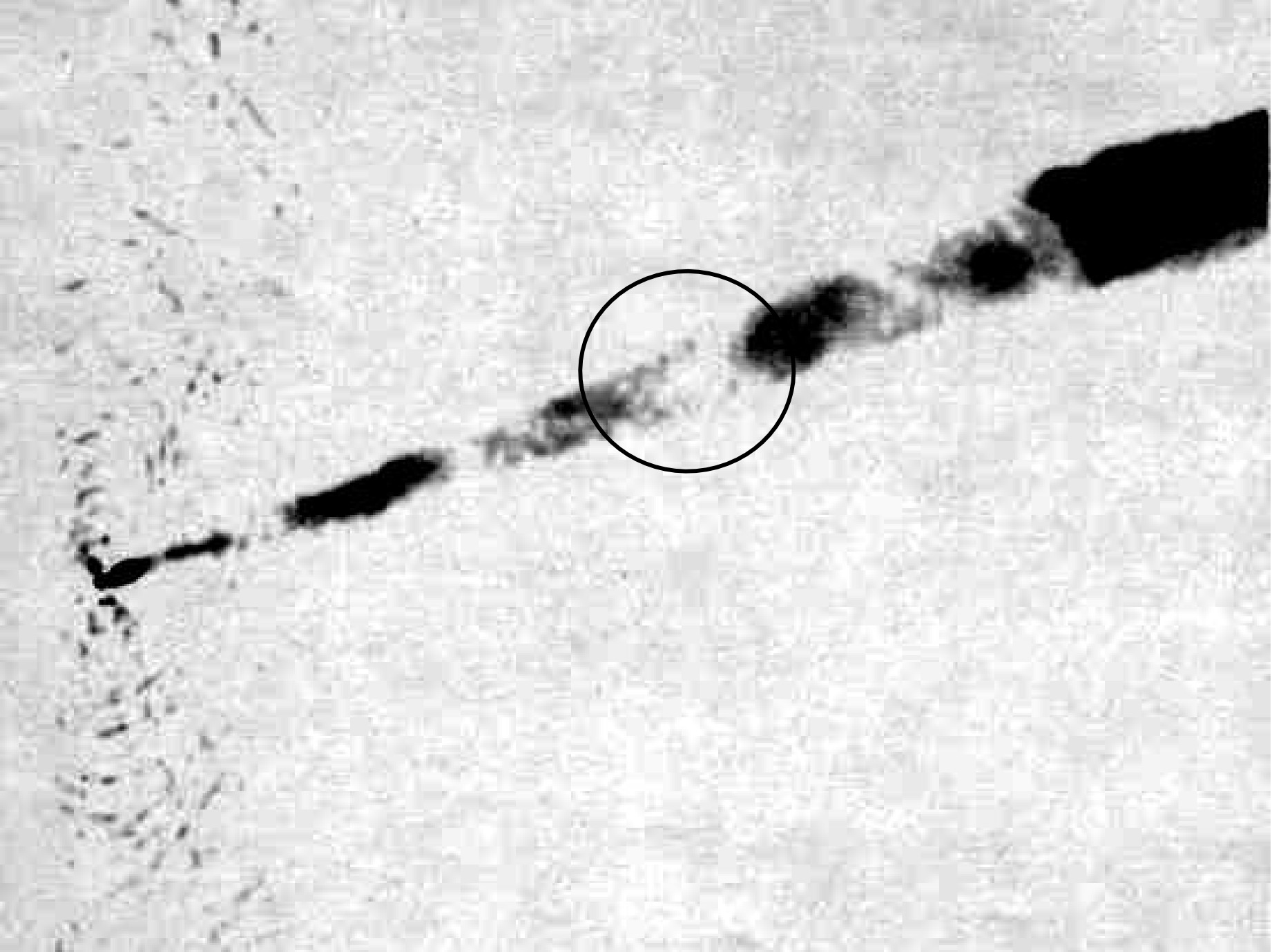} 
\caption[VLA image of M87 with limb-brightened feature]
{VLA image of M87 with limb-brightened feature (circled). 
Figure reproduced from Owen et al. \cite{1989v340p698}}
\label{fig1:m87_limb}
\end{center}
\end{figure}

Another feature seen in some AGN jets is \emph{limb-brightening} at pc scales. 
High resolution radio maps of these jets show the edges  
are brighter than the central spine of the jet (Figure \ref{fig1:m87_limb}). 
Such features are commonly observed in few FR I sources and blazars \cite{2003v47p551}. 
This feature is usually explained by the ``spine-sheath'' model where the 
velocity at the jet spine is larger compared to the velocity at the boundary. 
Such a radial stratification of velocity across the jet arises when jet moves 
through the ambient medium and the viscosity involved causes a shear at 
the boundary. The difference in the flow velocity of the jet between the spine 
and the boundary can cause differential Doppler boosting and this in turn 
can produce a \emph{limb-brightened} jet. 
Alternatively, velocity stratification at the jet boundary can accelerate the 
particles via shear acceleration \cite{1981v33p399,1981v7p352} 
(see chapter \ref{chap:mkn501}). These particles can then emit radiation via synchrotron 
process giving rise to a \emph{limb-brightened} feature. Particles at the jet 
boundary can also be accelerated by turbulent waves initiated by the instabilities 
\cite{1982v254p472}.  

\chapter{Particle Acceleration and Radiative Processes}
\label{chap:acc_n_rad}
The observed very high energy $\gamma$-ray emission (up to GeV-TeV energies) 
from blazars suggests the presence of extremely relativistic particles 
in AGN jets. Also the
power-law spectra observed over a broadband starting from radio-to-$\gamma$-ray 
indicates the jet emission to be dominated by non-thermal processes.
The promising mechanism by which particles can be accelerated to very high 
energies in AGN jets is ``Fermi mechanism'' suggested by Enrico Fermi 
\cite{1949v75p1169} to explain the power-law nature of the cosmic ray spectrum.

\section{Fermi Acceleration}\label{sec3:accln}
\begin{figure}[tp]
\begin{center}
\includegraphics[width=125.2mm, height=209.6mm,bb=0 0 354 594]{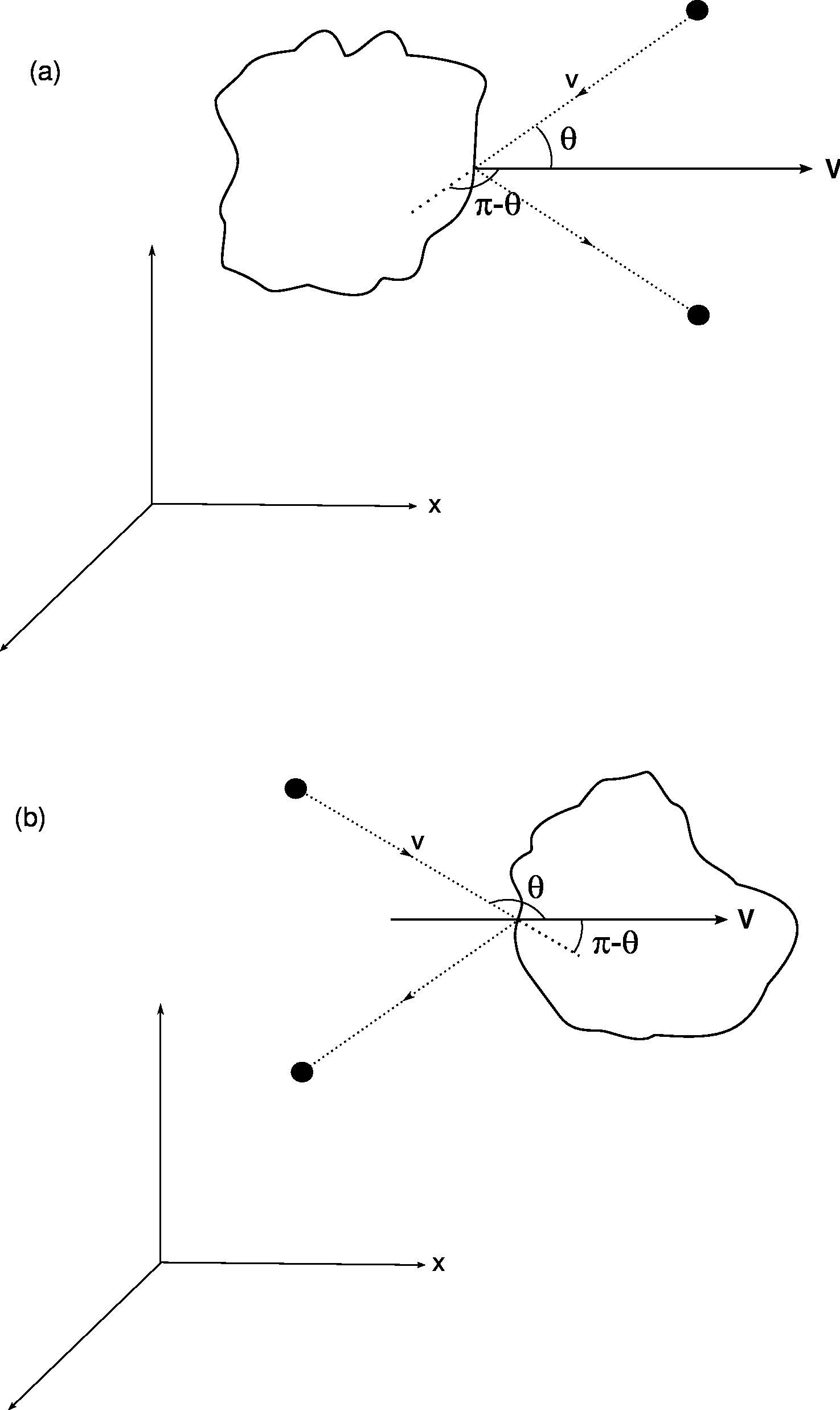} 
\caption[Illustration of Fermi acceleration mechanism]
{Illustration of Fermi acceleration mechanism: Collision between a particle with 
velocity $v$ and a massive cloud moving with velocity {\bf V}. (a) Head-on collision;
(b) Follow-on collision.}
\label{fig2:fermiacc}
\end{center}
\end{figure}

In 1949, Fermi \cite{1949v75p1169} proposed that particles can be accelerated 
to high energies when they are scattered by magnetic irregularities 
(\emph{magnetic mirrors}) associated with moving clouds in the 
interstellar medium. 
Consider a cloud moving with a velocity $V$ along the x-axis in the 
observer's frame (Figure \ref{fig2:fermiacc}). Let us
assume the cloud to be massive in comparison with the scattered particle 
and hence the centre of mass frame (CM-frame) is the frame of the cloud itself.
The energy and the momentum of the particle before scattering in the 
CM-frame can be obtained using Lorentz transformation as \cite{book:rybicki}
\begin{align}
\label{eq3:eprime}
E^\prime &= \Gamma_c\, (\,E - V p_x\,)\\
\label{eq3:pxprime}
p_x^\prime&= \Gamma_c \,\left(\,p_x - \frac{VE}{c^2}\,\right)
\end{align}
where $E$ and $E^\prime$ are the energy of the particle in the observer's 
frame and CM-frame respectively and $p_x$ and $p_x^{\prime}$ are 
the corresponding momenta 
along x-axis (we represent the quantities in CM-frame with prime).
The Lorentz factor of 
the moving cloud $\Gamma_c$ is given by
\begin{align}
\Gamma_c &= \left(\,1-\frac{V^2}{c^2}\,\right)^{-1/2}
\end{align}
Since the CM-frame is the frame of scatterer itself, the energy and the 
x-component of momentum after collision will be 
\begin{align}
E_s^\prime &= E^\prime \\
p_{x,s}^\prime &= -p_x^\prime
\end{align}
Transforming the scattered energy back into the observer frame we get
\begin{align}
\label{eq3:es}
E_s&= \Gamma_c\,(\,E^\prime-Vp_x^\prime\,) \nonumber \\
&=\Gamma_c^2\left[E\left(1+\frac{V^2}{c^2}\right)-2\,Vp_x\right]
\end{align}
where we have used equations (\ref{eq3:eprime}) and (\ref{eq3:pxprime}). If we 
express the momentum along x-axis in terms of energy 
\begin{align}
p_x = \frac{E\;v_x}{c^2}
\end{align}
The scattered energy can then be written as
\begin{align}
\label{eq3:e_s}
E_s = \Gamma_c^2E\left(1+\frac{V^2}{c^2}-\frac{2\;Vv_x}{c^2}\right)
\end{align}
Here $v_x$ is the velocity of the particle along x-axis. 
For $V \ll c$, the change in the particle energy due to collision 
can be obtained using equations (\ref{eq3:e_s}) as
\begin{align}
\Delta E &= E_s - E \nonumber \\
&\approx E\left[2\left(\frac{V}{c}\right)^2-\frac{2\;Vv_x}{c^2}\right]
\end{align} 
where we have retained the terms only up to second order in $V/c$.
If the particle arrives at an angle $\theta$ with respect to the velocity of 
the scatterer as shown in Figure \ref{fig2:fermiacc}, then $v_x = -v\cos\theta$ 
and we get  
\begin{align}
\label{eq3:energygain}
\frac{\Delta E}{E} = 2\;\frac{V}{c^2}\;\left[\,v \cos \theta+V \,\right] 
\end{align} 
From equation (\ref{eq3:energygain}) it is evident that the particle  
gains energy in head-on collisions. On the other hand, for follow-on collisions 
$\cos\theta$ is negative, and the particle  
loses energy. Since in the scatterer's frame the velocity of the 
approaching particle is more than the receding one due to addition of 
velocities, the probability for head-on collision is more than that of 
follow-on collision. Hence there will be a net gain in particle energy. 

\subsection{Stochastic Acceleration - Second Order Fermi Mechanism}
Let us consider a situation where the scattering centers move randomly, 
then the net gain is 
obtained by averaging equation (\ref{eq3:energygain}) over the angle $\theta$. 
The acceleration process is then referred to stochastic acceleration.
An example of stochastic acceleration process is \emph{turbulent acceleration},
where particles are energized via scattering by moving magnetic 
inhomogeneities associated with turbulence in a flowing fluid. Since the 
collision probability between the particle and the scatterer is anisotropic,
we need to know the probability of a collision happening at an angle $\theta$
to perform the averaging. If we assume $v\approx c$, then the rate of 
collision in the scatterer's frame will be greater by a factor
$\Gamma_c[1+(V/c)\cos\theta]$ \cite{book:longairv2}. Hence the probability 
of collision at angle $\theta$ will be proportional to this factor and
the average will be 
\begin{align}
\avg{\frac{2V\cos\theta}{c}}&\approx\left(\frac{2V}{c}\right)
\frac{\int\limits_{-1}^{+1}\mu\,[1+(V/c)\mu]\;d\mu}
{\int\limits_{-1}^{+1}[1+(V/c)\mu]\;d\mu}\nonumber \\
&=\frac{2}{3}\left(\frac{V}{c}\right)^2
\end{align}
where $\mu = \cos\theta$. From equation (\ref{eq3:energygain}), the 
average energy gain will then be 
\begin{align}
\left<\frac{\Delta E}{E}\right> \approx \frac{8}{3}\left(\frac{V}{c}\right)^2
\end{align}
Since the average increase in energy is of the order $(V/c)^2$, this process is 
called as \emph{second order Fermi acceleration} mechanism.

\subsection{Shock Acceleration - First Order Fermi Mechanism}\label{sec3:shock_accln}
If we consider only head-on collisions in Fermi acceleration, then 
the dominant term in 
equation (\ref{eq3:energygain}) will be of the order of $(V/c)$ 
(for $v\approx c$) and the acceleration process is called as 
\emph{first order Fermi mechanism}. Acceleration of charged particles at a shock front
is an example of this mechanism. 
A shock is a discontinuity
initiated by perturbation in a supersonic fluid flow. Fluid on either side of
the shock will be in different state of equilibrium and are connected by the 
conservation equations. The strength of a shock is determined by its 
Mach number $M$ defined as 
\begin{align}
M = \frac{v_1}{c_s}
\end{align}
where $v_1$ is the velocity of the upstream fluid with respect to the shock front 
and $c_s$ is the local speed of sound in the upstream region. Here, we denote upstream
as the fluid ahead of the shock front and downstream as the one behind. 
Strong shocks are the one with $M\gg1$.
Particles  in a magneto hydrodynamic fluid (plasma) are scattered by magnetic 
inhomogeneities associated with turbulence in the flowing fluid. In presence
of a shock in the fluid they get
energised by crossing the shock front from upstream to downstream or vice versa. 

Let us consider the case of non-relativistic strong shock 
where the shock velocity $U \gg c_s$ and $U\ll c$. 
In the shock frame the plasma will pass through it with an upstream
velocity $v_1(=U)$ and the downstream velocity $v_2$. 
From the mass conservation we get 
\begin{align}
\rho_1\, v_1 = \rho_2 \,v_2
\end{align} 
where $\rho_1$ and $\rho_2$ are the mass density of the upstream and 
downstream plasma. For a strong shock in the limit of $M\rightarrow\infty$ we 
can write \cite{book:parks}
\begin{align}
\frac{\rho_1}{\rho_2} = \frac{\gamma_s +1}{\gamma_s -1}
\end{align}
where $\gamma_s$ is the ratio of the specific heats at constant pressure 
and volume. For a fully ionised gas we have $\gamma_s = \frac{5}{3}$ and hence
$\frac{\rho_2}{\rho_1} = 4$ and $v_2 = \frac{1}{4}U$. Hence the plasma on either 
side of the shock (upstream or downstream) will see the plasma approaching 
from the other side of the shock (downstream or upstream) with velocity 
$\frac{3}{4}U$. In the proper frame of the plasma (upstream or downstream), 
the particle distribution is isotropic due to scattering. Let us consider 
a particle of energy $E$ with x-component momentum $p_x$ in the upstream 
plasma. We have chosen a coordinate system where $x$ axis normal to the shock front. 
The energy of the particle $E^\prime$ in the frame of downstream plasma will be
\begin{align}
E^\prime = E+p_x\,V
\end{align} 
where $V= \frac{3}{4}U$ is the velocity of the downstream plasma with respect 
to the upstream plasma. 
Since the particles are relativistic, we can write $p \approx E/c$ and 
$p_x \approx \frac{E}{c}\cos\theta$. Here $\theta$ is the angle between the 
particle momentum and the shock normal. Hence the particle enters the 
downstream with an energy increment
\begin{align}
\Delta E = E \left(\frac{V}{c}\right)\cos\theta
\end{align}
The probability of the particle crossing the shock front and entering 
into the downstream
within the angle interval $\theta$ to $\theta+d\theta$ is proportional to
$\sin\theta \cos\theta d\theta$. Hence the average increase in energy in 
crossing the shock once is
\begin{align}
\avg{\frac{\Delta E}{E}}=\frac{2}{3}\left(\frac{V}{c}\right)
\end{align}
A similar situation happens for the particle crossing the shock front 
from downstream to upstream plasma. Hence the average fractional energy
gain in making one round trip is 
\begin{align} 
\avg{\frac{\Delta E}{E}}=\frac{4}{3}\left(\frac{V}{c}\right)
\end{align}
Thus a particle in the vicinity of a shock is scattered by magnetic
inhomogeneities and gets accelerated to higher 
energy by crossing the shock front multiple number of times.
Also, since the average energy gain is of the order of $(V/c)$, shock
acceleration is efficient than the stochastic acceleration.

\subsection{Acceleration at Shear Layers}\label{sec2:shear_accln}
\begin{figure}[tp]
\begin{center}
\includegraphics[width=144.4mm, height=82.0mm,bb=0 0 409 233]{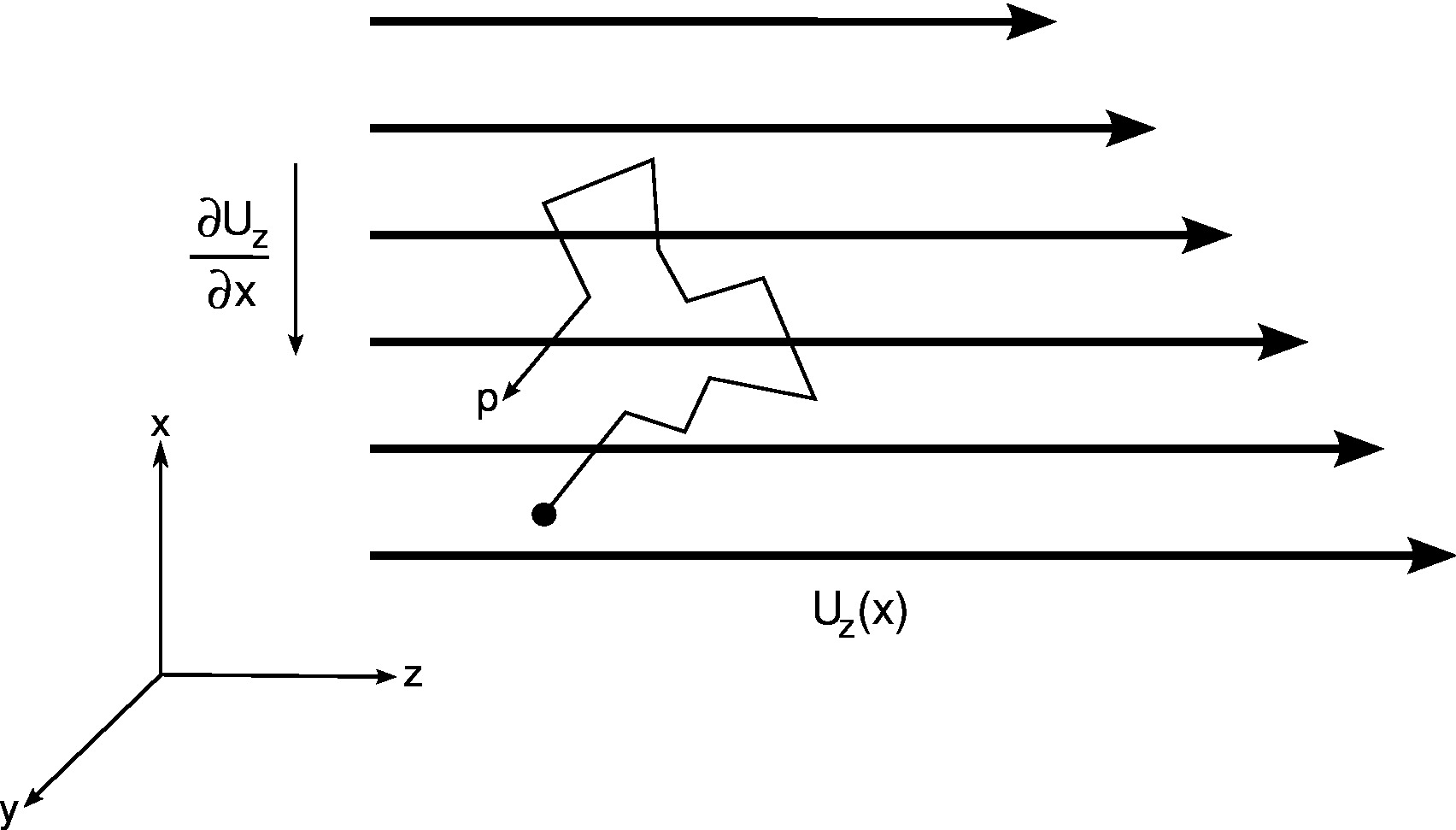} 
\caption{Particle acceleration at sheared flow} 
\label{fig2:shearacc}
\end{center}
\end{figure}

A velocity shear can arise in a plasma flow if it passes through a 
viscous medium. Particles can then get accelerated when they are scattered 
between different velocity layers by magnetic irregularities 
\cite{1981v33p399}(Figure \ref{fig2:shearacc}). Let us consider the case 
of a gradual shear, where the 
mean free path of the particle is much smaller than the transverse width of 
the sheared velocity layer. 
Suppose the flow is non-relativistic and we choose a reference  
frame in which the local fluid is at rest and the flow velocity is along
z-axis, ${\bf U}=U_z(x){\bf \hat{e}_z}$. If $\tau$ is the mean scattering time, 
then the distance travelled by the particle along x-axis before getting scattered 
will be 
\begin{align}
\delta x = \frac{p_1}{m}\,\cos\theta\, \tau
\end{align}
where $p_1$ is the momentum of the particle, $m$ is its mass and $\theta$ 
is the angle between the particle momentum and x-axis. The change in 
the fluid velocity due to this displacement will be 
\begin{align}
{\bf \delta u} = \left(\pd{U_z}{x}\right)\delta x \,{\bf \hat{e}_z}
\end{align}
where $\left(\pd{U_z}{x}\right)$ is the shear velocity gradient. The 
momentum of the particle $p_2$ relative to the local fluid will then be 
\cite{1990v356p255}
\begin{align}
\label{eq3:shearmom}
p_2^2 = p_1^2\left(1+2\,\frac{m\,\delta u}{p_1}\,\sin\theta \,\cos\phi+
         \frac{m^2\,\delta u^2}{p_1^2}\right)
\end{align}
where $\phi$ is the angle between y-axis and the projection of the particle
momentum in yz-plane. Since $E=\frac{p^2}{2m}$, the average fractional energy 
gain due to shear acceleration will be
\begin{align}
\avg{\frac{\Delta E}{E_1}}\propto\left(\pd{U_z}{x}\right)^2\tau^2
\end{align}
The first order term in equation (\ref{eq3:shearmom}) will get cancelled 
because for every forward scattering of particle between two velocity layers 
there will be an equivalent reverse scattering with negative ${\bf \delta u}$. 

\subsection{Accelerated particle distribution}
The Fermi acceleration process leads to a power-law distribution of 
particles \cite{book:gaisser}.
Consider after each encounter the particle energy increases by a factor
$\xi$. i.e $\Delta E = \xi E$. Let us assume a particle with initial 
energy $E_o$ enters the acceleration region. After $n$ encounters the 
particle energy will be
\begin{align}
\label{eq3:etot}
E = E_o\,(1+\xi)^n
\end{align} 
Also, let the probability of escape from the acceleration region after an encounter
be $P_{esc}$. Hence, after $n$ encounters the probability of the particle to remain
in the acceleration region will be $(1-P_{esc})^n$. Thus the number of particles
in the acceleration region with energies greater than $E$ will be 
\begin{align}
\label{eq3:partspec}
N(\ge E) &\propto \sum\limits^\infty_{j=n}(1-P_{esc})^j \nonumber\\
  &= \frac{(1-P_{esc})^n}{P_{esc}}
\end{align}  
From equation (\ref{eq3:etot}), $n$ can be written as
\begin{align}
\label{eq3:nval}
n = \frac{\ln(E/E_o)}{\ln(1+\xi)}
\end{align}
Substituting equation (\ref{eq3:nval}) in equation (\ref{eq3:partspec}), we get
\begin{align}
N(\ge E) \propto \frac{1}{P_{esc}}\left(\frac{E}{E_o}\right)^{-p}
\end{align}
where the index $p$ is given by
\begin{align}
p = \frac{\ln(\frac{1}{1-P_{esc}})}{\ln(1+\xi)}
\end{align}

\section{Radiation Emission Mechanisms}
\label{sec3:radmech}
\subsection{Blackbody Radiation}
The radiation emitted by a distribution of charged particle which is in thermal equilibrium 
with itself as well as the emitted radiation is called as \emph{blackbody radiation}.
The specific energy density of a blackbody spectrum is given by Planck spectrum
\begin{align}
\label{eq3:bb1}
u_\nu(\Omega) = \frac{2h\nu^3/c^3}{\exp(h\nu/k_BT)-1} \quad 
\text{ergs cm$^{-3}$ Hz$^{-1}$ sr$^{-1}$}
\end{align}
where $\nu$ is the observed photon frequency, $\Omega$ the solid angle, $h$ 
the Planck constant, $k_B$ the Boltzmann constant and $T$ is the blackbody temperature. 
The peak photon frequency
at which the energy density is maximum for a given temperature $T$ is given
by Wien's law, $\nu_{peak}= 2.82 (k_B/h)T$. Integrating equation (\ref{eq3:bb1})
over the entire photon frequency and the solid angle we get 
the \emph{Stefan-Boltzmann} law for the energy density of a blackbody spectrum 
\begin{align}
u(T) = a\, T^4
\end{align}
where $a = 4\sigma_{SB}/c$ and $\sigma_{SB}$ is the 
\emph{Stefan-Boltzmann} constant.
 
\subsection{Bremsstrahlung Radiation}
\begin{figure}[tp]
\begin{center}
\includegraphics[width=127.0mm, height=88.9mm,bb=0 0 358 250]{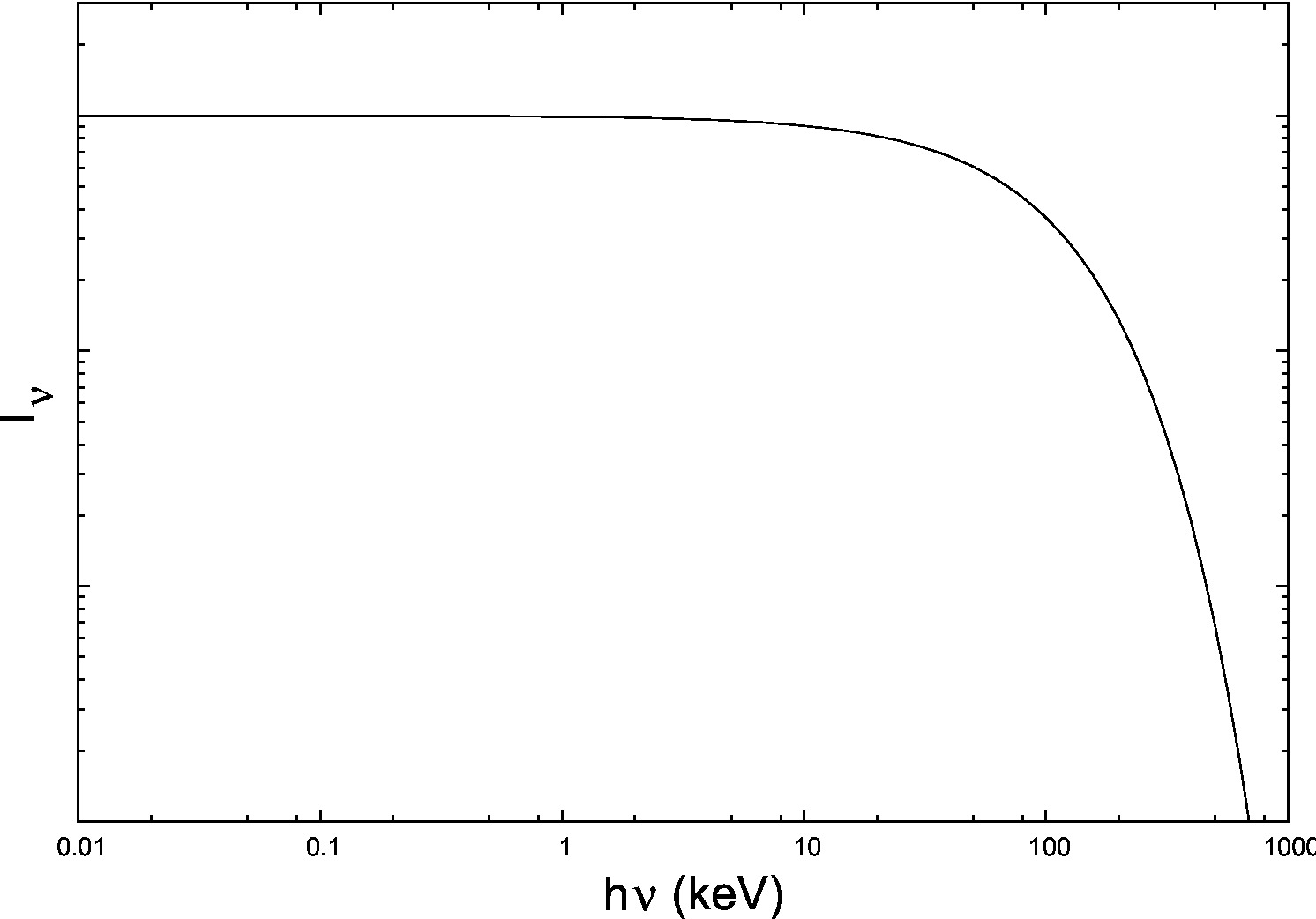} 
\caption[Intensity spectrum of bremsstrahlung radiation]
{Intensity spectrum (arbitrary units) of bremsstrahlung radiation due to a 
thermal electron distribution at temperature 100 keV.} 
\label{fig2:bremspec}
\end{center}
\end{figure}

When a charged particle moves in a Coulomb field it gets accelerated and emits 
electromagnetic spectrum. This radiation is known as \emph{Bremsstrahlung}
radiation. Let us consider the motion of an electron in the Coulomb field 
of an ion. We will assume small angle scattering where the deviation of the 
electron path from a straight line is negligible. The variation in the 
dipole moment of the electron-ion system gives rise to dipole radiation
and it can be shown that the total energy emitted per frequency by the 
electron is \cite{book:rybicki}
\begin{align}
\frac{dW(b)}{d\omega}=\left\{ \begin{array}{ll}
\frac{8\, Z^2\,e^6}{3\pi\, c^3\,m_e^2\,v^2\,b^2}\,, & b\ll v/\omega \\
0\,,& b\ll v/\omega
\end{array}\right.
\end{align}
where $\omega = 2\pi \nu$ is the angular frequency of the emitted photon, $b$
is the impact parameter, $Ze$ is the charge of ion, $v$ the electron velocity
and $m_e$ and $e$ are the electron mass and its charge respectively. The total power emitted
per frequency per volume for a medium with ion density $n_i$ and electron
density $n_e$ assuming a fixed electron speed $v$ will be \cite{book:rybicki} 
\begin{align}
\frac{dW}{d\omega dV dt} = \frac{16\pi\, e^6}{3\sqrt{3}\;c^3\,m_e^2\,v}\;
n_e\;n_i\;Z^2\;\mathrm{g}_{ff}(v,\omega)
\end{align}
where $\mathrm{g}_{ff}$ is known as Gaunt factor which is a function of energy of the 
electron and the frequency of the emitted photon. For a thermal distribution, 
the particles follow a Maxwellian velocity distribution and we obtain
the power per frequency per volume as \cite{book:rybicki}
\begin{align}
\label{eq3:brem1}
\frac{dW}{d\omega dV dt} = \frac{2^5\pi\, e^6}{3\,m_e\,c^3}
\left(\frac{2\pi}{3\,k\,m_e}\right)^{1/2}T^{-1}Z^2\;n_e\;n_i\;e^{-h\nu/kT}
\;\bar{\mathrm{g}}_{ff}
\end{align}
where $\bar{\mathrm{g}}_{ff}$ is the velocity averaged Gaunt factor. From 
equation (\ref{eq3:brem1}) we find that the bremsstrah\-lung emission from a thermal 
particle distribution gives rise to a flat spectrum with an exponential cutoff at about
$h\nu\sim kT$ (Figure \ref{fig2:bremspec}).

\subsection{Synchrotron Radiation}\label{sec3:synrad}
The radiation emitted due to the helical motion of a relativistic 
charged particle along a magnetic field is know as 
\emph{synchrotron radiation} \cite{book:rybicki,book:frankshu}. 
The radiative power emitted by a relativistic electron with velocity 
$\beta(=v/c)$
moving at an angle $\alpha$ (pitch angle) with respect to an uniform
magnetic field $B$ is given by \cite{book:rybicki}
\begin{align}
\label{eq3:synpow0}
P_{syn}=\frac{2}{3}\,r_e^2\,c\,\gamma^2\,\beta^2\,B^2\,\sin^2\alpha
\end{align}
where $r_e=e^2/m_e\,c^2$ is the classical electron radius and $\gamma$ is 
the Lorentz factor of the electron, $\gamma = (1-\beta^2)^{-1}$. For an
isotropic distribution of mono energetic electrons we need to average 
equation (\ref{eq3:synpow0}) over all angles and we get
\begin{align}
\label{eq3:synpower}
P_{syn}= \frac{4}{3}\,\beta^2\,\gamma^2\,c\,\sigma_T\, U_B
\end{align}
where $\sigma_T$ is the Thomson cross section and $U_B=B^2/8\pi$ is the 
magnetic field energy density. Equation (\ref{eq3:synpower}) is also 
equal to the average energy lost by an electron via synchrotron 
process.   

\begin{figure}[tp]
\begin{center}
\includegraphics[width=127.0mm, height=88.9mm,bb=0 0 347 246]{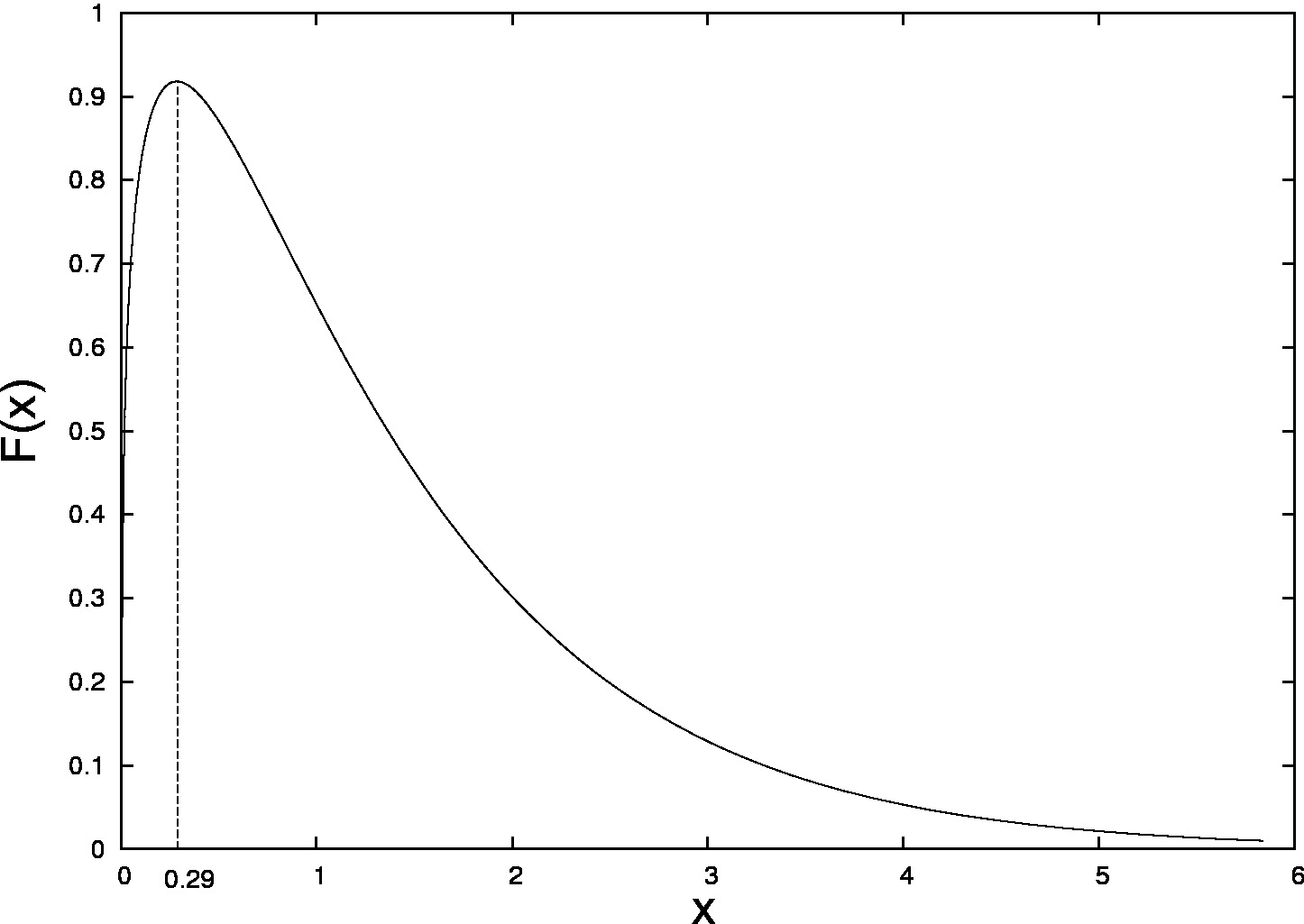} 
\caption{Synchrotron power function}
\label{fig2:syn_pow}
\end{center}
\end{figure}

The synchrotron spectrum emitted by an electron of energy 
$\gamma\, m_e\,c^2$ moving with pitch angle $\alpha$ can be written as
\cite{1970v42p237}
\begin{align}
\label{eq3:synspe}
P_{syn}(\gamma,\nu)=\frac{\sqrt{3}\;e^3B\sin \alpha}{m_e\, c^2}\;F\left(\frac{\nu}{\nu_c}\right)
\end{align}
where $\nu$ is the frequency of the emitted photon and  
\begin{align}
\nu_c = \frac{3\,e\,B\,\gamma^2}{4\pi\, m_e\,c}\;\sin\alpha
\end{align}
is the critical frequency.
The synchrotron power function $F(x)$ is defined as
\begin{align}
F(x)= x\int\limits^\infty_xK_{5/3}(\xi)\;d\xi
\end{align}
where
$K_{5/3}$ is the modified Bessel function of order $5/3$. The shape of the 
spectrum is determined by $F(x)$ with a 
peak located at 
$\approx 0.29 (\nu/\nu_c)$ (Figure \ref{fig2:syn_pow}). Alternatively one can 
write the single particle
emission spectrum using equation (\ref{eq3:synpower}) as \cite{book:frankshu}
\begin{align}
\label{eq3:synpow1}
P_{syn}(\gamma,\nu) = \frac{4}{3}\,\beta^2\,\gamma^2\,c\,\sigma_T\, U_B\, \phi_\nu(\gamma)
\end{align}
where $\phi$ is a function of $\nu$ and $\gamma$ satisfying the relation
\begin{align}
\int\limits_0^\infty \phi_\nu(\gamma)\;d\nu = 1
\end{align}
Considering the shape of the spectrum (Figure \ref{fig2:syn_pow}), one can approximate the function
$\phi$ as a Dirac delta function (to the orders of unity)
\begin{align}
\label{eq3:synphi}
\phi_\nu(\gamma)\to\delta(\nu - \gamma^2\nu_L)
\end{align}
where $\nu_L = eB/2\pi m_e\,c$ is the Larmor frequency.

For an isotropic power-law electron distribution given by
\begin{align}
\label{eq3:powpar}
N(\gamma)=k\,\gamma^{-p}\quad \gamma_{min}<\gamma<\gamma_{max}
\end{align}
the radiation energy emitted per second per frequency for 
$\gamma_{max}\gg \gamma_{min}$ can be shown as \nolinebreak \cite{1970v42p237}
\begin{equation}
\label{eq3:synspec}
\frac{dW}{d\nu dt}\approx \frac{4\pi\, k\,e^3\,B^{(p+1)/2}}{m_e\,c^2}
\,\left(\frac{3\,e}{4\pi\, m_e\,c}\right)^{(p-1)/2}a(p)\,\nu^{-(p-1)/2}
\end{equation} 
where $a(p)$ is a function of particle spectral index. Hence the emitted 
synchrotron spectrum is a power-law with index $(p-1)/2$.

Moreover, a charged particle in a magnetic field besides emitting synchrotron radiation,
can absorb a photon and get energized. This absorption process is called 
as \emph{Synchrotron Self Absorption}. Also a photon can induce a charged 
particle to emit in a direction and at a frequency of the photon itself 
(\emph{stimulated emission} or \emph{negative absorption}). The absorption
coefficient for synchrotron self absorption process is given 
by \cite{1991v252p313}
\begin{align}
\label{eq3:ssacoeff}
\kappa_\nu = -\frac{1}{8\pi\, m_e\, \nu^2}\int \frac{N(\gamma)}
{\gamma\,(\gamma^2-1)^{1/2}}\;\frac{d}{d\gamma}\;[\,\gamma(\gamma^2-1)^{1/2}
\;P_{syn}(\gamma,\nu)]
\end{align}
The specific intensity ($I_\nu$) of the synchrotron radiation can then be found using 
the radiative transfer equation as \cite{book:frankshu}
\begin{align}
I_\nu = S_\nu(1-e^{-\tau_\nu})
\end{align}
where we have assumed a source with uniform properties and no background illumination.
Here $S_\nu$($=j_\nu/\kappa_\nu$) is the synchrotron source function
and $\tau_\nu$ is the optical depth defined over a distance $s$ as 
\begin{align}
\tau_\nu = \int\limits_s \kappa_\nu\; ds^\prime
\end{align}
The synchrotron emissivity $j_\nu$ for an isotropic emission 
is given by
\begin{align}
\label{eq3:synemiss}
j_\nu=\frac{1}{4\pi}\int\limits_1^\infty P_{syn}(\gamma,\nu)\,N(\gamma)\;d\gamma
\end{align} 
For a power-law distribution of particle (equation (\ref{eq3:powpar})),
the synchrotron self absorption coefficient will be \cite{book:frankshu}
\begin{align}
\label{eq3:synabscoeff}
\kappa_\nu\propto B^{(p+2)/2}\;\nu^{-(p+4)/2}
\end{align}
Hence it may be possible for a source to be optically thick ($\tau_\nu>1$) at
low frequencies but optically thin ($\tau_\nu<1$) at high frequencies. From equations
(\ref{eq3:synpow1}), (\ref{eq3:synemiss}) and (\ref{eq3:synabscoeff}) one can find that
the source function $S_\nu$ for the optically thick region will be a power-law 
of the form
\begin{align}
S_\nu\propto\nu^{5/2}
\end{align}
The total synchrotron spectrum will then be a broken power-law with spectrum changing
from $\nu^{5/2}$ at lower frequencies to $\nu^{-(p-1)/2}$ at high frequencies. 
The frequency at which the 
index changes is called as synchrotron self absorption frequency. For AGN jet 
emission this frequency is observed to be within a range of few gigahertz.

\subsection{Inverse Compton Radiation} \label{sec3:icrad}
Scattering of low energy electrons by high energy (or hard) photons 
is called as \emph{Compton scattering} and the reverse process where high 
energy electrons scatter off low energy (or soft) photons 
is called as \emph{inverse Compton scattering}. 
The spectrum obtained due to the scattering
of soft photons by relativistic electrons is called as \emph{inverse
Compton spectrum}. If the energy of the incident photon in electron's 
rest frame is much smaller than the electron rest mass energy, then one can 
ignore the recoil of electron and the scattering process leave the photon 
energy unchanged in electron's rest frame. In such case the scattering
process is described by Thomson cross section with differential cross section
given by 
\begin{align}
\frac{d\sigma_T}{d\Omega}=\frac{1}{2}\,r_e^2\,(1+\cos^2\theta)
\end{align}
Here $r_e$ is the classical electron radius and $\theta$ is the angle between
the incident and the scattered photon directions. On the other hand if the
recoil of the electron becomes considerable, then the scattering cross section
is described by Klein-Nishina cross section. The differential cross section in 
this case will include the quantum effects and is given by
\begin{align}
\frac{d\sigma_T}{d\Omega}=\frac{r_e^2}{2}\;\frac{\epsilon_s^2}{\epsilon^2}
\;\left(\frac{\epsilon}{\epsilon_s}+\frac{\epsilon_s}{\epsilon}-\sin^2\theta\right)
\end{align}
where $\epsilon$ and $\epsilon_s$ are the energies of the incident and the scattered
photon. 
These conditions can be expressed in terms of 
$\epsilon$
and the Lorentz factor of the relativistic electron $\gamma$ as, 
$\gamma \epsilon \ll m_e\, c^2$ for scattering in Thomson regime and 
$\gamma \epsilon \gg m_e\, c^2$ for scattering in Klein-Nishina regime
\cite{book:rybicki,1970v42p237}.
Also, in the Thomson limit the scattered photon energy ($\epsilon_s$) can be 
shown as \cite{1970v42p237} 
\begin{align}
\epsilon_s \approx \gamma^2 \,\epsilon
\end{align}
Hence for large $\gamma$ the photon energy gain is very large. Nevertheless 
this gain 
is quite small compared with the electron energy and the electron loses
only a small fraction of energy in each scattering. However in Klein-Nishina
regime the electron loses almost its entire energy to the photon in a single
scattering. Hence the scattered photon energy in this case will be 
\begin{align}
\epsilon_s \approx \gamma\, m_e\, c^2
\end{align}
The radiative power emitted (or the power lost by an electron) due to 
inverse Compton scattering of an isotropic soft photon distribution 
can be shown as \cite{book:rybicki}
\begin{align}
\label{eq3:compower}
P_{com}= \frac{4}{3}\;\beta^2\,\gamma^2\,c\,\sigma_T\, U_{ph}
\end{align}
where $U_{ph}$ is the energy density of the soft target photon distribution 
and the scattering is assumed to be in Thomson regime. Comparison of
equation (\ref{eq3:compower}) with the power emitted due to synchrotron 
emission (equation (\ref{eq3:synpower})) one finds 
\begin{align}
\frac{P_{syn}}{P_{com}}= \frac{U_B}{U_{ph}}
\end{align}
When the scattering happens in Klein-Nishina regime the power lost by an
electron can be computed using
\begin{align}
P_{com,kn} = \int\int (\epsilon_s-\epsilon)\; \frac{dN}{dtd\epsilon_s}\;d\epsilon_s
\end{align}
where $\frac{dN}{dtd\epsilon_s}$ is the emission rate of the 
scattered photon per frequency given by \cite{1970v42p237} 
\begin{align}
\frac{dN}{dtd\epsilon_s} &= \frac{2\pi \,r_e^2\,c}{\gamma^2}
\;\frac{n(\epsilon)\,d\epsilon}{\epsilon} \nonumber \\
&\times\left[2\,q\, \ln q+(1+2\,q)(1-q)+
\frac{1}{2}\;\frac{(\Gamma_\epsilon\, q)^2}{(1+\Gamma_\epsilon\, q)}\; (1-q)\right]
\end{align}
Here $n(\epsilon)$ is the number density of the soft target photons and the 
quantities $\Gamma_\epsilon$ and $q$ are defined as
\begin{align}
\Gamma_\epsilon &= \frac{4\,\epsilon\,\gamma}{m_e\,c^2}\\
q &= \frac{\epsilon_s}{\Gamma_\epsilon \;(\,\gamma\, m_e\, c^2-\epsilon_s)}
\end{align}

For a power-law distribution of particle given by equation (\ref{eq3:powpar})
the emitted photon spectrum when the scattering happens in Thomson regime
will be \cite{1970v42p237}
\begin{align}
\label{eq3:comspec}
\frac{dW}{dtd\epsilon_s}&=\pi\, r_e^2\,c\,k\,2^{p+3}
\,\frac{p^2+4\,p+11}{(p+3)^2\,(p+1)(p+5)}\,\epsilon_s^{-(p-1)/2} \nonumber \\
&\times \int \epsilon^{(p-1)/2}\,n(\epsilon)\;d\epsilon
\end{align}
From equations (\ref{eq3:synspec}) and (\ref{eq3:comspec}) we find that
the spectrum  emitted by both synchrotron and inverse Compton process can be 
represented by a  
power-law with same index $(p-1)/2$. In case of extreme Klein-Nishina limit
the emitted spectrum will be 
\begin{align}
\label{eq3:comspeckn}
\frac{dW}{dtd\epsilon_s}&=\pi\, r_e^2\,c\,k\,(m_e\,c^2)^{p+1}
\,\epsilon_s^{-p} \nonumber \\
&\times \int\frac{d\epsilon}{\epsilon}\;n(\epsilon)\left(
\ln\frac{\epsilon\,\epsilon_s}{m_e^2\,c^4}+C(p)\right)
\end{align}
where $C(p)$ is a parameter of order unity \cite{1970v42p237}. Thus we see 
the inverse Compton spectrum in extreme Klein-Nishina limit is much steeper
than that of Thomson limit (equation (\ref{eq3:comspec})).

\subsubsection{Synchrotron Self Compton}\label{sec2:ssc}
In many astrophysical systems, the high energy emission is explained as a result of 
the inverse Compton scattering of synchrotron photons. Here, the same electron 
population which is responsible for the synchrotron emission will scatter off 
these photons to higher energies.
This process is commonly referred as \emph{synchrotron self-Compton} (or SSC) 
mechanism. For example, the ratio of the  
X-ray to TeV $\gamma$-ray fluxes obtained during simultaneous observation of 
BL Lac objects, both in quiescent and flaring state, can be explained in the 
context of SSC process \cite{1995v449p99}.

Let us assume that the radiation emitting plasma of AGN jet be confined in a spherical
region with tangled magnetic field. This plasma moves down the jet at relativistic 
speed. The distribution of electrons in the emission region is assumed to be a 
power-law  described by equation (\ref{eq3:powpar}). Therefore, the optically 
thin synchrotron 
spectrum will be a power-law with index $\alpha=(p-1)/2$ 
(equation (\ref{eq3:synspec})). The SSC flux at the photon energy $\epsilon_s$ 
can then be predicted from the observed synchrotron flux as \cite{1987p280,1979v76p306}
\begin{align}
F^{SSC}(\epsilon_s) \approx d(\alpha)\,\theta_d^{-2(2\alpha+3)}\,\nu_m^{-(3\alpha+5)}\,
(F_m^{Syn})^{2(\alpha+2)}\,\epsilon_s^{-\alpha}\,\ln\left(\frac{\nu_{max}}{\nu_m}\right)
\left(\frac{1+z}{\delta}\right)^{2(\alpha+2)}
\end{align}
where $\theta_d$ is the angular size of the source, $\nu_m$ is the synchrotron self 
absorption frequency, $F_m^{Syn}$ is the synchrotron flux at frequency $\nu_m$,
$\nu_{max}$ is synchrotron high frequency cutoff corresponding to the high energy cutoff
in the particle spectrum, $z$ is the redshift of the source and $\delta$ is the 
Doppler factor of the jet. Here $d(\alpha)$ is a function depending only on $\alpha$ 
and has values $d(0.25) = 130$, $d(0.50) = 43$,
$d(0.75) = 18$ and $d(1.00) = 9.1$. If we assume the X-ray emission from AGN jets as 
due to SSC process, then by comparing the predicted SSC flux at X-ray energy with the 
observed flux at that 
energy one can estimate the Doppler factor of the jet \cite{1993v407p65}.

\subsubsection{External Compton}
In external Compton mechanism, the photons which are produced outside the emission 
region are scattered off to high energies by inverse Compton process. A general case
assuming the scattering of an isotropic distribution of soft target photons to high energies
by a distribution of relativistic electrons can be studied using equations 
(\ref{eq3:comspec}) and/or (\ref{eq3:comspeckn}). However in case of AGN jet
these equations are invalid. Here as the emission region moves down the jet at 
relativistic speed, the plasma see an anisotropic distribution of target photons due 
to Doppler boosting. Consider the case of a spherical emission region moving down the jet 
with bulk Lorentz factor $\Gamma$. Let the inclination angle of the jet to the line of 
sight of the observer be $\theta_o$. For simplicity let us assume the external target 
photon distribution be monochromatic and isotropic in the frame of the central source.
The energy density of these target photons in the emission region frame will then be 
$\approx \Gamma^2 u_{iso}^\star$, where $u_{iso}^\star$ is the energy density of the 
isotropic external radiation field. The target photon distribution in the frame of 
the emission region will peak at energy $\epsilon^{\prime}\approx\Gamma \epsilon^{\star}$,
where $\epsilon^{\star}$ is the energy of the external photon distribution. The resultant 
inverse Compton emissivity $j_c(\epsilon,\Omega)$ at photon energy $\epsilon$ emitted in
a direction $\Omega$, due to a power-law distribution of particles described by 
equation (\ref{eq3:powpar}), is given by \cite{1995v446p63}
\begin{align}
\label{eq3:ecemiss}
j_c(\epsilon,\Omega) \approx \frac{c\,\sigma_T\,u_{iso}^\star \,K}{8\pi\,\epsilon^{\star}}
\;[\,\Gamma\,(1+\mu)\,]^{1+\alpha}\left(\frac{\epsilon}{\epsilon^{\star}}\right)^{-\alpha} \quad
\text{ergs cm$^{-3}$ s$^{-1}$ $\epsilon^{-1}$ sr$^{-1}$}
\end{align}
where $\alpha = (p-1)/2$ and $\mu=\cos\;\theta_z$, with $\theta_z$ being the angle between the 
emitted photon and the jet axis. Since the scattering is assumed to happen in Thomson regime, 
equation (\ref{eq3:ecemiss}) is valid only for the scattered photon energies satisfying 
the relation 
\begin{align}
\gamma_{min}^2\le \frac{\epsilon}{\Gamma \epsilon^{\star}(1+\mu)}\le \gamma_{max}^2
\end{align}
The  quantity $\Gamma(1+\mu)$ can be written in terms of the viewing angle $\theta_o$ 
and the Doppler factor of the jet $\delta$ as 
\begin{align}
\Gamma (1+\mu) =  \delta \;\frac{1+\cos \;\theta_o}{1+v/c}
\end{align}
where $v$ is the velocity of the emission region along the jet.

\section{Equipartition Magnetic Field}\label{sec2:equi}
Consider a spherical source of volume $V$ with a power-law distribution of 
electrons described by equation (\ref{eq3:powpar}) cooling in a 
magnetic field $B$. The total energy of the electrons will be 
\begin{align}
U_e &= V m_e\, c^2\int\limits_{\gamma_{min}}^{\gamma_{max}}
\gamma \,N(\gamma)\;d\gamma \nonumber \\
&=\frac{Vm_e \,kc\,^2}{2-p} \;(\gamma_{max}^{2-p}-\gamma_{min}^{2-p})
\end{align}
The total synchrotron luminosity of the source using equation 
(\ref{eq3:synpower}) (assuming $\beta\approx 1$) will be  
\begin{align}
L &= V\int\limits_{\gamma_{min}}^{\gamma_{max}}P_{syn}(\gamma)\,N(\gamma)\;d\gamma\\
&= \frac{4\,\sigma_T\,c\,U_B\,k\,V}{3\,(3-p)}(\gamma_{max}^{3-p}-\gamma_{min}^{3-p})
\end{align}
We can write $\gamma_{min}$ and $\gamma_{max}$ in terms of characteristic 
synchrotron photon frequency $\nu_{min}$ and $\nu_{max}$ using 
equation (\ref{eq3:synphi}) as $\gamma_{min}=(\nu_{min}/\nu_L)^{1/2}$ and 
$\gamma_{max}=(\nu_{max}/\nu_L)^{1/2}$. Then the ratio of total energy of the 
electrons to the synchrotron luminosity will be
\begin{align}
\frac{U_e}{L}=\frac{A}{B^{3/2}}
\end{align}
where $A$ is a constant that depends only on the particle spectral index $p$.
If the source had other particles like hadrons along with electrons then the total
particle energy will be $U_p=\nolinebreak a\,U_e$, where $a>1$. Then the total energy of the source 
will be 
\begin{align}
U_{tot}&=U_p+U_B \\
		&=\frac{aAL}{B^{3/2}}+V\frac{B^2}{8\pi}
\end{align}
The magnetic field $B_{min}$ for which the total energy of the system is minimum 
can be obtained by solving  
\begin{align}
\left(\pd{U_{tot}}{B}\right)_{B=B_{min}} = 0
\end{align}
and we get
\begin{align}
\label{eq3:bmin}
B_{min}=\left(\frac{6\pi\, aAL}{V}\right)^{2/7}
\end{align}
On the other hand, the equipartition magnetic field $B_{eq}$ obtained from the relation $U_p=U_B$
will be 
\begin{align}
\label{eq3:beq}
B_{eq}=\left(\frac{8\pi\, aAL}{V}\right)^{2/7}
\end{align}
Equations (\ref{eq3:bmin}) and (\ref{eq3:beq}) differ by an factor less than 
$10$ percent and the total energy corresponding to $B_{eq}$ and 
$B_{min}$ will be 
\begin{align}
U_{tot}(B_{eq}) &= 2V\left(\frac{B_{eq}^2}{8\pi}\right) \quad \textrm{and} \\
U_{tot}(B_{min}) &= \frac{7}{3}V\left(\frac{B_{min}^2}{8\pi}\right) 
\end{align}
or
\begin{align}
U_{tot}(B_{eq}) = \frac{6}{7}\left(\frac{B_{eq}}{B_{min}}\right)^2 U_{tot}(B_{min}) \approx 1.01\, U_{tot}(B_{min})
\end{align}
Hence it is customary to use equipartition value for the magnetic field while
modelling the sources to ensure a minimum energy condition 
\cite{1959v129p849,book:khembhavi_narlikar}.

\section{Hadronic Processes}\label{sec2:hadronicprocess}
If protons are accelerated to high energies they can lose their energy through 
synchrotron emission and hadronic interactions. The main hadronic interactions by 
which an energetic proton can lose its energy are the following:
\begin{itemize}
\item Bethe-Heitler process:
\begin{align}
p+h\nu \rightarrow p + e^+ + e^-
\end{align}  
where $h\nu$ is a photon. The photon threshold energy in the rest frame of 
proton for the Bethe-Heitler process is the sum of rest mass energies of 
the electron and positron (i.e. $1.022$ MeV). The 
cross section for this process in case of an ultra relativistic proton ($\beta\approx1$)
can be expressed in terms of the photon momentum $k^\prime$ when 
$2\le k^\prime\le 4$ as \nolinebreak \cite{1992v400p181}
\begin{align}
\sigma_{BH}(k^\prime)\approx \frac{2\pi}{3}\,\alpha_{fs}\,r_e^2\,Z^2
\left(\frac{k^\prime-2}{k^\prime}\right)^3 \left(1+\frac{1}{2}\,\eta+\frac{23}{40}\,\eta^2+
\frac{37}{120}\,\eta^3+\frac{61}{192}\,\eta^4\right)
\end{align}
where $\alpha_{fs}$ is the fine structure constant, $r_e$ is the classical electron radius,
$Z$ is the charge of the ion in units of $e$ and $\eta=(k^\prime-2)/(k^\prime+2)$.
For $k^\prime>4$ the cross section can be approximated as \cite{1992v400p181}
\begin{align}
\sigma_{BH}(k^\prime)\approx \alpha_{fs}\,r_e^2\,Z^2\,\bigg\{&\frac{28}{9}\,\ln\,2k^\prime-\frac{218}{27}
\nonumber \\ &+\left(\frac{2}{k^\prime}\right)^2 
\bigg[6\,\ln\,k^\prime-\frac{7}{2}+\frac{2}{3}\,\ln^2\,2k^\prime
\nonumber \\ &\qquad\qquad\quad-\frac{1}{3}\pi^2\,\ln\,2k^\prime 
+2\zeta(3)+\frac{\pi^2}{6}\bigg] \nonumber \\
&-\left(\frac{2}{k^\prime}\right)^4\left(\frac{3}{16}\,\ln\,2k^\prime+\frac{1}{8}\right)
\nonumber \\ &-\left(\frac{2}{k^\prime}\right)^6
\left(\frac{29}{9.256}\,\ln\,2k^\prime-\frac{77}{27.512}\right)\bigg\}
\end{align}
\item proton-proton collision:
\begin{align}\label{eq3:ppcoll}
p+p \rightarrow X + \sum_{i=1}^{m} \pi_{i} \quad \text{where}\quad X\rightarrow\text{hadrons}
\end{align}  
Here $m$ is the multiplicity of secondary pions. 
The threshold energy for this reaction is
$E_{th}=2m_\pi c^2(1+m_\pi/4m_p)\approx 280$ MeV, where $m_\pi$ and $m_p$ are the masses 
of the $\pi^0$-meson and the proton. When the incident proton is in the GeV to TeV 
energy region, the total cross section can be approximated by \cite{book:aharonian} 
\begin{align}
\sigma_{pp}(E_p) \approx 30\,[0.95+0.06\;\ln\;(E_{kin}/1\text{GeV})] \quad \text{mbarn}
\end{align}
where $E_p$ is the initial proton energy and $E_{kin} = E_p - m_pc^2$, 
and $E_{kin}\ge 1$ GeV. Here it is assumed that $\sigma_{pp}=0$ at lower energies.
\item photo-meson process:
\begin{align}
p+h\nu \rightarrow X + \sum_{i=1}^{m} \pi_{i} \quad\text{where}\quad X\rightarrow\text{hadrons} 
\end{align}  
The cross section for the photo-meson process increases starting from the threshold energy
of the photons $E_{th}=150$ MeV (in the rest frame of protons) reaching their maximum 
value $\sim 3\times10^{-28}$ cm$^2$ at $E\sim300-400$ MeV, and then decrease 
\cite{1992v257p465}.
\end{itemize}

Decay modes for the pions produced in these reactions are as follows
\begin{align}
\label{eq3:pd1}
\pi^{0} \; &\rightarrow \; 2 \gamma \\
\label{eq3:pd2}
\pi^+(\pi^-) \;&\rightarrow \; \mu^+(\mu^-)+ \nu_{\mu}(\bar{\nu}_{\mu}) \\
\label{eq3:pd3}
\mu^+(\mu^-)\; &\rightarrow \; e^+( e^-) + \nu_e(\bar{\nu}_e) + \bar{\nu}_{\mu}(\nu_{\mu}) 
\end{align} 
The decay products, $\gamma$-rays and pairs, can then initiate an 
electromagnetic cascade by causing the production of further pairs and $\gamma$-rays.

\section{Emission Models}
The radio-to-UV/X-ray radiation from AGN jets are generally attributed to synchrotron 
emission due to cooling of relativistic non-thermal electrons in a magnetic field. 
However there are two different approaches concerning the high energy emission viz. 
\emph{leptonic} model and \emph{hadronic} model. High energy radiation ranging from 
MeV to TeV energies are generally observed from blazars. Hence these models are 
discussed in the context of these sources.
In leptonic models, the high energy radiation will be dominated by the inverse Compton
emission from the same ultra relativistic electrons producing synchrotron radiation
\cite{1985v298p114,1992v397p5,1992v256p27,1994v421p153}. 
On the other hand, in hadronic models, the high energy radiation is mainly due to pair cascades 
initiated by the interaction of relativistic protons with photons and proton synchrotron
radiation \cite{2000v515p149,1993v269p67}.

\subsection{Leptonic Models} \label{sec2:leptmod}
Leptonic models assume the electrons in the jets are accelerated to ultra relativistic
velocities via Fermi acceleration process. Whereas protons are not sufficiently accelerated  
and their energies remain lower than the threshold energy required to initiate 
the hadronic interactions. Some models assume the hadrons are cold and mainly provide 
the inertia required for the jet to reach up to kpc/Mpc scales 
\cite{1997v286p415,2000v543p535}. The 
accelerated electrons beside emitting synchrotron
radiation also scatter off soft target photons to hard X-ray and $\gamma$-ray energies by
inverse Compton process. The possible target photons for this process are the synchrotron
photons produced within the jet (SSC) (\S \ref{sec2:ssc}) and/or the external photons 
entering into the jet. The sources of external photons which can play an important role in 
explaining the high energy radiation in AGN jet are 
\begin{itemize}
\item the radiation from the accretion disk around the central massive object 
\item the reprocessed accretion disk radiation from the broad line emitting region 
\item the infra-red photons from the dusty torus. 
\end{itemize}
Some parameters of the leptonic model can be constrained from the relativistic Doppler 
boosting required for the high energy radiation to be transparent against the pair 
production opacity with soft photons (\S \ref{sec1:gamma_trans}). However this
effect may be non-negligible at very high energies and the resultant spectrum will 
be hard. Also, the synchrotron radiation from the secondary electrons may become 
important \cite{2008v387p1206}. Moreover, the detection of subluminal velocities 
($\beta_{app}<1$) in the sub-pc scale jets of few TeV blazars suggest that the
relativistic jets of these sources decelerate. The varying Doppler factor due to
this deceleration will have a significant impact on the observed properties of the 
blazars \cite{2003v594p27}.

In simplistic approaches, the underlying 
electron distribution is either a single or broken power-law with index/indices 
inferred from the observed photon spectral index/indices.
A reasonable estimate of the parameters can then be obtained from the
observations. For example, the underlying magnetic field can be estimated considering
equipartition between the electrons and the magnetic field energy densities 
(\S \ref{sec2:equi}). As discussed earlier, the Doppler factor can be 
estimated from the observed 
superluminal motion of the knots or using $\gamma$-ray transparency 
(\S \ref{sec1:relativistic}). Also, from the measured variability time-scale, the 
size of the emission region can be constrained. While these simplistic models have been
successful in reproducing the spectrum of AGN jets, they lack a self-consistent basis for 
the shape of the electron distribution. 

A more realistic approach consists of the solution of a kinetic equation involving 
acceleration of the particles and radiative as well as non-radiative cooling mechanisms
\cite{1962v6p317}. 
One zone models assume the observed emission as a result of efficient cooling of
non-thermal electrons from a region with tangled magnetic field. The evolution of the 
particle distribution $N(\gamma,t)$ in this region can be described in its simplest form
by the kinetic equation as  
\begin{align}
\label{eq3:kinetic}
{\partial N(\gamma,t) \over \partial t} + 
{\partial  \over \partial \gamma} [P(\gamma,t) N(\gamma,t)] = Q (\gamma,t)
\end{align}
where $\gamma$ is the Lorentz factor of the electron, $P(\gamma,t)$ is the energy loss rate
and $Q(\gamma,t)$ is the injection rate of non-thermal particles. The injection can be
a single burst of non-thermal particles injected at time $t=0$ 
(\emph{one-time} injection models)
or a continuous injection of non-thermal particles. If the losses  
are mainly due to synchrotron and inverse Compton processes, then for a 
power-law distribution of particles similar to the one given in equation 
(\ref{eq3:powpar}), the resultant photon spectrum will be a power-law with 
index $(p-1)/2$. However since the energy loss rate is proportional to $\gamma^2$
for these processes (\S \ref{sec3:synrad} and \S \ref{sec3:icrad}), the high energy 
particles cool more efficiently than the low energy ones. This leads to a
depletion of high energy particles in case of one time injection models and 
gives rise to a 
time-dependent high energy cut off in the non-thermal particle distribution.
The emitted photon spectrum will then exponentially decrease at high energies 
corresponding to this cut off energy in the particle spectrum. Alternatively one can
infer the age of the emission region by translating this exponentially decreasing
feature in the photon spectrum to the high energy cut off in the particle 
distribution. On the other hand, in the continuous injection models, the depleted high energy
electrons are continuously replenished. This gives rise to a broken 
power-law particle distribution with a break at energy for which the cooling 
time-scale is equal to the age of the emission region. The resultant photon
spectrum will then be a broken power-law instead of one with an exponential 
cut off.

Two zone models are more involved than one zone models where acceleration 
of particles are also considered along with the cooling mechanisms.
According to these models particles are accelerated to 
relativistic energies in an acceleration region. These high energy particles
are then injected into 
a cooling region where they lose most of their energies by radiative 
and non-radiative processes. 
The evolution of the particles are governed by the kinetic
equations corresponding to acceleration region and cooling region. These
equations can be written in their simplest form as 
\begin{align}\label{eq3:kin_ar}
{\partial n(\gamma,t) \over \partial t} + 
{\partial  \over \partial \gamma} \left[\left(P_{AR}(\gamma,t) 
+ \dot{\gamma}_{acc}\right)n(\gamma,t)\right] 
+ \frac{n(\gamma,t)}{t_{esc}} = Q (\gamma,t) 
\end{align}

\begin{align}\label{eq3:kin_cr}
{\partial N(\gamma,t) \over \partial t} + 
{\partial  \over \partial \gamma} [P_{CR}(\gamma,t) N(\gamma,t)] = \frac{n(\gamma,t)}{t_{esc}}
\end{align}
where the equation (\ref{eq3:kin_ar}) governs the evolution in the acceleration region 
and the equation (\ref{eq3:kin_cr}) in the cooling region.
Here $n(\gamma,t)$ and $N(\gamma,t)$ are the particle distribution in the acceleration
region and the cooling region, $P_{AR}(\gamma,t)$ and $P_{CR}(\gamma,t)$ are the 
respective energy loss rates, $\dot{\gamma}_{acc}$ is the particle acceleration rate, 
$t_{esc}$ is the particle escape time-scale in acceleration region and $Q (\gamma,t)$
is the particle injection rate. The injection into the acceleration region can be 
mono energetic  electrons or a residual particle distribution of an earlier 
acceleration process. If the acceleration happens at a shock front then the 
acceleration rate $\dot{\gamma}_{acc}$ can be approximated as 
\begin{align}
\dot{\gamma}_{acc} \approx \frac{\gamma}{t_{acc}}
\end{align}
where $t_{acc}$ is the acceleration time-scale and can be estimated from the 
theory of diffusive shock acceleration \cite{1983v46p973}. The index of
the particle spectrum in acceleration region is governed by the acceleration 
and escape time-scales. Whereas the maximum energy to which the particles can be 
accelerated is determined by the acceleration and cooling time-scales. One zone
and two zone models involve many parameters even in their simplistic form and are 
often cannot be constrained. However these models can throw light on the 
underlying physics of the source and its predictions compared with 
future observations will help us to understand these sources better.

Sambruna et al. \cite{2002v571p206} used a one-time injection model to
explain the broadband emission from the knots of several AGN detected in 
radio, optical and X-ray.
The spectra modelled by them predicts an exponential cut off at
optical/UV or X-ray energies. Whereas  Liu \& Shen \cite{2007v668p23} 
proposed a two zone model to explain the X-ray emission from the 
knots of M87 (a nearby FRI radio galaxy). One zone models failed to 
reproduce the observed flux and/or the spectral index from the knots of 
this source. 

In case of blazars the picture is different. One-zone models are used by 
various authors to explain the emission from  
blazars \cite{1996v463p555,1997v320p19}. These models assume a homogeneous
distribution of particles and magnetic field throughout the emission region.
The broadband spectra of the blazars are successfully explained by this model. 
However the time lags observed between the flares at different frequencies 
cannot be explained under this model. Also this model requires the light 
travel time to be shorter than the synchrotron cooling time-scales in order 
to satisfy the homogeneity of the particle distribution. This introduces a
constraint on the size of the emission region. 
Two zone models offers more 
insight into these sources \cite{1998v333p452,2000v536p299}. Kirk et al.
\cite{1998v333p452} solved the kinetic equation considering the temporal
as well as the spatial variation of the particle distribution in the 
cooling region. They showed that the time lags between the flares at different 
frequencies can be an outcome of the difference between cooling 
and acceleration time-scales.

\subsection{Hadronic Models}
In hadronic models, the protons are accelerated along with electrons to ultra-relativistic 
energies and cool off mainly through proton synchrotron and photo-meson 
interactions (synchrot\-ron-proton blazar model) \cite{1993v269p67}. 
The dominant channels for the photo-meson process 
are \cite{1989v221p211}
\begin{align*}
p+\gamma &\rightarrow \pi^0+p \\
p+\gamma &\rightarrow \pi^{+}+n \\
p+\gamma &\rightarrow \pi^+ + \pi^- + p 
\end{align*}
The pions decay as shown in equations (\ref{eq3:pd1}), (\ref{eq3:pd2}) and (\ref{eq3:pd3})
and decay products initiate an electromagnetic cascade.
Hadronic models explain the high energy emission as a result of these cascades and 
proton synchrotron emission. The type of the resulting cascade spectrum depends upon the 
compactness parameter $l$ which measures the optical depth with respect to 
pair creation (equation (\ref{eq1:compactness})). For 
emission region with small compactness parameter $l\ll 60$
the electromagnetic cascade terminates after few generations and hence the photon luminosity 
is concentrated at high energies. On the other hand, as $l$ increases, 
more and more generations 
shift power towards lower energies. 

The target photons for photo-meson process can be the
electron synchrotron radiation and/or the external photons. 
M{\"u}cke \& Protheroe \cite{2001v15p121} using Monte Carlo technique simulated the
proton interactions and the subsequent cascades. They considered
the co-acceleration of protons along with electrons while the 
synchrotron emission from the latter is responsible for the 
low energy emission from blazars. These photons serve as 
target photons for the p$\gamma$ interactions.
They showed that the cascades initiated by the $\pi^0$ decay and 
$\pi^\pm$ decay generate a featureless $\gamma$-ray 
spectra. In contrast, the proton synchrotron cascades and $\mu^\pm$ synchrotron cascades 
produce a two-component $\gamma$-ray spectrum
commonly observed in flaring blazars.
In general, direct proton
and $\mu^\pm$ synchrotron radiation is mainly responsible for the high energy bump in 
blazars, whereas the low energy bump is dominated by synchrotron radiation from the primary 
electrons, with a contribution from the secondary electrons \cite{2003v18p593}. 

Unlike leptonic models, hadronic blazar models result in neutrino emission through
the production and decay of charged mesons (equations (\ref{eq3:pd2}) and 
(\ref{eq3:pd3})).
Another important source of high energy neutrinos is the production and decay of
charged kaons. In the case of p$\gamma$ interactions positively charged kaons are 
produced \cite{1999v16p160,2000v124p290}. They decay into muons and direct high
energy muon-neutrinos. 
These muon-neutrinos will not have suffered energy losses through $\pi^\pm$ and 
$\mu^\pm$ synchrotron radiation unlike the ones originating from $\pi^\pm$ and 
$\mu^\pm$ decay. 
Therefore they appear as an excess in comparison to the remaining neutrino
flavors at the high energy end of the emerging neutrino spectrum.
Detection of predicted neutrino spectrum play an important role in validating the
hadronic models.

Investigation of time-dependent hadronic models 
is very difficult because of time consuming Monte-Carlo cascade
simulations. Also 
it is difficult to reconcile their rapid variability observed in blazars 
($< 1$ hour) with the radiative cooling time-scales of protons \cite{2000v5p377}. 

\begin{figure}[htp]
\begin{center}
\includegraphics[width=125.2mm,bb= 0 0 82 71]{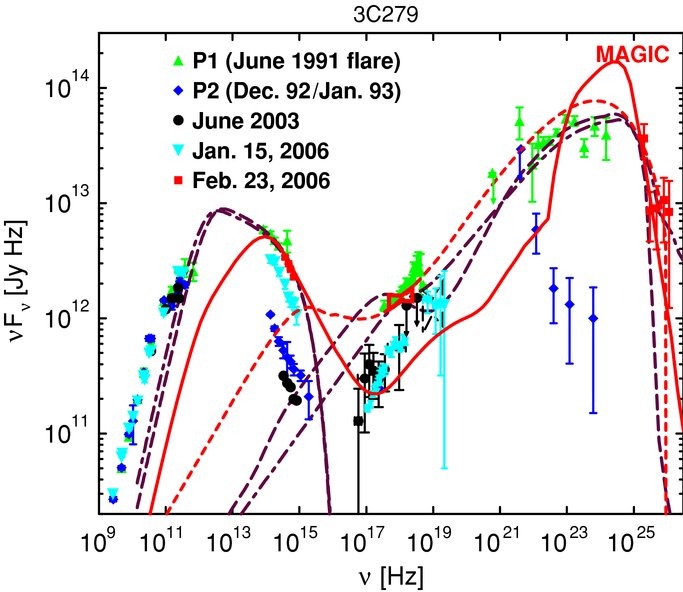} 
\caption[Spectral fits of 3C279 using leptonic and hadronic models]
{Spectral fits to the spectral energy distribution of 3C279 using a leptonic 
external-Compton model (solid (red)); 
leptonic SSC model (short-dashed (red)); hadronic model with electron synchyrotron photons as 
targets for the photo-meson process (dot-dashed (maroon)) and hadronic model with electron 
synchrotron + external photons as targets for photo-meson process (long-dashed (maroon)).
Figure reproduced from B\"{o}ttcher et al. \cite{2009v703p1168}}
\label{fig2:lept_had}
\end{center}
\end{figure}
B\"{o}ttcher et al.\cite{2009v703p1168} studied the simultaneous multi wavelength 
observation of the very high energy (VHE) blazar 3C279 using the 
leptonic and hadronic models. Leptonic one zone model requires unrealistic 
parameters to explain the observed spectrum. 
Whereas, the hadronic synchrotron-proton blazar model is able to fit the 
broadband spectrum successfully. They also considered the contribution from the 
external target photons for hadronic interactions in order to reduce the 
energy loss time-scale. 
In Figure \ref{fig2:lept_had} we show their fit to the spectral energy distribution of 3C279 
based on leptonic and hadronic models.

\section{Aim of the Thesis}
Despite the availability of enormous amount of information about AGN by virtue of 
high resolution and high sensitivity experiments at present,  
there exist a large amount of uncertainties regarding the physics of various 
observed features \cite{2006v44p463,book:hughes,1997rjaproc,1997v35p607}. 
In the work presented in this thesis we shall attempt to understand certain 
features of AGN jets in view of the recent observations. 

\begin{itemize}

\item
In {\bf Chapter \ref{chap:agn_knots}}, we describe our works to interpret the 
underlying physics of the AGN knots. 
The on-board X-ray satellite \emph{Chandra}\footnote {\emph{Chandra} is
a satellite borne X-ray telescope launched by NASA.} studied the knots of 
several AGN for which radio and optical informations are already available
\cite{2000v542p655,2001v549p161,2002v571p206,2000v544p23}. 
The X-ray emission from these knots can be either due to synchrotron emission or 
inverse Compton emission \cite{2000v542p655,2001v549p161,2000v544p23,
2001v326p7,2002v571p206,2001v556p79}.
If the X-ray flux lies below the extrapolation of radio-to-optical flux 
then the synchrotron origin is plausible else the emission may be due to 
inverse Compton process. Sambruna et al. \cite{2002v571p206} used one 
zone model to explain the 
X-ray emission from these knots. 
This model gives rise to a spectrum with a time-dependent exponential 
high-frequency cutoff. Hence for  
the knots with synchrotron origin of X-ray, 
their model predicted an exponentially decreasing steep spectrum at this energy.
However the photon spectral index measured from the short duration observations
of the knots at X-ray energies
contradicts this prediction \cite{2001v556p79}. On the other hand, the 
acceleration process may exist for 
a longer duration such that a continuous injection of non-thermal particles 
may be viable. 
We model the observed radio-optical-X-ray spectra of the knots of the AGN
1136-135, 1150+497, 1354+195 and 3C 371 by using a continuous injection
plasma model. We assume the knot to be a uniform expanding sphere with 
continuous injection of non-thermal particles. The electron
distribution and resultant radiation spectrum is computed by taking into
account synchrotron cooling, inverse Compton scattering of cosmic microwave 
background and adiabatic cooling due to the expansion of the sphere. 
The continuous injection of particles will generate a break in the electron
distribution at energy where the cooling time-scale is equal to the age of 
the emission region (cooling break). The energy at which the index 
changes is time dependent, and 
synchrotron/adiabatic cooling is dominant for the particles with energy greater 
than this break energy. 
We show that this model can successfully reproduce the observed spectrum
from the knots of these sources.

We also interpret the knots as an outcome
of an internal shock which we discuss in \S \ref{sec4:internalshock}.
An internal shock interpretation for the knots was first suggested by 
Rees \cite{1978v184p61} to explain the knots of M87. 
We study the jet dynamics and the 
viability of the above mentioned continuous injection model using a simple 
internal shock model.
The central engine of the AGN is assumed to emits blobs of matter sporadically 
at relativistic speeds. In the process, the fast moving blobs will collide
with the previously ejected slow moving ones thereby forming a shock. 
We implement the model for the knots of AGN and compute the time-evolution 
of the non-thermal particles
produced. Also we compare the results obtained with the broadband fluxes
from knots of several AGN jets and their observed positions. The motivation
here is to find quantitative values of the model parameters by demanding 
that the model can self consistently explain the observation. 

It is also noted that the observed X-ray flux from the knots in the 
jet of the nearby galaxy M87 
cannot be explained by considering simple one zone models involving continuous 
injection or one-time injection of non-thermal particles \cite{2005v627p140}. 
Perlman \& Wilson \cite{2005v627p140} proposed a modified CI model where the 
volume within which particle acceleration occurs is energy dependent. Using 
this phenomenological model they observed that the particle acceleration 
takes place in a larger fraction of the jet volume in the inner jet than the 
outer jet. Also particle acceleration region occupy a smaller fraction of the 
jet volume at higher energies. Liu \& Shen \cite{2007v668p23} proposed a 
two-zone model with the acceleration region and cooling region spatially 
separated. The advection of particles from the acceleration region to the 
cooling region introduces a break in the particle spectrum which along with 
the cooling break in the cooling region produces a double broken power law.
The synchrotron emission from such a particle distribution is used to fit the 
observed spectra. We propose an alternate two-zone model to explain the 
observed X-ray flux from these knots and
which we discuss in \S \ref{sec4:m87}. Here we 
consider the acceleration of power-law distribution of electrons in an 
acceleration region which are then injected into a cooling region and 
cool via radiative processes. We show that one will obtain a broken
power-law particle spectrum when the initial spectrum injected into the 
acceleration region is flatter than the characteristic spectrum of the 
acceleration region.
This particle distribution will then develop an additional cooling break 
depending upon the age of the knot while cooling in the cooling region.
The synchrotron emission from the resultant particle distribution in the 
cooling region is used to reproduce the broadband spectrum
of the knots of M87 jet.

\item
In {\bf Chapter \ref{chap:mkn501}}, we discuss a model to explain
the observed limb-brightened structure of the blazar MKN 501. 
The high resolution image of the nearby BL Lac object MKN 501 in radio show a 
transverse jet structure with the edges being brighter than the central spine.
Such a feature is commonly referred as ``limb-brightened'' structure. This 
feature is usually explained by the ``spine-sheath'' model where the velocity
at the spine is larger than the velocity at the boundary. The limb-brightened 
structure can then be explained by considering the differential Doppler 
boosting for a proper combination of the bulk Lorentz factors and the 
viewing angle. 
The viewing angle deduced for MKN 501 based on this model is $\gtrsim 15^\circ$
\cite{2004v600p127}. However the high-energy studies of MKN 501 demand the viewing angle 
to be $\sim5^\circ$. Since the high-energy emission is originated from the inner part 
of the jet close to the nucleus, a possible bending of the jet was suggested by
the earlier work \cite{2004v600p127}. We explain
the observed limb-brightened structure of MKN 501 jet as an outcome of efficient 
particle acceleration process at the jet boundary.  Here the particles can be 
accelerated by shear acceleration or turbulent acceleration. We deduce
the required condition for the shear acceleration to be dominant over 
turbulent acceleration and discuss the diffusion of particles accelerated 
at the boundary into the jet medium. Also we derive the spectral index of the 
particle distribution accelerated via shear acceleration process and turbulent 
acceleration process and show the observed index at the boundary of MKN 501 jet
supports the former. The bending of the jet is not a requirement for this
interpretation unlike the explanation based on differential Doppler boosting.

\item
In {\bf Chapter \ref{chap:blazar}}, we study the temporal and 
spectral behaviour of the non-thermal emission from blazars.
The flux variability observed in blazars have been studied by several 
authors using one zone and two zone models 
\cite{1996v463p555,1998v333p452,1999v306p551,2000v536p299}. 
Kirk et al. \cite{1998v333p452} and Kusunose et al.  
\cite{2000v536p299} assumed a two zone model where particles are accelerated in a 
region presumably by a shock and escape into emission region where they lose 
their energy by radiative processes. 
Chiaberge \& Ghisellini \cite{1999v306p551} explained the short time 
variability observed in MKN 421 by dividing the emission region
into thin slices. 
We study the spectral and temporal behaviour of blazars by considering a 
two zone model under two different scenarios of acceleration process. Mono energetic 
particles are assumed to be accelerated in 
the acceleration region and then injected into the cooling region where they lose
most of their energy via radiative processes. The behaviour of the resultant spectrum
is then studied for two cases of particle acceleration process in the acceleration region. 
In the first case, the rate of particle acceleration is assumed to be energy 
dependent and in the second case, it is independent of the same. The model is then 
applied on the BL Lac object MKN 421 and its flare characteristics are studied 
under these two cases. It is found that, detailed information about the temporal 
behaviour of blazars can throw light on the underlying particle acceleration
mechanism. 

\item
Finally in {\bf Chapter \ref{chap:summary}}, we summarize the work presented 
in the thesis and discuss the possible future work.
\end{itemize}

\chapter{Emission Models and the Dynamics of AGN knots} \label{chap:agn_knots}
Knots in kpc scale jets of several AGN are recently studied in 
X-ray by the on-board X-ray satellite \emph{Chandra}
\cite{2000v542p655,2001v549p161,2002v571p206,2000v544p23}. 
\emph{Chandra} due to its excellent spatial 
resolution is able to resolve bright X-ray knots and in most of the cases it
coincides with their radio/optical counterparts \cite{2001v549p161,
2002v571p206,2001v556p79}. The radio-to-optical emission from these knots are 
generally accepted to be of synchrotron origin, whereas the X-ray emission 
could be due to synchrotron \cite{2001v326p7,2002v571p206,2001v556p79} 
or inverse Compton processes depending on its radio-to-optical ($\alpha_{RO}$)
and optical-to-X-ray index ($\alpha_{OX}$)
\cite{2001v549p161,2002v571p206,2001v556p79,
2000v544p23,2000v544p23,2000v542p655}. 
If $\alpha_{RO} > \alpha_{OX}$, then the X-ray flux lies above the extrapolation of
radio-to-optical flux and hence  a single emission mechanism may not 
explain the observed fluxes. In such a case, the X-ray emission may be due to inverse 
Compton process or it may arise from a different electron population other than the one 
responsible for the radio/optical emission. However, it should be noted here that such
a spectrum can still be a resultant synchrotron radiation from an electron distribution 
modified due to inverse Compton scattering happening at extreme 
Klein-Nishina regime \cite{2002v568p81}. The synchrotron spectrum in this case can 
satisfy the spectral indices requirement, $\alpha_{RO} > \alpha_{OX}$. 
On the other hand, if $\alpha_{RO} < \alpha_{OX}$, the X-ray flux lies below
the extrapolation of radio-to-optical flux and
synchrotron origin of X-ray is acceptable \cite{2000v544p23,2002v571p206,
2001v556p79}. The synchrotron origin of X-rays for knots was strengthened
in case of the knots of $3C 271$ since the alternate inverse Compton model would require 
exceptionally large Doppler factors \cite{2001v556p79}.

When the X-ray emission can be attributed to the inverse Compton process, 
the possible choices of target photons are radio/optical synchrotron 
photons (SSC) \cite{2000v540p69} or external photons. The source of external
photons which can be dominant at kpc scale jet is  
the cosmic microwave background\footnote {According to 
the \emph{Big Bang} theory, cosmic microwave 
background (CMB) radiation is the relic radiation left over from the formation 
of the universe.} (IC/CMB) \cite{2000v544p23,2001v549p161,
2002v571p206,2001v556p79}. The SSC interpretation of X-ray emission would require 
large jet powers and magnetic fields much lower
than the equipartition values whereas IC/CMB requires relatively low 
jet power and near equipartition magnetic fields \cite{2000v544p23}.

These possible radiative process identifications have to be
associated with (and confirmed by) dynamical models regarding the
origin and subsequent evolution of the radiating non-thermal particles.
In many models, these non-thermal particles are assumed to be 
generated by a short duration acceleration process and the particle
distribution is determined by radiative losses (one-time injection) 
\cite{2001v549p161,2002v571p206,2000v544p23,1973v26p423,1962v6p317,
book:pacholczyk}. The high energy particles cool more efficiently (\S \ref{sec3:synrad}
 and \S \ref{sec3:icrad}) and hence get depleted in time.
This give rise to a time-dependent 
high energy cut off in the non-thermal particle distribution.
If the X-ray emission is attributed to synchrotron emission
by these particles, then these models predict an
exponentially decreasing X-ray spectrum \cite{2001v549p161,2002v571p206,
2000v544p23,2001v556p79}.
This can be translated to a high energy cutoff in the electron 
distribution which in turn gives an estimate of the age of the knot.
These non-thermal electrons move with a bulk speed $v \approx c$ along the
jet. Thus from the age of the knot, one can determine the location
in the jet where the short duration acceleration process occurred. 
The distance of the knot from
the central object and the short duration of acceleration 
(much less than the age of the knot) may
naturally put strong constraints on any models of the acceleration
process. Also the photon spectral slope measured during the
short duration observations of 3C 371 is $\alpha_X \approx 1.7 \pm 0.4$ 
which is  in apparent contradiction to the predicted exponential X-ray 
spectrum by this model \cite{2001v556p79}. 
Moreover, the model requires the coincidence that the age of the knot 
be equal to the time
required for X-ray emitting electrons to cool. A larger 
survey of X-ray jets have to be sampled to confirm whether
this is statistically plausible.

On the other hand, it may also be possible that the acceleration process
exists for a duration longer than the age of the knot, and hence, 
there is a continuous
injection of non-thermal particles. The acceleration process may be due to internal
shocks formed as a result of collision between sporadically ejected relativistic 
blobs of matter from the central engine \cite{2001v325p1559}.
In such a case the duration of acceleration can be roughly the 
cross-over time of the collided blobs. If we assume a typical blob 
size $\sim$ kpc (which is roughly the size of the knots seen in radio) 
and its velocity $\sim c$, the acceleration duration can be as large as 
$\sim 10^{11}$s. In the next section we study the case where the knots of AGN are 
modelled as an spherical 
expanding emission region with a continuous injection of 
non-thermal particles and discuss the results obtained. In \S\ref{sec4:internalshock}
we interpret the same considering an internal shock scenario and study its dynamic 
properties.
Though 
one-time injection and continuous injection models can successfully 
reproduce the observed spectrum of knots of many AGN, they failed to
explain the same for the knots of a nearby AGN, M87. The X-ray flux of M87 knots 
when compared with radio-to-optical flux suggests a synchrotron origin but its flux
and/or spectral index cannot be reproduced by these simple models. In 
\S\ref{sec4:m87} we propose a modified synchrotron model to understand the 
broadband emission from the knots of M87.
 
\section{A Continuous Injection Plasma Model} \label{sec4:cont_inj}

We consider the knot of the AGN as a plasma moving relativistically 
along the jet with a bulk Lorentz factor $\Gamma$. In the rest frame, 
it is assumed that the plasma uniformly occupies an expanding sphere 
with radius $R (t) = R_o + \beta_{exp} c t$, where $R_o$ is the
initial size of the sphere and $\beta_{exp} c$ is the expansion velocity. 
Initially at $t = 0$ there are no non-thermal 
particles in the system. A continuous and constant particle injection rate 
for $t > 0$ is assumed, with a power-law distribution of energy,
\begin{align}
Q (\gamma)\; d\gamma = q_o\, \gamma^{-p}\; d\gamma \;\;\; \hbox  {for}\;\;\; \gamma > \gamma_{min}
\end{align}
where $\gamma$ is the Lorentz factor of the electrons.
The evolution of the total number of non-thermal particles in the 
system, $N(\gamma,t)$, can be conveniently described by a kinetic equation 
of form
\begin{align}
\label{eq4:kinetic}
{\partial N(\gamma,t) \over \partial t} + {\partial  \over \partial \gamma} [P(\gamma,t) N(\gamma,t)] = Q (\gamma)
\end{align}
Here $P(\gamma,t)$ is the particle energy loss rate given by
\begin{align}
P(\gamma,t) = -(\dot \gamma_{S} (t) + \dot \gamma_{IC} (t) + \dot \gamma_{A} (t))
\end{align}
where $\dot \gamma_{S} (t)$, $\dot \gamma_{IC} (t)$ and $\dot \gamma_{A} (t)$ are
the cooling rates due to synchrotron, inverse Compton scattering of the cosmic microwave 
background (IC/CMB) radiation and adiabatic expansion respectively. These 
cooling rates are given by
\begin{align}
\label{eq4:synloss}
\dot \gamma_{S} (t) &= {4 \over 3}\, {\sigma_T \over m_e\, c}\, {B^2(t) \over 8 \pi}\, 
\gamma^2\\ \nonumber\\
\label{eq4:icloss}
\dot \gamma_{IC} (t) &= {16 \over 3}\, {\sigma_T \over m_e\, c^2}\, \Gamma^2 
\,\sigma_{SB} T^4_{cmb} (z)\, \gamma^2 \\ \nonumber\\
\label{eq4:adbloss}
\dot \gamma_{A} (t) &= {\beta_{exp}\, c\, \gamma \over R(t)}
\end{align}
The former two cooling rates are associated with radiative losses and the latter 
is a non-radiative energy loss associated with the loss in internal energy used up 
for expansion \cite{1962v6p317}. Here the evolving magnetic field is parameterized to be 
$B(t) = B_0 (R(t)/R_0)^m$ and $T_{cmb} (z) = 2.73 (1 + z)$ is the temperature of the 
CMB radiation at the redshift $z$ of the source. Note that the time $t$ and other 
quantities in the above equations are in the rest frame of the plasma.

We solved the equation (\ref{eq4:kinetic}) numerically for $N(\gamma, t)$ using the 
finite difference scheme described by Chang \& Cooper
\cite{1970v6p1} and the resultant synchrotron
and inverse Compton spectra are computed at an observing time $t = t_o$. 
Finally, the flux at the Earth is computed taking into account the Doppler 
boosting \cite{1984v56p255}, characterized by the Doppler factor 
$\delta \equiv [\Gamma(1 - \beta\cos\theta)]^{-1}$, 
where $\beta c$ is the bulk velocity of the plasma moving down the jet and 
$\theta$ is the
angle between the jet and the line of sight of the 
observer\footnote{Here and everywhere else in this thesis 
$H_o = 75$ km s$^{-1}$ Mpc$^{-1}$ and $q_0 = 0.5$ are adopted}.

While the total non-thermal particle distribution has to be computed
numerically, a qualitative description is possible by comparing cooling
time-scales with the observation time $t_o$. The cooling time-scale
due to synchrotron and inverse Compton cooling at a given time $t$ and
Lorentz factor $\gamma$ is $t_c (t,\gamma) \approx 
\gamma/(\dot \gamma_{S} + \dot \gamma_{IC})$. Then, $\gamma_c$, defined
as the $\gamma$ for which this cooling time-scale is equal to
the observation time, $t_c (t_o,\gamma_c) \approx t_o$, becomes,
\begin{align}
\gamma_c \approx {m_e\, c \over \sigma_T}\, {t_o^{-1} \over [ {B^2 (t_o) \over 8 \pi} + 4\,
\Gamma^2 \, c \,\sigma_{SB}\, T^4_{cmb}\, (z)]}
\end{align}
The adiabatic cooling time-scale $t_a$ also turns out to be $\approx t_o$, since 
$t_a \approx R(t)/\beta_{exp} c \approx t_o$ for  $R_o \ll R(t_o)$.
Thus, the non-thermal particle distribution
at time $t = t_o$ can be divided into three distinct regions \cite{1962v6p317}:

\begin{enumerate}
\item In the regime $\gamma \ll \gamma_c$, radiative cooling is
not important and $N(\gamma,t_o) \approx q_o \gamma^{-p} t_o$. The corresponding
spectral index for both synchrotron and inverse Compton emission are $\alpha = (p-1)/2$.

\item In the regime $\gamma \gg \gamma_c$, either synchrotron
or inverse Compton cooling is dominant and $N(\gamma, t_o) \propto \gamma^{-(p+1)}$. 
The corresponding spectral index for both synchrotron and inverse Compton 
emission are $\alpha = p/2$.

\item In the regime $\gamma \approx \gamma_c$, either synchrotron
or inverse Compton cooling as well as adiabatic cooling are important and
the spectral slope is in the range $(p-1)/2 > \alpha > p/2$. 

\end{enumerate}

The computed spectrum depends on the following ten parameters: the observation
time $t_o$, the magnetic field at the time of observation 
$B_f = B (t = t_o)$, the magnetic field
variation index $m$, the
radius of the knot at the time of observation, $R_f = R(t=t_o)$, the index $p$, the minimum
Lorentz factor $\gamma_{min}$, the Doppler factor $\delta$, the bulk
Lorentz factor $\Gamma$, expansion velocity $\beta_{exp}c$ and the normalization
of the injection rate $q_o$. On the other hand, 
there are only three observational
points, namely, the radio, optical and X-ray fluxes. Clearly, the parameters
are under constrained, and it is not possible to extract meaningful quantitative
estimates.  However, the motivation here is to show that this model can
explain the observed data with reasonable values of the above parameters.

\subsection{Results and Discussion}
\begin{figure}[tp]
\begin{center}
\includegraphics[width=99.3mm, height=90.6mm,bb=0 0 563 514]{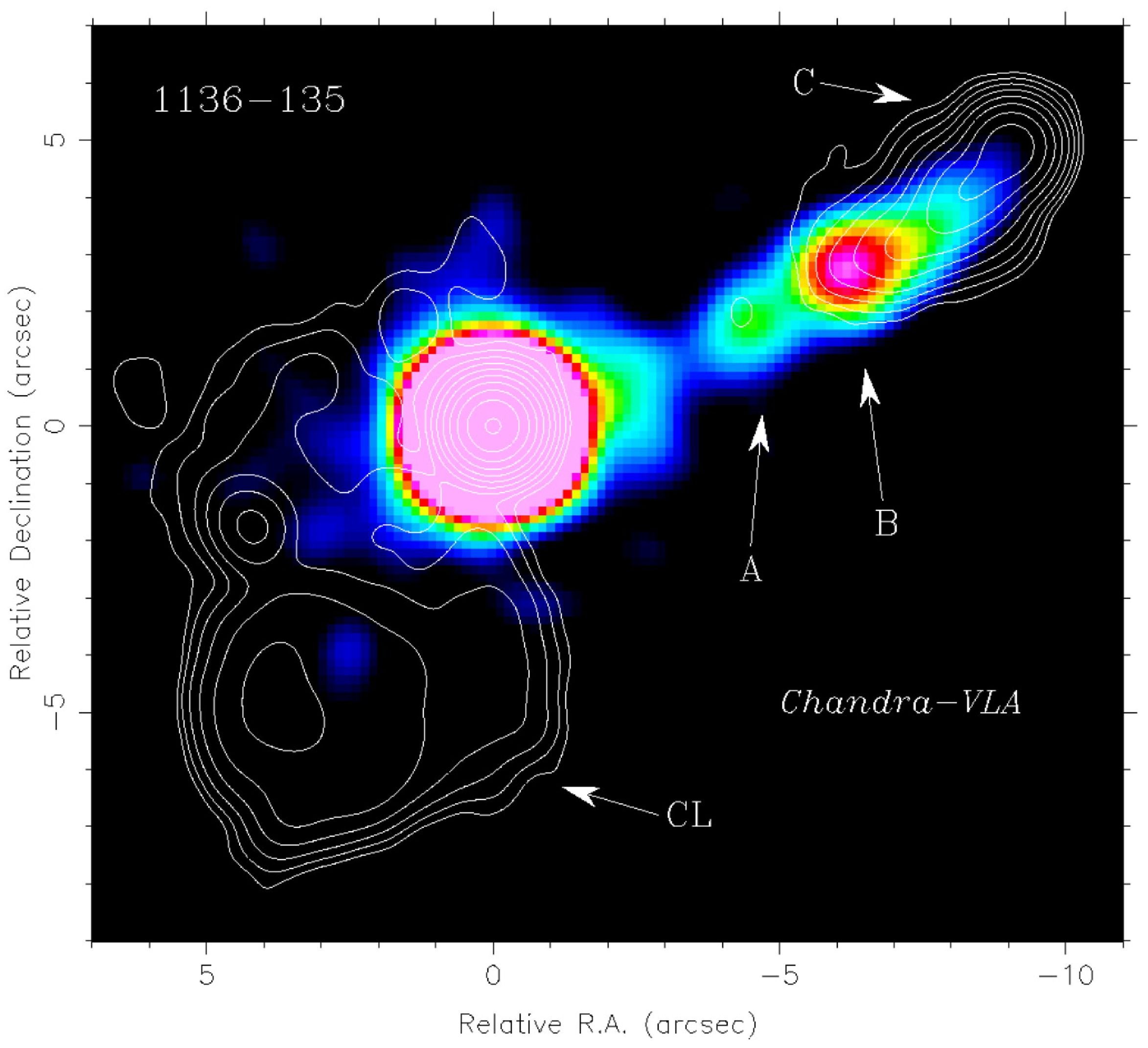} 
\caption[Multi wavelength image of 1136-135]
{\emph{Chandra} image of 1136-135 overlaid with VLA radio contours.
Figure reproduced from Sambruna et al. \cite{2002v571p206}}
\label{fig4:1136}
\end{center}
\end{figure}

\begin{figure}[tp]
\begin{center}
\includegraphics[width=158.31mm, height=140.34mm,bb= 0 0 476 422]{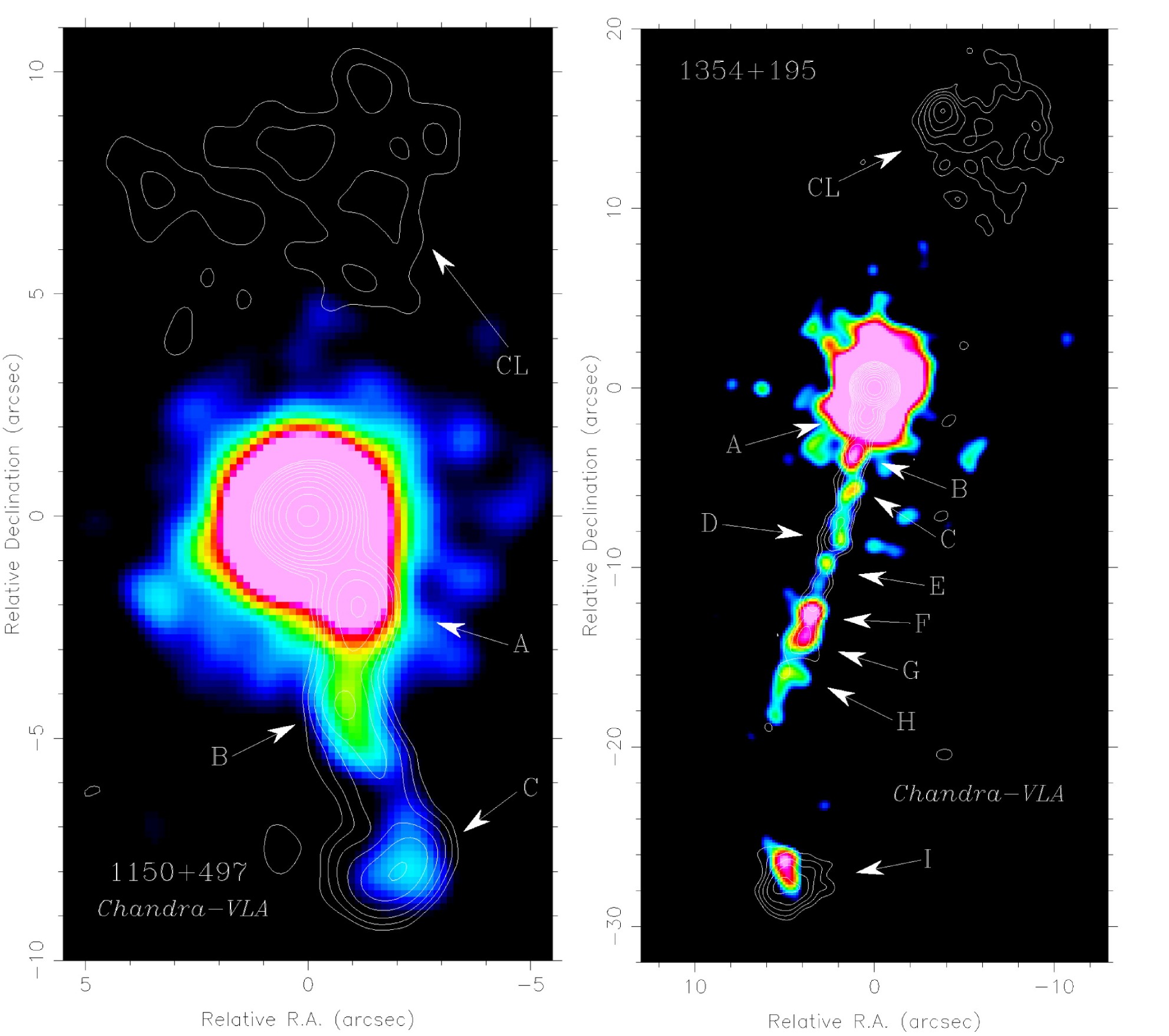} 
\caption[Multi wavelength image of 1150+497 and 1354+195]
{\emph{Chandra} image of 1150+497 and 1354+195 overlaid with VLA 
radio contours. Figure reproduced from 
Sambruna et al. \cite{2002v571p206}}
\label{fig4:1150_1354}
\end{center}
\end{figure}

\begin{figure}[tp]
\begin{center}
\includegraphics[width=87.63mm, height=134.0mm,bb=0 0 248 380]{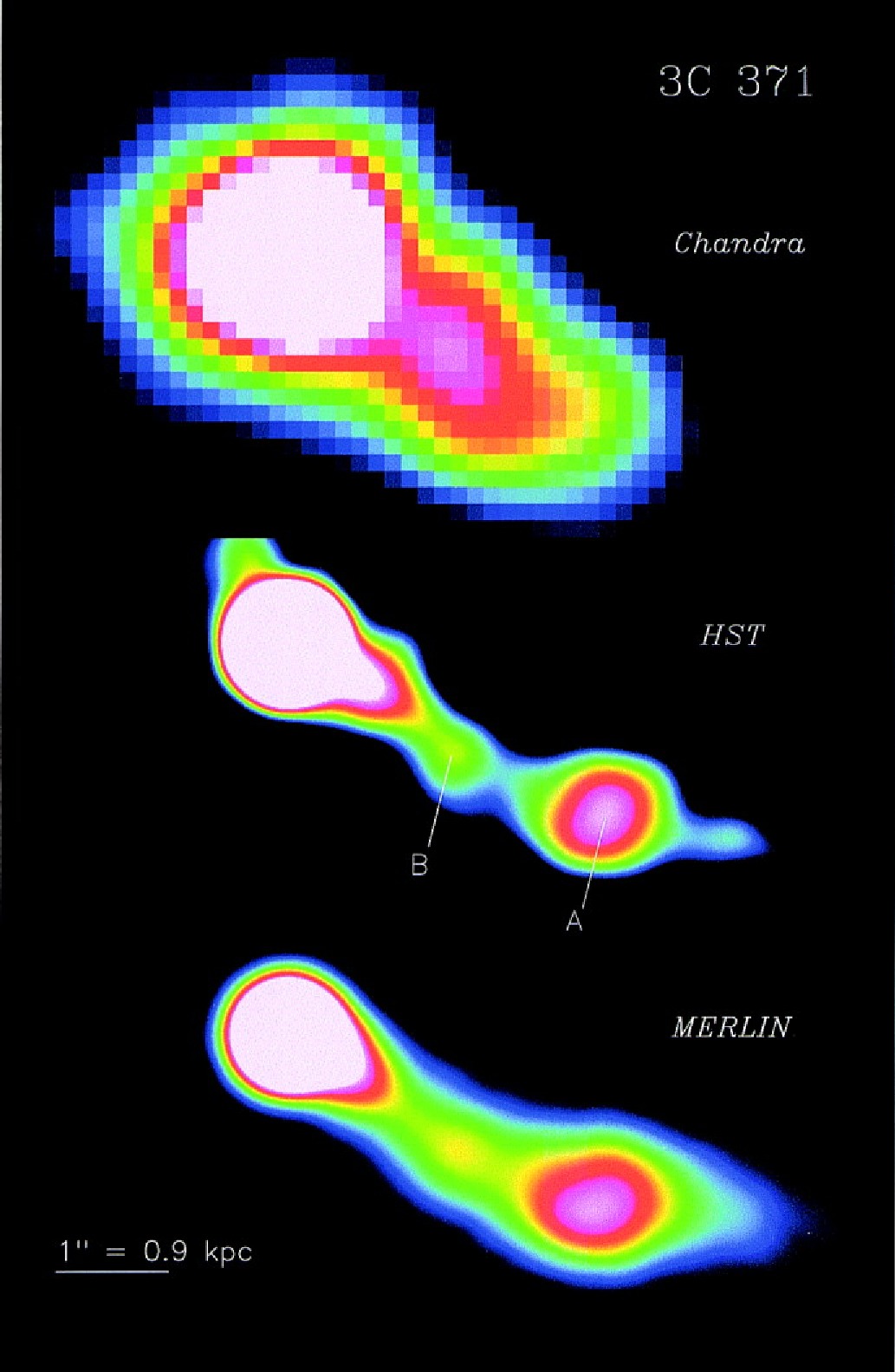} 
\caption[Multi wavelength image of 3C371]
{Multi wavelength image of 3C371 in X-ray by \emph{Chandra}(top),
in optical by \emph{HST} (middle) and in radio by \emph{Merlin} (radio). 
Figure reproduced from Pesce et al. \cite{2001v556p79}}
\label{fig4:3c371}
\end{center}
\end{figure}

\begin{figure}[tp]
\begin{center}
\includegraphics[width=150.0mm,bb=0 0 547 536]{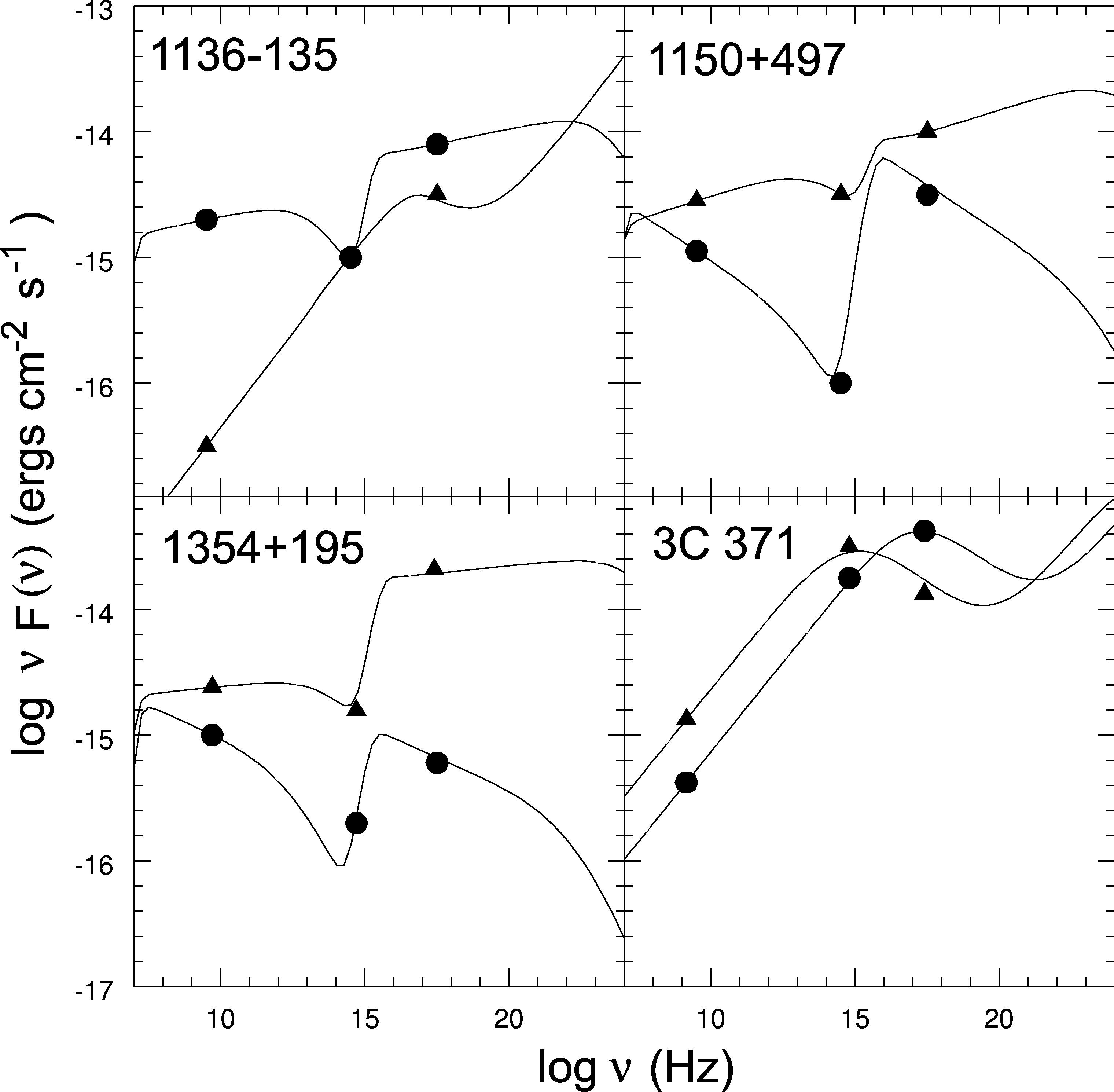} 
\caption[Spectral fit for the AGN knots using continuous injection plasma model]
{The observed fluxes in radio, optical and X-ray compared with model spectrum 
using parameters given in table \ref{ch4:table1}. The data for 3C 371 is taken from 
Pesce et al. \cite{2001v556p79}, while the rest are taken from 
Sambruna et al. \cite{2002v571p206}. Triangles correspond to knot A and circles 
to knot \nolinebreak B.}
\label{fig4:plot1}
\end{center}
\end{figure}

We applied the above model to the knots of the AGN observed by 
\emph{Chandra} along with the information available in radio and optical
energies (Figure \ref{fig4:1136}, \ref{fig4:1150_1354}, \ref{fig4:3c371}). 
In Figure \ref{fig4:plot1}, the computed spectra are compared with the data for different knots
for four sources namely 1136-135, 1150+497, 1354+195 and 3C 371. 
The values of the parameters used are tabulated
in Table \ref{ch4:table1}. The injected power in non-thermal particles in the rest
frame of the knot can be written as
\begin{align}
P_{inj} = \int_{\gamma_{min}}^\infty (\gamma m_e c^2)\, Q (\gamma)\; d \gamma = 
q_o\, {m_e\, c^2 \over p-2}\, \gamma_{min}^{-(p-2)}
\end{align}
while the total jet power can be approximated as \cite{1997v286p415}
\begin{align}
P_{jet} = \pi R^2\, \Gamma^2 \,\beta\, c\, (U_p + U_e + U_B)
\end{align}
where $U_p$, $U_e$ and $U_B$ are the energy densities of the protons,
electrons and the magnetic field respectively. Here it has been assumed that the 
protons are cold
and the number of protons is equal to the number of electrons. 
We found the jet power ranges from $10^{46}$ to
$2 \times 10^{48}$ ergs s$^{-1}$, while the injected power is 
generally three orders of magnitude lower. This means that the
non-thermal acceleration process is inefficient and most of
the jet power is expected to be  carried to the lobes. The magnetic
field $B_f$ is nearly equal to the equipartition values.

The X-ray emission for the knots in 3C371 and Knot A of 1136-135 are
identified as being due to synchrotron emission which is consistent 
with earlier works \cite{2002v571p206,2001v556p79}. 
However, in this
case the predicted X-ray spectral index is $\alpha_X = \alpha_{R} + 1/2$
instead of being exponential. Note that this relation between the
spectral indices is independent of the parameters used to fit the
data. For the rest of the sources the X-ray emission is attributed to 
IC/CMB which is again consistent with the results obtained from the earlier 
works \cite{2002v571p206}.
However, for some of the
sources the optical spectral index is now $\alpha_O = \alpha_{R} + 1/2$
instead of being exponential. 

 \begin{table}[tp]\footnotesize
\caption[Continuous Injection Plasma Model Parameters]
{Parameters for model fittings}
\label{ch4:table1}
   \begin{tabular}{lcccccccccc}
   \toprule
Source/knot& $B_f$ & $\gamma_{min}$ & $p$ & $t_o$ & $\delta$ & 
$\Gamma$ & $\beta_{exp}$ & $P_{inj}$ & $P_{jet}$ & $B_f/B_{equ}$ \\
& {\scriptsize ($\times 10^{-5}$ G)} &  &  
& {\scriptsize ($\times 10^{11}$ s)} &  & 
 &  & {\scriptsize (ergs s$^{-1}$)} & 
{\scriptsize (ergs s$^{-1}$)} &  \\
   \midrule
1136-135A & 0.9& 2.0& 2.4& 0.2& 5 & 5 & 0.8 & 44.7 & 47.5 & 0.4 \\
1136-135B & 4.0& 20.0& 2.9& 9& 5 & 5 & 0.1 & 44.2 & 47.9 & 0.5 \\
1150+497A & 2.5& 30.0& 2.85& 9& 5 & 3.5 & 0.1 & 44.2 & 47.3 & 0.4 \\
1150+497B & 4.3& 30.0& 3.3& 9& 5 & 3.5 & 0.1 & 44.1 & 47.3 & 0.75 \\
1354+195A & 1.7& 40.0& 3.0& 9& 3.5 & 2 & 0.1 & 45.8 & 48.4 & 0.04 \\
1354+195B & 8.0& 25.0& 3.2& 9& 3.5 & 2 & 0.1 & 44.7 & 47.5 & 0.63 \\
3C 371A & 1.3& 10.0& 2.4& 12& 3.5 & 3.5 & 0.1 & 42.8 & 46.4 & 0.8 \\
3C 371B & 1.0& 10.0& 2.4& 1& 3.5 & 3.5 & 0.5 & 43.5 & 46.0 & 0.9 \\
\bottomrule
\end{tabular}
\vskip 0.5cm
{\bf Columns}: (1) Source and knot name taken from Pesce et al. 
\cite{2001v556p79} for 3C371 and the rest from Sambruna et al. 
\cite{2002v571p206}; (2) Magnetic field at the observation 
time, $B=B(t=t_o)$; (3) Minimum Lorentz factor $\gamma_{min}$; 
(4) Power-law index of the injected non-thermal particle $p$; (5) Observation 
time $t_o$; (6) Doppler factor $\delta$; (7) Bulk Lorentz factor $\Gamma$; 
(8) Velocity of expansion $\beta_{exp}$ (in units of $c$); (9) Log of the 
injected power 
$P_{inj}$; (10) Log of the total jet power $P_{jet}$; (10) Ratio of the 
magnetic field to the equipartition value. For all cases, the magnetic field 
variation index $m$ and the size of the source at $t = t_o$ is fixed 
at $1.5$ and $5 \times 10^{21}$ cm, respectively. 
 \end{table}

The present model can be confirmed (or ruled out) vis-$\grave{a}$-vis 
one-time injection models, 
by future measurements of the radio $\alpha_R$, optical $\alpha_O$ 
and X-ray $\alpha_X$ spectral indices. In particular, the following
cases are possible:

{\flushleft
\begin{enumerate}
\item In the case $\alpha_R \approx \alpha_X$, the X-ray emission is probably 
due to IC/CMB. Both the continuous injection and one-time injection models 
are equally viable.

\item In the case $\alpha_X \approx \alpha_R +1/2$, the X-ray emission 
would be due to synchrotron emission
from electrons in the cooling dominated region. The continuous injection
scenario will be favored in such case.

\item In the case $\alpha_X > \alpha_R +1/2$, when the X-ray emission is 
exponentially decreasing, it
should be attributed to the high energy
cutoff in the electron distribution. The one-time injection
scenario will be favored.

\item In the case $\alpha_X < \alpha_R$, when the X-ray emission is 
exponentially increasing, it should be attributed to the low energy
cutoff ($\gamma_{min}$) in the electron distribution and the X-ray emission
should be due to IC/CMB. Both the continuous
injection and one-time injection models are equally viable.
\end{enumerate}}

Similar arguments can be put forth for the optical spectral index $\alpha_O$
as compared to the radio. It should be noted that for some older systems the
one-time injection would be the natural scenario, while for younger systems the
continuous injection would be more probable. The technique described above will
be able to differentiate between the two, and a generic constraint on
the acceleration time-scales and typical age of the knots may be obtained.
A generic model where the injection rate decays in time may then be
used to fit the observations. The measurement of spectral indices in
different wave-lengths will also reduce the number of unconstrained 
parameters in the model fitting, leading to reliable estimates of
the system parameters.

\section{Internal Shock Interpretation}
\label{sec4:internalshock}

The possibility of a continuous injection of non-thermal particles into the 
knots of AGN as well as its dynamical properties can be studied using a specific 
injection model.  
While there is no consensus on the
origin of these non-thermal particles, one of the standard model
is the internal shock scenario \cite{1978v184p61,2001v325p1559}. 
Here the particles are energized by Fermi acceleration in shocks produced
during the interaction of relativistically moving blobs ejected
from the central engines with different speeds. 
A detailed description of the shock formation
and subsequent electron acceleration is complicated and would require
numerically difficult magneto hydrodynamic simulations.
Moreover from the limited number of observables, which can be obtained
from the featureless spectrum in two or three different energy bands,
one may not be able to constrain the various assumptions and/or the
initial conditions of such a detailed study. Nevertheless, 
a qualitative idea as to whether the internal shock model is
consistent with the present observations (and if so, qualitative
estimates of the model parameters) would be desirable. Such an
estimate would provide insight into the temporal behavior of
the central engine.

We implement an internal shock model with simplifying
assumptions and compute the time evolution of the non-thermal particles
produced. We also compare the results obtained with the broadband fluxes
from knots of several AGN jets and their observed positions. The motivation
here is to find a consistent set of model parameters that can explain the
observations and thereby make qualitative estimates of their values.
Apart from the fluxes at different energy bands, the spectral indices
in each band can also provide important diagnostic information about
the nature of these sources. Hence, we have
analyzed long ($ > 40$ ks)
\emph{Chandra} observations of three AGN and present the constrains that
were obtained on the
X-ray spectral indices of the individual knots.

\subsection{\emph{Chandra} X-ray Data Analysis}\label{sec4:dataanaly}
Long-exposure Chandra observations of the sources PKS 1136-135, 
PKS 1150+497, and 3C 371 were performed with the Advanced
CCD Imaging Spectrometer (ACIS-S) with the source at
the aim point of the S3 chip.
The Observation ID (ObsID) and the
exposure time of the observation are given in Table \ref{ch4:table2}. Earlier
shorter duration observations of these sources revealed two
bright knots for each source, whose positions from the nucleus
are given in Table \ref{ch4:table3}. These longer duration observations allow
for better constraint on the X-ray spectral indices of these
knots.  

The data from \emph{Chandra} X-ray 
observatory\footnote{http://cxc.harvard.edu/cda/}
were analyzed using the \emph{Chandra} data analysis software 
CIAO\footnote{http://cxc.harvard.edu/ciao/} and 
the latest calibration files were used to produce the spectrum.
The X-ray
counts from each individual knot was extracted using a circular region
centered at the knot. 
The background was estimated from the counts obtained from 
same size regions located at the same 
distance from the nucleus but at  different 
azimuth angles. The radius of the circular region was chosen to be 
$0.74^{\prime \prime}$
for the sources $1136-135$ and $1150+497$, while for $3C 371$
a smaller radius of $0.6^{\prime \prime}$ was used. These sizes were chosen
to minimize any possible contamination from the nucleus and/or the
other knot. 

Spectral fits were undertaken on the data 
using the XSPEC 
package\footnote{http://heasarc.gsfc.nasa.gov/docs/xanadu/xspec/} 
in C statistic mode, which is the appropriate
statistic when the total counts are low. The flux and the energy
spectral indices obtained are tabulated in Table \ref{ch4:table2}.
The quoted spectral indices are not very meaningful due to large errors.

\begin{table}[tp] \footnotesize
\centering
\begin{minipage}{110mm}
\caption[Details of \emph{Chandra} Observations of AGN jet]
{\emph{Chandra} Observations}
\label{ch4:table2}
   \begin{tabular}{lcccccccc}
    \toprule
Source name & Obs Id  & Exposure & Knots & $F_{0.3-3.0}$ & $\alpha_X$\\
& & {\scriptsize (ks)} & & {\scriptsize (ergs cm$^{-2}$ s$^{-1}$)} & \\
\midrule
 $1136-135$  & $3973$ & $77.37$ & A & $0.63$ & $1.24^{+1.51}_{-0.66}$ \\ 
   & & & B &  $1.56$ & $0.65^{+0.69}_{-0.28}$\\  
 $1150+497$ & $3974$ & $68.50$ & A & $3.15$ & $0.66^{+0.28}_{-0.26}$  \\
  & & & B & $0.61$ & $0.90^{+2.22}_{-1.02}$  \\
 $3C 371$  & $2959$ & $40.86$ & A & $3.32$ & $1.43^{+0.85}_{-0.76}$  \\
   & & & B & $7.84$ & $1.07^{+0.26}_{-0.23}$  \\
\bottomrule 
\end{tabular}
\vskip 0.5cm
{\bf Columns}: (1) Source name; (2) \emph{Chandra} Observation Id; 
(3) Exposure time; (4) Knots prominent in X-ray; (5) Flux in $0.3-3.0$ keV 
energy band; (6) X-ray energy spectral index.
\end{minipage}
\end{table}

\begin{table}[tp] \footnotesize
\centering
\begin{minipage}{110mm}
\caption{Observed Knot Features}
\label{ch4:table3}
   \begin{tabular}{lcccccccc}
    \toprule
Source name & Type & z & Knot & Position & $\alpha_{RO}$ & $\alpha_{OX}$ & Ref\\
&  &  &  & {\scriptsize (arcsec)} & & & \\
\midrule
 $1136-135$ & FSRQ & $0.554$ & A & $4.5$ & $0.73$ & $0.83$ &  \cite{2002v571p206} \\ 
 &  &  & B & $6.7$ & $1.04$ & $0.68$ &\\ 
 $1150+497$& FSRQ & $0.334$ & A & $2.1$ & $0.99$ & $0.83$ &  \cite{2002v571p206} \\ 
 &  &  & B & $4.3$ & $1.24$ & $0.44$ & \\ 
 $1354+195$& FSRQ & $0.720$ & A & $1.7$ & $1.05$ & $0.6$ &  \cite{2002v571p206}\\ 
 &  &  & B & $3.6$ & $1.15$ & $0.68$ & \\ 
 $3C 273$& QSO & $0.158$ & A & $13$ & $0.86$ & $0.61$ & \cite{2001v549p161}\\ 
 &  &  & B & $15$ & $0.9$ & $0.73$ & \\ 
 $3C 371$& Bl Lac & $0.051$ & A & $1.7$ & $0.9$ & $1.28$ & \cite{2001v556p79}\\
  &  &  & B & $3.1$ & $0.76$ & $1.14$ & \\ 
\bottomrule
\end{tabular}
\vskip 0.5cm
{\bf Columns}: (1) Source name; (2) Type of the source; (3) Redshift;
(4) Knots prominent in X-ray (The nomenclature is same for all the knots as 
they are in the literature except for $3C 371$ where A and B are reversed); 
(5) Position of the knot; (6) Radio-to-Optical index; (7) Optical-to-X-ray 
index; (8) References: Sambruna et al. \cite{2002v571p206}, 
Sambruna et al. \cite{2001v549p161}, Pesce et al. 
\cite{2001v556p79}. 
\end{minipage}
\end{table}

\subsection [The Model]{Model}

In the internal shock model framework, temporal variations of the
ejection process produces density fluctuations (moving with
different velocities), which collide at some distance from
the source to produce an observable knot. This distance will
depend on the time-scale over which the variation takes place.
In general the system will
exhibit variations over a wide range of time-scales and
knots like features would be produced at different
distance scales. Here, we consider large-scale jets
(with deprojected distances $\approx 100$ kpc) which are
expected to arise from variability occurring on a corresponding large 
time-scale. Variations
on smaller time-scales would produce knot structures on 
smaller distance scales, for example, pc scale or even smaller jets,
which would be unresolved for these sources.
These smaller time-scale variabilities will be smoothed out
at large distances, and hence one expects  
the jet structure of these sources to be determined by 
variations over a single characteristic time-scale. To further simplify
the model, we approximate the density and velocity fluctuations
as two discrete 
blobs with equal masses, $M_1 = M_2 = M$, 
having Lorentz factors,
$\Gamma_1$ and $\Gamma_2$ that are ejected one after the other,
from the central engine with a time delay of $\Delta t_{12}$. The
collision of the blobs is considered to be completely inelastic; i.e. 
the blobs coalesce and
move as a single cloud, which is identified with the observed knot. 
From conservation of momentum, the Lorentz factor of the knot is
\begin{align}
\Gamma = \sqrt{\left(\frac {\Gamma_1\, \beta_1 +\Gamma_2\, \beta_2 }{2}\right)^2+1}
\end{align}
where $\beta_{1,2}=v_{1,2}/c$ are the velocities of the blobs.
Since the collision is inelastic, a fraction of the bulk kinetic energy
is dissipated.
Denoting all quantities in the rest frame of the knot by subscript $K$,
this dissipated energy $\Delta E_{K}$ can be estimated as
\begin{align}
\Delta E_K = [\,\Gamma_{1K} + (\Gamma_{2K}-2)\,] M c^2
\end{align}
The Lorentz factors of the blobs in the knot's rest frame 
$\Gamma_{1K,2K}=(1-\beta^2_{1K,2K})^{-1/2}$ are computed using
\begin{align}
\beta_{1K,2K} = \frac {\beta_{1,2}-\beta}{1-\beta_{1,2}\,\beta}
\end{align}
where $\beta = (1-1/\Gamma^2)^{1/2}$.
The time-scale on which this energy will be dissipated can
be approximated to be the crossing-over time of the two blobs,
\begin{align}
T_{ON,K} \approx \frac {2 \, \Delta x_{K}}{c\,(\beta_{2K}-\beta_{1K})}
\label{eq4:TON}
\end{align}
where, $\Delta x_{K}$ is the average size of the two blobs in the rest frame
of the knot.  
It is assumed that  this dissipated bulk 
kinetic energy, $\Delta E_K$, gets converted efficiently  to the energy
of the non-thermal particles produced during the collision. 
The number of non-thermal particles injected per unit
time into the knot is taken to be,
\begin{align}
Q_K (\gamma)\; d\gamma = A \,\gamma^{-p}\; d\gamma \;\;\; \hbox  {for}\;\;\; 
\gamma > \gamma_{min}{,} 
\end{align}
where $\gamma$ is the Lorentz factor of the electrons, 
$p$ is the particle index and $A$ is the normalization constant 
given by,
\begin{align}
A = \frac{\Delta E_K}{T_{ON,K}}\,\frac{(p-2)}{m_ec^2}\,\gamma_{min}^{(p-2)} 
\end{align}
Here the injection is assumed to be uniformly occurring for a
time $T_{ON,K}$.  The cloud is assumed to be permeated with a tangled 
magnetic field, $B_K$. 

We used the kinetic equation (\ref{eq4:kinetic}) to study the 
evolution of the total number of non-thermal particles in the system. 
However for the total loss rate we considered only the synchrotron and
IC/CMB (equations (\ref{eq4:synloss}) and (\ref{eq4:icloss})). 
Using equation (\ref{eq4:adbloss}), 
the adiabatic cooling time-scale can be written as  
$t_{a}\approx \gamma/\dot \gamma_A=R(t)/(\beta_{exp}\,c)$. And for any given 
time of observation, $t_{O,K}$, the size of the blob will be  
$R(t_{O,K}) = R_o + \beta_{exp}\,c\,t_{O,K}$. From the above one finds that  
$t_{a}$ will be always larger than $t_{O,K}$ and hence the adiabatic cooling 
can be neglected. For $t_{O,K}<T_{ON,K}$, the resultant particle distribution 
will be a broken power-law giving rise to a composite synchrotron spectrum with 
a spectral break (\S \ref{sec4:cont_inj}).

The predicted spectrum and size of a knot depends on 
nine parameters, which are
the mass of the blobs $M$, their average size $\Delta x_{K}$, the
Lorentz factors $\Gamma_1$ and $\Gamma_2$, the particle injection index
$p$, the minimum Lorentz factor $\gamma_{min}$, the  
magnetic field $B_0$,
the inclination angle of the jet $\theta$, and the observation time $t_{O,K}$.

From these parameters and the location of the knot in the sky plane, one
can infer the time delay $\Delta t_{12}$ between the ejection of the two blobs.
The projected distance of the knot from the source $S$ can be written as
\begin{align}
S =  c \,( \beta_1\, t_c + \beta\, t_{O})\, \sin\theta
\label{eq4:seqn}
\end{align}
where $t_{O} = \Gamma t_{O,K}$ is the time of the observation after
the formation of the knot in the source frame. The time elapsed $t_c$, 
after the ejection of the first blob and the start of the collision,
is given by,
\begin{align}
t_c = \frac{v_2 \,\Delta t_{12} - \Delta x_1}{v_2 - v_1}
\end{align}
where $\Delta x_1$ is the size of the first blob 
$\approx \Gamma_{1,K} \Delta x_{K}/\Gamma$.
Thus $\Delta t_{12}$ can be estimated using the above equation, where $t_c$ is
given by equation (\ref{eq4:seqn}), and it essentially depends on 
four parameters, $\theta$, $\Gamma_1$, $\Gamma_2$ and $t_{O,K}$.

A total time, $t_{tot}$, can be defined to be the time that has elapsed
between the ejection of the first blob and the observation, $t_{tot} = 
t_c + t_{O}$. 
For two knots, A and B, the time difference between the ejection of their 
first blobs, $t^{AB}$ is then
\begin{align}
t^{AB} = t^A_{tot} - t^B_{tot} - t_{LT}
\end{align}
where $t_{LT}$ is the light travel time difference between the two knots,
which is approximated to be
\begin{align}
t_{LT} \approx \frac{S^A - S^B}{c \,{\tan} \theta} 
\end{align}

The power of the jet, can be defined in two different ways. The instantaneous
power, which is the power when the system is active, can be defined to
be the average energy of the blobs divided by the time-scale on which
the blobs are ejected. This power can be 
estimated for each knot to be
\begin{align}
P_{ins} \approx \frac {M c^2 (\Gamma_1 + \Gamma_2)/2}{\Delta t_{12}}
\end{align}
On the other hand,
the time averaged power of the jet can be defined as the 
typical energy ejected
during active periods divided by the time-scale on which such activity
occurs. For two knots, A and B, this can be approximated to be
\begin{align}
P_{ave} \approx \frac {[M^A c^2 (\Gamma^A_1 + \Gamma^A_2) + M^B c^2 (\Gamma^B_1 + \Gamma^B_2)]/2}{\Delta t^{AB}}
\end{align}

\subsection{Results and Discussion}

The model has been applied to those knots of kpc scale jets, which have
been detected by \emph{Chandra} and for which radio and optical data are available.
This criterion was satisfied by the two brightest knots of the AGN:
 $1136-135$ (Figure \ref{fig4:1136}), $1150+497$ (Figure \ref{fig4:1150_1354}), 
$1354+195$ (Figure \ref{fig4:1150_1354}), $3C 273$ (Figure \ref{fig1:3c273}) 
and $3C371$ (Figure \ref{fig4:3c371}). In this
work, the knot closer to the nucleus is referred to Knot A and the
further one as Knot B. For these sources this nomenclature is same as 
in the literature \cite{2002v571p206,2001v549p161} except for $3C 371$, for which the 
farther one has been referred to Knot A \cite{2001v556p79}. For three of these 
sources, the X-ray spectral
indices were constrained using long exposure observations as described
in \S \ref{sec4:dataanaly}. The observed properties of the sources 
and the knots 
are tabulated in Table \ref{ch4:table2} and \ref{ch4:table3}.   

\begin{figure}[tp]
\begin{center}
\includegraphics[width=150.0mm, height=150.0mm,bb=5 13 551 543]{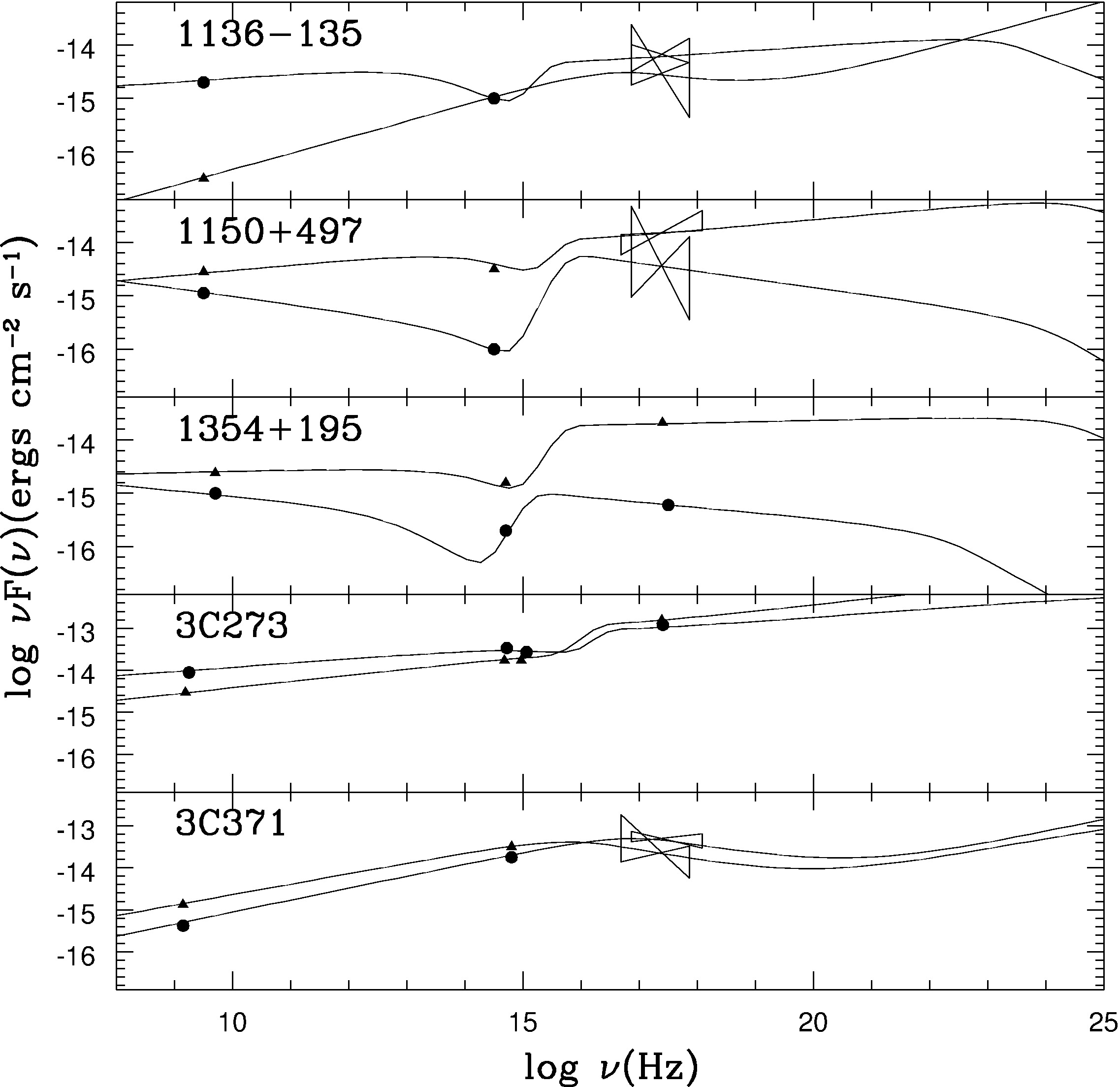} 
\caption[Spectral fits for the knots using internal shock model] 
{The observed fluxes in radio, optical and X-ray compared with model spectrum 
using parameters given in Table \ref{ch4:table4}.  Knot A fluxes are represented by 
filled triangles, and knot B fluxes are represented by filled circles.}
\label{fig4:is_plot}
\end{center}
\end{figure}
\begin{table}[tp]\footnotesize
\caption[Internal Shock Model Parameters] 
{Internal Shock Model Parameters} 
\label{ch4:table4}
 \begin{tabular}{lcccccccccccc}
\toprule
Source & Knot  & $\theta$ & $\Gamma_1$ & $\Gamma_2$ & $\Gamma^\ast$ & 
$t_{O,K}$ & $t_c^\ast$ & $t_{tot}^\ast$ & $\log\,$M & $\gamma_{min}$ & p & $B$ \\
& & {\scriptsize (deg)} & & & & {\scriptsize ($10^{11}$s)} & 
{\scriptsize ($10^{12}$s)} & {\scriptsize ($10^{12}$s)} & {\scriptsize (g)}& & 
& {\scriptsize ($10^{-5}$G)} \\
\midrule
1136-135 & A & 11.5 & 4.6 & 5.4 & 5.0 & 0.25 & 11.8 &  11.6 & 38.0 & 2 & 2.4 & 1.1 \\ 
 & B &\ldots  & 4.1 & 5.9 & 5.0 & 8.5 & 13.1 & 17.3 & 36.4 & 20 & 2.9 & 4.5\\ 
1150+497 & A & 10.2 & 2.5 & 3.0 & 2.8 & 16.0 & 1.2 & 5.1 & 37.5 & 30 & 2.8 & 1.4 \\ 
& B &\ldots & 2.6 & 3.5 & 3.0 & 6.2 & 8.8 & 10.0 & 36.8 & 30 & 3.3 & 4.0\\ 
1354+195 & A & 8.21 & 1.7 & 2.3 & 2.0 & 9.0 & 8.1 & 8.1 & 38.1 & 37 & 3.0 & 1.7\\ 
& B &\ldots & 1.7 & 2.3 & 2.0 & 8.7 & 15.0 & 17.0 & 37.1 & 20 & 3.2 & 8.0\\ 
3C 273 & A & 8.23 & 2.2 & 4.0 & 3.1 & 2.1 & 27.0 & 24.0 &  36.8 & 50 & 2.7 & 0.4\\ 
& B &\ldots & 1.7 & 2.3 & 2.0 & 8.0 & 30.0 & 32.0 & 37.8 & 80 & 2.8 & 0.6\\ 
3C 371 & A & 15.8 & 1.6 & 2.4 & 2.0 & 5.9 & 0.18 & 1.4 &  35.5 & 20 & 2.5 & 1.0\\ 
& B &\ldots & 2.0 & 2.4 & 2.2 & 2.3 & 0.39 & 0.75 & 36.8 & 10 & 2.4 & 0.6\\ 
\bottomrule
\end{tabular}
\vskip 0.5cm
Eight model parameters and derived quantities. 
The ninth parameter is 
$\Delta x_K = 5.0 \times 10^{21}$ cm for all sources. Columns marked
with an asterisk are derived quantities and not parameters.\\
{\bf Columns}: (1) Source name; (2) Knot; (3) Viewing angle; (4) Lorentz 
factor of the first blob; (5) Lorentz factor of the second blob; 
(6) Lorentz factor of the Knot; (7) Observation time; (8) Collision time; 
(9) Total time; (10)  Mass of the blobs; (11) Minimum Lorentz factor of the 
particle injected into the knot; (12) Injected particle spectral index; 
(13) Magnetic field. 
\end{table}

\begin{table}[tp]\footnotesize
\caption{Knot/Jet Properties}\label{ch4:table5}
 \begin{tabular}{lcccccccccc}
\toprule
Source & Knot & $\Delta t_{12}$ & $t^{AB}$ & $\frac{B}{B_{equ}}$ & 
$\frac{N_{nth}}{N_K}$ & $\log\;P_{ins}$ & $\log\; P_{ave}$ & $T_{ON,K}$ & 
$\frac{t_{O,K}}{T_{ON,K}}$ & D\\
 && {\scriptsize ($10^{11}$s)} & {\scriptsize ($10^{11}$s)} &  & & 
{\scriptsize (ergs s$^{-1}$)} & {\scriptsize (ergs s$^{-1}$)} & {\scriptsize ($10^{11}$s)} &  & {\scriptsize (kpc)}\\
\midrule
1136-135 & A & 1.1 & 3.1 & 0.55 & 0.88 & 48.60 & 48.17 & 20.4 & 0.01 & 112.8 \\ 
  & B & 2.5 & & 0.74 & 0.74 & 46.65 & & 9.1 & 0.93 & 163.4 \\ 
1150+497 & A & 0.9 & 5.0 & 0.12 & 0.13 & 47.98 & 47.31 & 17.0 & 0.94 & 50.5 \\ 
 & B & 3.9 & & 0.65 & 0.43 & 46.63 & & 10.7 & 0.58 & 96.1 \\ 
1354+195 & A & 7.5 & 18.0 & 0.04 & 0.38 & 47.49 & 47.15 & 9.6 & 0.94 & 78.7 \\ 
 & B & 17.0& & 0.64 & 0.78 & 46.10 & & 9.6 & 0.90 & 135.5 \\ 
3C 273 & A & 20.0 & 34.0 & 0.03 & 0.80 & 46.00 & 46.63 & 5.4 & 0.39 & 240.7 \\ 
 & B & 31.6 & & 0.02 & 0.16 & 46.60 & & 9.6 & 0.84 & 248.5\\ 
3C 371 & A & 1.4 & 1.5& 0.39 & 0.88 & 45.59 & 46.98 & 7.2 & 0.83 & 11.2 \\ 
 & B & 1.0 & & 0.28 & 0.29 & 47.14 & & 16.3 & 0.14 & 7.7 \\ 
\bottomrule
\end{tabular}
\vskip 0.5cm
{\bf Columns}: (1) Source name; (2) Knot; (3) Time delay between the ejection of 
the blobs; (4) Time delay between the ejection of the first and the third blob;
(5) Ratio of the magnetic field to equipartition magnetic field; (6) Ratio of 
non-thermal electrons to the total number of protons; (7) The instantaneous power;
(8) Time-averaged power; (9) Time-scale over which non-thermal particles are injected;
(10) Ratio of the observation time to particle injection time-scale; (11) Deprojected 
distance.
\end{table}

Figure \ref{fig4:is_plot} shows the observed radio, optical and X-ray fluxes 
of these knots
along with the computed spectrum corresponding to model parameters that  
are given in Table \ref{ch4:table4}. Since the number of parameters is large as compared
to the observables, a unique set of parameter values cannot be obtained.
Two consistency checks have been imposed on the parameter values: 
that the number of non-thermal electrons which will be injected into
the system, $N_{nth}$, is smaller than the total number of protons, $N_k$,
and that the
magnetic field, $B$ should not deviate far from the 
equipartition value, $B_{equ}$. 
Both of these conditions
are satisfied by the parameter sets as shown in Table \ref{ch4:table5}, 
where the
ratios $B/B_{equ}$ and $N_{nth}/N_K$ are given.

For each source, 
the time delay $\Delta t_{12}$ between
the ejection of the two blobs that form the knots are
nearly equal to the time difference between the ejection of the first blobs
of Knot A and Knot B, $t^{AB}$. This gives an overall \emph{single} time-scale of
activity for each source which ranges from $10^{11} - 10^{12}$ s and
can reproduce the knot properties, as had been assumed in the
development of the simple internal shock model. 
This result is important since if it had not been true, a more
complex temporal behavior would have to be proposed, wherein
the jet structure is due to variability of the source at two
different time-scales, the first being the time difference between
the ejection of two blobs that form a knot, and the second being
the time difference between the activities that produced the two
knots.

The cross-over
time, $T_{ON,K}$, is determined here by equation (\ref{eq4:TON}). Since,
$\beta_{2K}-\beta_{1K} \approx (\Gamma_1 - \Gamma_2)/\Gamma \approx 0.3$,
$T_{ON,K} \approx 10^{12} (\Delta x_K / 5 \times 10^{21} \hbox {cm})$ s.
During this time, i.e., the time when there is injection of particles
into the system, the knot would travel a distance $\approx c\, T_{ON} 
\approx c\, \Gamma \,T_{ON,K} \approx 50 \,(\Gamma/5) (\Delta x_K / 5 \times 10^{21} \hbox {cm})$ 
kpc. This is a significant fraction of the total
observable distance traveled by the knot, from formation 
$\approx c\, t_c \approx 100$ kpc (see Table \ref{ch4:table5}) to the termination
of the jet in the radio lobe, $\approx 200$ kpc. Hence, as
shown in Table \ref{ch4:table5}, it is possible to fit the spectrum of all
the knots with an observation time $t_{O,K}$, which is less than 
the crossing-over time $T_{ON,K}$, implying that there is  continuous 
injection of particles into the system.
In this scenario, the
synchrotron and IC spectra will have a break corresponding to a Lorentz
factor $\gamma_c$ where the cooling time-scale equals the 
observation time \S \ref{sec4:cont_inj}. 

The knots of 3C371 and 
Knot A of 1136-135 are unique in this sample, since 
their the X-ray flux lies below the extrapolation of
the radio-to-optical spectrum to X-ray wavelengths. This allows
for the interpretation that the X-ray flux is due to synchrotron
emission \cite{2002v571p206,2001v556p79}. For Knot A of 1136-135 and
Knot B of 3C371, this implies that the spectral break 
for the synchrotron emission occurs at the X-ray regime 
(Figure \ref{fig4:is_plot}), which in
turn indicates that these sources are relatively younger. Indeed,
the ratio of the observation time to the injection time, 
$t_{O,K}/T_{ON,K}$, for these sources are smallest (Table \ref{ch4:table5}). 
On the other hand, for Knot A of 3C371, the spectral break
can occur at the optical band even if the X-ray flux is interpreted
 as being due to synchrotron emission (Figure \ref{fig4:is_plot}) and hence 
this source
need not be relatively young. However, this is only possible if
$t_{O,K} < T_{ON,K}$ and there is continuous injection
of particles. Otherwise, a sharp cutoff in the spectrum at the optical 
band would have occurred and the X-ray emission would not be due to
synchrotron emission.

Figure \ref{fig4:is_plot} shows that the radio, optical and X-ray spectral
indices for different knots may vary and highlights the
need for more spectral measurements in all bands. A definite
prediction of this model is that for most knots, the X-ray
spectral index should be equal to the radio spectral index, indicating
that the X-ray flux is due IC/CMB. Such spectral constrains would
be  particularly important since,
although it has been demonstrated here that the internal
shock model can explain the broadband spectrum of these
sources, there could be other models that may be physically
and observationally more favorable. Analysis of the Very Large array 
(VLA) and \emph{Hubble Space Telescope} (HST) images of $3C 273$ 
have shown that the optical and radio spectral indices are different,
indicating the presence of an additional emission mechanism
for the source \cite{2005v431p477}. Also, the X-ray flux may be 
due to a second population of non-thermal electrons, rather than being the 
IC/CMB spectrum of the same distribution that produces the radio and optical
emission \cite{2004v613p151}. An argument in support of this is that since  
in the IC/CMB model the X-ray emission
is due to electrons that are only a factor ten more energetic than 
those which produce the radio, a source in which  
the X-ray flux falls rapidly from the center should also exhibit
a similar decrease in radio emission. However this feature 
is not observed
(e.g. 3C273 and 1354+195). Moreover, the jet power required
in the IC/CMB model can be very large, $\approx 10^{48}$ ergs/s,
which may be larger than the power inferred from the giant radio
lobes ($\lesssim 10^{47}$ ergs/s). While the former argument
may not strictly be applicable to the internal shock model (since
each knot is a separate entity and the distance from the source
is not a measure of the age of the source), the power requirement
for some sources  may indeed be very large, for e.g. 
Knot A of 1136-135 requires $P_{ave} \approx 2\times 10^{48}$ ergs/s 
(Table \ref{ch4:table5}). However, the energy requirement may be decreased if
the magnetic field is sub-equipartition e.g. 3C273 (Table \ref{ch4:table5}).
Thus it is desirable to obtain direct observational signatures,
such as spectral indices, to discriminate between models.

A realistic description of the knots is more complicated than the
simple model considered here. For example, the forward and reverse
shocks that should form when the blobs collide, may provide 
different injection rates and at different
locations within the Knot. However, the physics of these shock formations
and the subsequent acceleration of particles is complicated and
unclear, especially if they are mediated by magnetic fields. In the
future, results from sophisticated numerical simulations could be
compared with higher resolution data (which can resolve the
internal structure of the knots) to prove (or disprove) the
internal shock model.

\section{A Two Zone Model for the knots of M87}\label{sec4:m87}
The models discussed in earlier sections can explain the broadband spectrum from 
the knots of many AGN jets. However they fail 
in case of M87, 
a nearby giant elliptical galaxy, at a distance = 16 Mpc \cite{1991v373p1}, 
possessing 
an one-sided jet with projected distance $\approx$ 2 kpc. The jet is bright in 
radio, optical and X-ray energies. The jet structure is very well studied in radio, 
infrared, optical and 
X-ray energies \cite{1989v340p698,1995v447p582,2001v561p51,1996v473p254,1996v307p61,
2005v627p140}.
The flux and the spectral indices at X-ray energies indicate a possible continuation 
of synchrotron emission of the radio-to-optical spectrum with the change in the 
spectral index beyond optical energies \cite{2002v564p683,2005v626p120}.
Simple theoretical models, namely the continuous injection 
model \cite{1962v6p317,1968v1p442,1987v225p335,1989v219p63} (\S \ref{sec4:cont_inj}) 
and the one-time injection model \cite{1973v26p423,1962v6p317,book:pacholczyk}
were unable to explain the observed X-ray flux and/or the spectral index. 
The X-ray flux predicted by the continuous injected model is more than the 
observed flux whereas the one-time injection model with pitch angle scattering 
under predicts the X-ray flux and the one without pitch angle scattering fails 
to predict the observed X-ray spectral index \cite{2005v627p140}.

We propose a two zone model to explain the non-thermal emission from the 
knots of M87 jet. In a two zone model, particles are accelerated in a region,
namely acceleration region (AR) (probably around a shock front), and 
subsequently 
cool off through radiative processes in an associated cooling region (CR).

\begin{figure}[tp]
\begin{center}
\includegraphics[width=120mm,bb=0 0 983 1742]{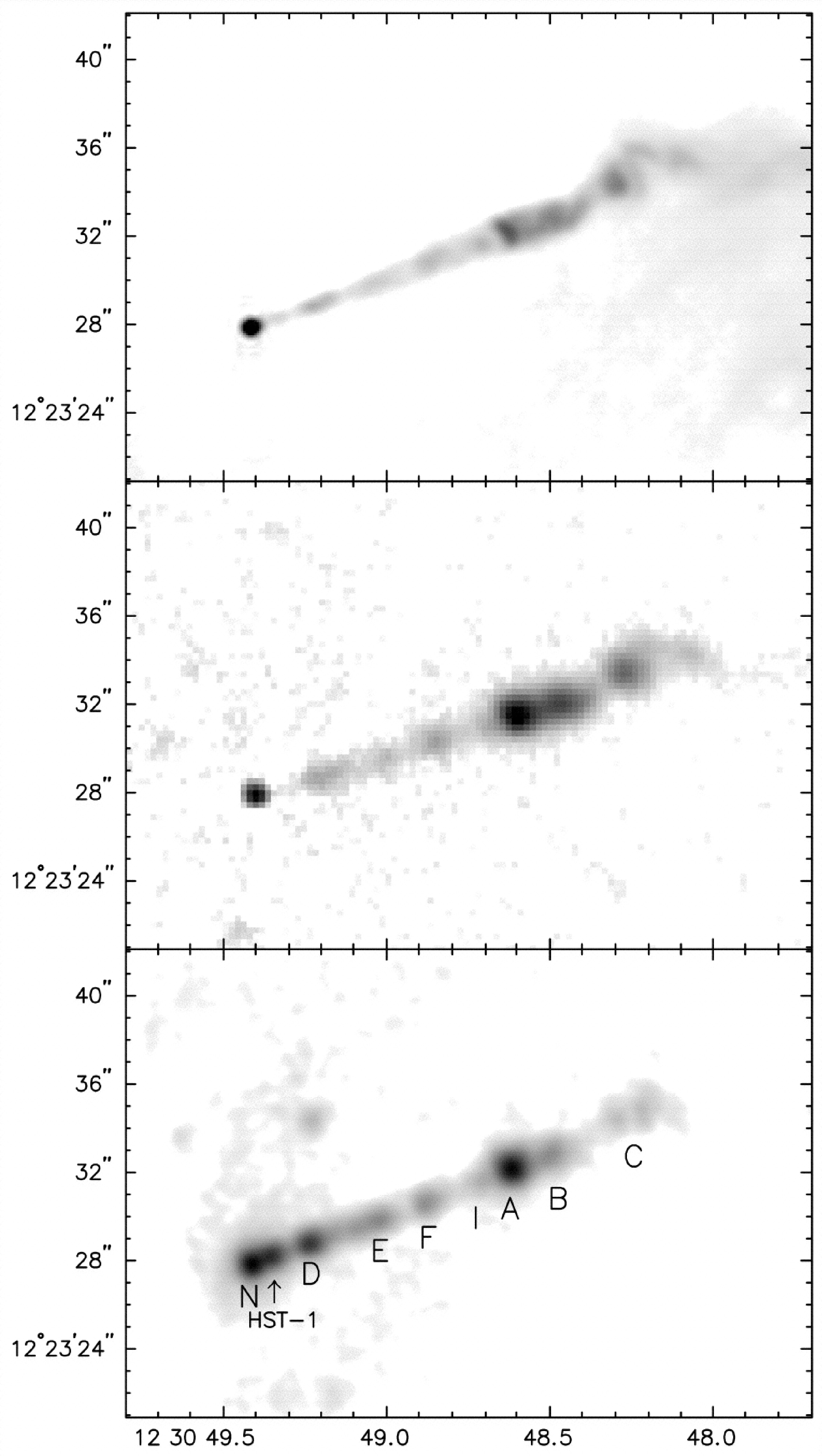} 
\caption[Multi wavelength image of M87]
{Grey-scale image of M87 in radio (top), optical (middle) and X-ray (bottom).
Figure reproduced from Wilson \& Yang \cite{2002v568p133}.}
\label{fig4:m87}
\end{center}
\end{figure}

\subsection{The Model}

We consider the acceleration of a power-law distribution of particles (which may 
be a relic of a past acceleration process) at a shock front and cooling via 
synchrotron radiation in a homogeneous magnetic field. We treat the present 
scenario as two zones: one around the shock front where the particles are 
accelerated (AR) and the downstream region where they lose most of their energy 
through the synchrotron process (CR). This model is then used to explain the 
radio-optical-X-ray spectrum of the knots in the M87 jet. We assume the CR 
to be a spherical blob of radius R with tangled magnetic field $B_{CR}$ and the AR 
is assumed to be a very thin region with magnetic field $B_{AR}$. A power-law 
distribution of electrons is continuously injected into the AR 
characterized by an acceleration time-scale $t_{acc}$. Particles are then 
accelerated at a rate $1/t_{acc}$ to a maximum energy determined by the loss processes. 
The AR is assumed to be compact and the emission from the CR mainly contributes the 
overall photon spectrum.

The kinetic equation governing the evolution of electrons in AR is given by

\begin{align}
\label{eq4:arkin}
{\partial n(\gamma,t) \over \partial t} = 
{\partial  \over \partial \gamma} \left[\left(\zeta_{AR}\, \gamma^2 
- \frac{\gamma}{t_{acc}}\right)n(\gamma,t)\right] 
- \frac{n(\gamma,t)}{t_{esc}} + Q (\gamma) 
\end{align}
where 
\begin{align}
Q (\gamma)\; d\gamma = q_o\, \gamma^{-p}\; d\gamma \;\;\; \hbox  {for}\;\;\; \gamma_{min}<\gamma<\gamma_b
\end{align}
Here $\gamma$ is the Lorentz factor of the electron, $t_{esc}$ is the escape time-scale and 
$\zeta_{AR} = \frac{\sigma_T}{6\pi\, m_e\, c}B_{AR}^2$.

Equation (\ref{eq4:arkin}) can be solved analytically using Green's function 
and the electron distribution for an energy-independent 
$t_{acc}$ and $t_{esc}$ at time $t$ is given by \cite{1999v302p253}
\begin{align}
\label{eq4:arnum}
n(\gamma,t) &= t_{acc} \,\gamma^{-(\alpha+1)} \left(1-\frac{\gamma}{\gamma_{max}}\right)^{\alpha-1}
\nonumber \\
&\times\int\limits_{x_o}^{\gamma} Q(x) \left[\frac{1}{x}-\frac{1}{\gamma_{max}}\right]^{-\alpha} dx
\end{align}
where $\alpha = t_{acc}/t_{esc}$ and the lower limit of integration $x_o$ is given by

\begin{align}
\label{eq4:arlimit}
x_o=\left[\frac{1}{\gamma_{max}}+\left(\frac{1}{\gamma}-\frac{1}{\gamma_{max}}\right)exp(t/t_{acc})\right]^{-1}
\end{align}
$\gamma_{max}=1/(\zeta_{AR}\,t_{acc})>\gamma_b$ is the maximum Lorentz factor an electron can 
attain in AR. For $t>>t_{acc}$ equation (\ref{eq4:arlimit}) can be approximated 
to be $\gamma_{min}$ 
as the injection term in equation (\ref{eq4:arnum}) vanishes for $x<\gamma_{min}$. 

The evolution of the electrons in CR is governed by the equation
\begin{align}
\label{eq4:crkin}
{\partial N(\gamma,t) \over \partial t} = 
{\partial  \over \partial \gamma} \left[\zeta_{CR}\, \gamma^2 N(\gamma,t)\right] 
+ Q_{AR}(\gamma) 
\end{align}
where $\zeta_{CR} = \frac{\sigma_T}{6\pi\, m_e\, c}B_{CR}^2$ and the last term is the injection 
from AR, $Q_{AR}(\gamma) = n(\gamma)/t_{esc}$. For $t>>t_{acc}$
\begin{align}
Q_{AR}(\gamma) &\approx q_o\,\alpha\, \gamma^{-(\alpha+1)}\left(1-\frac{\gamma}{\gamma_{max}}\right)^{\alpha-1} 
\nonumber\\
&\times\int\limits_{\gamma_{min}}^{MIN(\gamma,\gamma_b)} x^{-p} \left[\frac{1}{x}-\frac{1}{\gamma_{max}}\right]^{-\alpha} dx
\end{align}

The distribution of electron at time $t$ in CR from equation (\ref{eq4:crkin}) is given by

\begin{align}
N(\gamma,t) = \frac{1}{\zeta_{CR}\,\gamma^{2}}\int\limits_{\gamma}^{\Gamma_o} Q_{AR}(x)\; dx
\end{align}
where $\Gamma_o = \gamma/(1-\gamma\,\zeta_{CR}\,t)$.

From equation (\ref{eq4:arnum}), it can be shown that the injection into CR 
(for $\alpha+1>p$) is a broken power-law
with index $-p$ for $\gamma<\gamma_b$ and $-(\alpha+1)$ for $\gamma>\gamma_b$. The synchrotron
losses in CR introduces an additional break $\gamma_c$ in the electron spectrum 
depending upon the age of CR ($t_{obs}$) and the $B_{CR}$ (\S \ref{sec4:cont_inj}). 

\begin{align}
\gamma_c = \frac{1}{\zeta_{CR}\;t_{obs}}
\end{align}

The electron spectrum in CR at $t_{obs}$ can 
then have two different spectral shapes depending on the location of $\gamma_c$ with respect
to $\gamma_b$.

\begin{enumerate}[(i)]
\item $\gamma_c > \gamma_b$: The final spectrum will have two breaks with indices 
\begin{align}
\label{eq4:crspec1}
N(\gamma,t_{obs}) \propto \left\{
\begin{array}{ll}
\gamma^{-p},&\mbox {~$\gamma_{min}<\gamma<\gamma_b$~} \\
\gamma^{-(\alpha+1)},&\mbox {~$\gamma_b<\gamma<\gamma_c$~} \\
\gamma^{-(\alpha+2)},&\mbox {~$\gamma_c<\gamma<\gamma_{max}$~} 
\end{array}
\right.
\end{align}
\item $\gamma_c < \gamma_b$: In this case the indices are  
\begin{align}
\label{eq4:crspec2}
N(\gamma,t_{obs}) \propto \left\{
\begin{array}{ll}
\gamma^{-p},&\mbox {~$\gamma_{min}<\gamma<\gamma_c$~} \\
\gamma^{-(p+1)},&\mbox {~$\gamma_c<\gamma<\gamma_b$~} \\
\gamma^{-(\alpha+2)},&\mbox {~$\gamma_c<\gamma<\gamma_{max}$~} 
\end{array}
\right.
\end{align}
\end{enumerate}

The resultant synchrotron emissivity can then be calculated by convolving 
$N(\gamma,t)$ with single particle emissivity averaged over an isotropic 
distribution of pitch angles (equation (\ref{eq3:synemiss})).

The predicted spectrum depends on nine parameters, which are $q_o$, $\alpha$,
$\gamma_{min}$, $\gamma_b$, $\gamma_{max}$, $p$, $B_{CR}$, $R$ and $t_{obs}$.
Here $\alpha$ and $p$ are estimated from the radio-to-optical and optical-to-X-ray 
spectral indices; $q_o$ and $B_{CR}$ are constrained using the observed luminosity 
and equipartition magnetic field. For $R$ we assume the physical sizes measured in 
radio \cite{1982v261p457}. The age of the knot $t_{obs}$
is chosen to introduce a break in the observed spectrum at optical band and $\gamma_b$ is fitted to
reproduce the observed X-ray flux. $\gamma_{min}$ and $\gamma_{max}$ are used as free parameters and
are fixed at $5$ and $10^8$.

\subsection{Results and Discussion} 

\begin{figure}[tp]
\begin{center}
\includegraphics[width=136.8mm,bb=0 0 507 698]{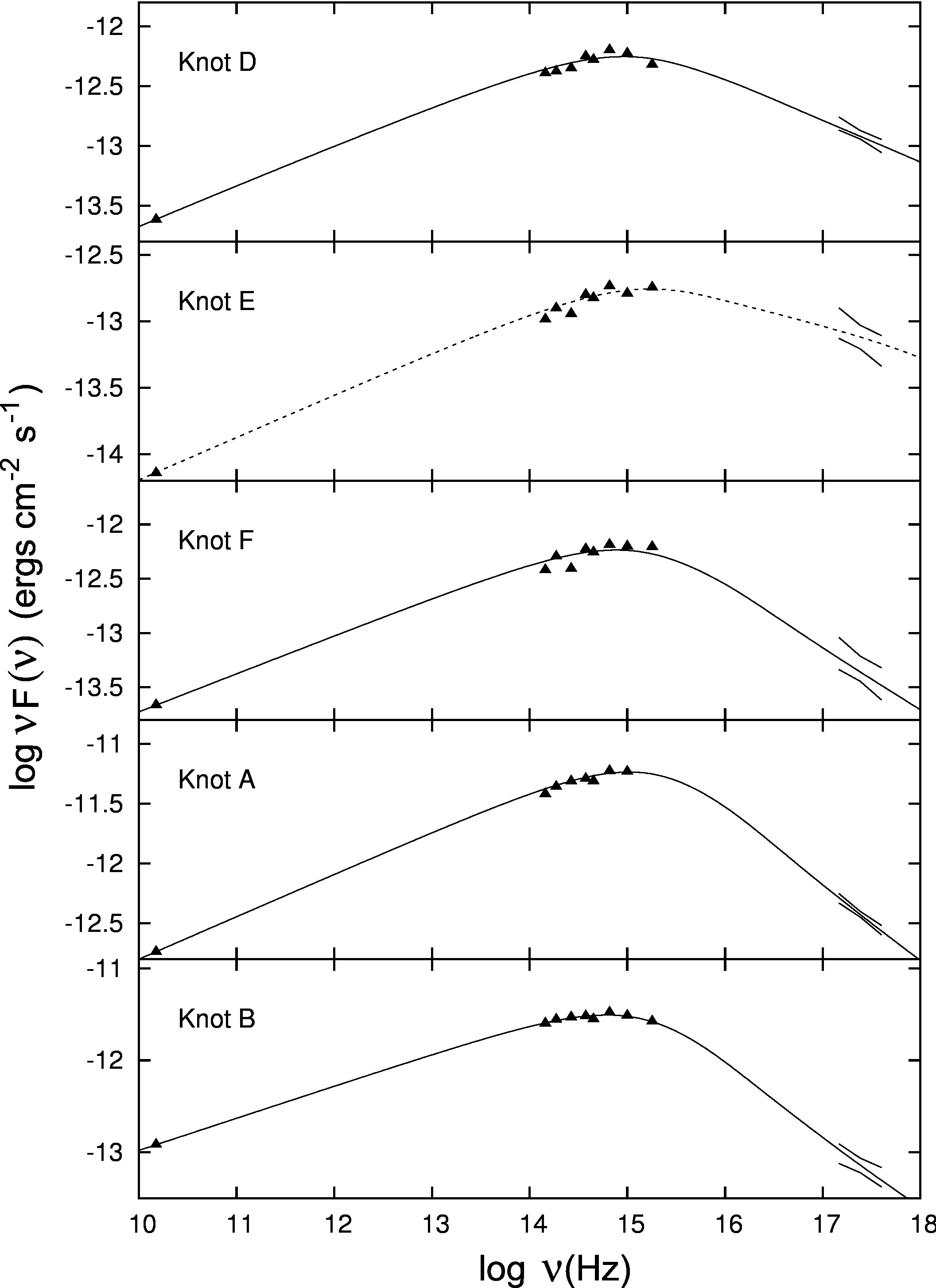} 
\caption[Spectral fit for the knots of M87 using two zone model]
{The spectral fit for the knots of M87 with the observed fluxes in radio, optical 
(Perlman et al. \cite{2001v551p206} and X-ray (Perlman \& Wilson 
\cite{2005v627p140}). Knot labels are the same as used in the literature \cite{2002v568p133}.}
\label{fig4:m87fit}
\end{center}
\end{figure}

We applied the above model to explain the knots D, F, A and B 
of M87 jet (Figure \ref{fig4:m87}). The results of the fitting are shown 
in Figure \ref{fig4:m87fit}
and the parameters used for the fit are given in Table \ref{ch4:table6}. 
The spectrum of knot E can be explained by simple continuous injection model and the parameters
we quote corresponds to continuous injection model. 
We did not model knot C
due to significant differences in X-ray-optical properties. The optical
image of knot C is a diffuse region with a single maximum
whereas in the X-ray image there exists two distinct maxima coincident 
with the diffuse optical knot \cite{2005v627p140}.

For all the fits shown in Figure \ref{fig4:m87fit}, $\gamma_c < \gamma_b$. However, one can fit 
the spectrum with $\gamma_c > \gamma_b$ with proper choices of the parameters $\alpha$, 
$\gamma_b$ and $\gamma_c$. This degeneracy arises due to the unavailability of the 
ultraviolet (UV) spectral index since the present model predicts the corresponding particle 
spectral index as  $-(\alpha+1)$ or $-(p+1)$  (equations (\ref{eq4:crspec1}) and 
(\ref{eq4:crspec2})) depending upon the above two conditions. Future observations at these 
photon energies may help in validating the present model and also will remove this degeneracy. 
Also, to obtain a precise values for $p$ and $\alpha$, spectral indices at radio, optical and 
X-ray energies should be known accurately. It should be noted here that in 
general $t_{acc}$ and $t_{esc}$ can be energy dependent. In such a situation 
the solution (equation (\ref{eq4:arnum})) may differ from its 
form and the index beyond $\gamma_b$ may not be the one discussed above.

 \begin{table}[tp]\footnotesize
   \caption[Two Zone Model Parameters for the Knots of M87]
   {Model Parameters for the Knots of M87}\label{ch4:table6}
   \begin{tabular}{p{0.5cm}cccccccc}
   \toprule
    Knot & $q_o$ & $\alpha$ & $p$ & $\gamma_b$ & $B_{CR}$ & $t_{obs}$ & $\gamma_c$ & $R$ \\
 & &{\scriptsize ($10^{-12}$)} & &{\scriptsize ($10^6$)} & {\scriptsize ($10^{-4}$ G)} & 
 {\scriptsize ($10^9$ s)} &{\scriptsize ($10^5$)} &{\scriptsize (pc)} \\
 \midrule
D & 7.5 & 1.75& 2.35& 1.8 & 9.3 & 1.6 & 5.7 & 12 \\ 
E & 4.7 & \ldots & 2.36& \ldots & 5.9 & 2.5 & 9.0 & 17 \\ 
F & 0.6 & 2.3 & 2.3 & 2.1 & 4.5 & 4.7 & 8.2 & 29 \\ 
A & 1.0 & 2.45& 2.29& 2.0 & 4.7 & 3.5 & 10.1& 55 \\ 
B & 0.9 & 2.75& 2.3 & 1.4 & 4.7 & 3.9 & 9.0 & 50 \\ 

\bottomrule
 \end{tabular}
\vskip 0.5cm
{\bf Columns}: (1) Knot name; (2) Normalisation of the power-law injected into AR 
(for knot E it is the normalisation of the power-law injected into CR (see text)); 
(3) Ratio between the acceleration time-scale and escape time-scale in AR; (4) Index of the 
power-law spectrum injected into AR; (5) Maximum energy of the electron Lorentz factor 
injected into AR; (6) CR magnetic field; (7) Time of observation; (8) Break Lorentz 
factor of the electrons due to synchrotron cooling in CR; (9) Size of CR measured in 
radio \cite{1982v261p457}.
  
For all cases, the minimum Lorentz factor ($\gamma_{min}$) injected into AR is $5$ and 
maximum Lorentz factor ($\gamma_{max}$) attained in AR is $10^8$. 

 \end{table}

A possible scenario of the present model is where the AR is a region around an internal shock 
following an external shock. The electrons injected into the AR can be those that are already 
accelerated by an external shock and are advected downstream to be accelerated further by the 
internal shock \cite{2008v2p247,1994v11v175}. Alternatively, re-acceleration of a power-law 
electron distribution by turbulence at boundary shear layers can also be another possible 
scenario \cite{1986v307p62,2003v47p521}. Inclusion of these scenarios in their exact form into 
the present model will make it more complex and is beyond the scope of the present work.

Perlman \& Wilson (2005) proposed a modified continuous injection model where the volume 
within which particle acceleration occurs is energy dependent \cite{2005v627p140}. This is 
expressed in terms of a filling factor $f_{acc}$ which is the ratio between the observed flux 
and the flux predicted by the simple continuous injection model. They found  $f_{acc}$ declining 
with increasing distance from the nucleus suggesting particle acceleration taking place in a 
larger fraction of the jet volume in the inner jet than the outer jet. The energy dependence 
of  $f_{acc}$ also indicates that particle acceleration regions occupy a smaller fraction of 
the jet volume at higher energies. Even though the model is phenomenological, it indicates that 
the process of high-energy emission from the knots is as complicated as their physical region. 
However, the mechanism responsible for the filling factor is not explained.

Liu \& Shen (2007) proposed a two zone model to explain the observed spectrum of the knots of 
the M87 jet. In their model electrons are accelerated to relativistic energies in the 
acceleration region (AR) and lose most of their energies in the cooling region (CR) through 
the synchrotron process. They considered that the AR and CR are spatially separated and 
introduced a break in the particle spectrum injected in the CR through the advection of 
particles from the AR to the CR. This along with the cooling break in the CR produces a double 
broken power-law with indices $-p$, $-(p+1)$ and $-(p+2)$ which is then used to fit the 
observed spectra. However, the present model assumes that the AR and CR are cospatial, 
supporting a more physical scenario where electrons accelerated by the shock cool in its vicinity.

We explored the possibility of the present model to reproduce the X-ray flux of other FR I 
galaxies (detected by Chandra) which are observed to have lower radio luminosity and relatively 
smaller jets when compared with FR II galaxies. 
The X-ray emission from the knots and/or the jets of the FR I galaxies, namely 3C 66B 
\cite{2001v326p1499}, 3C 346 \cite{2005v360p926}, Cen A \cite{2006v368p15} and 3C 296 
\cite{2005v358p843}, listed in the online catalogue of extragalactic X-ray jets 
XJET\footnote{http://hea-www.harvard.edu/XJET/}, which are not explained by synchrotron 
emission from simple one zone models, can be reproduced by the present model.


\chapter{Boundary Shear Acceleration in the Jet of MKN 501}
\label{chap:mkn501}
\begin{figure}[tp]
\begin{center}
\includegraphics[width=136.8mm,bb=0 0 590 552]{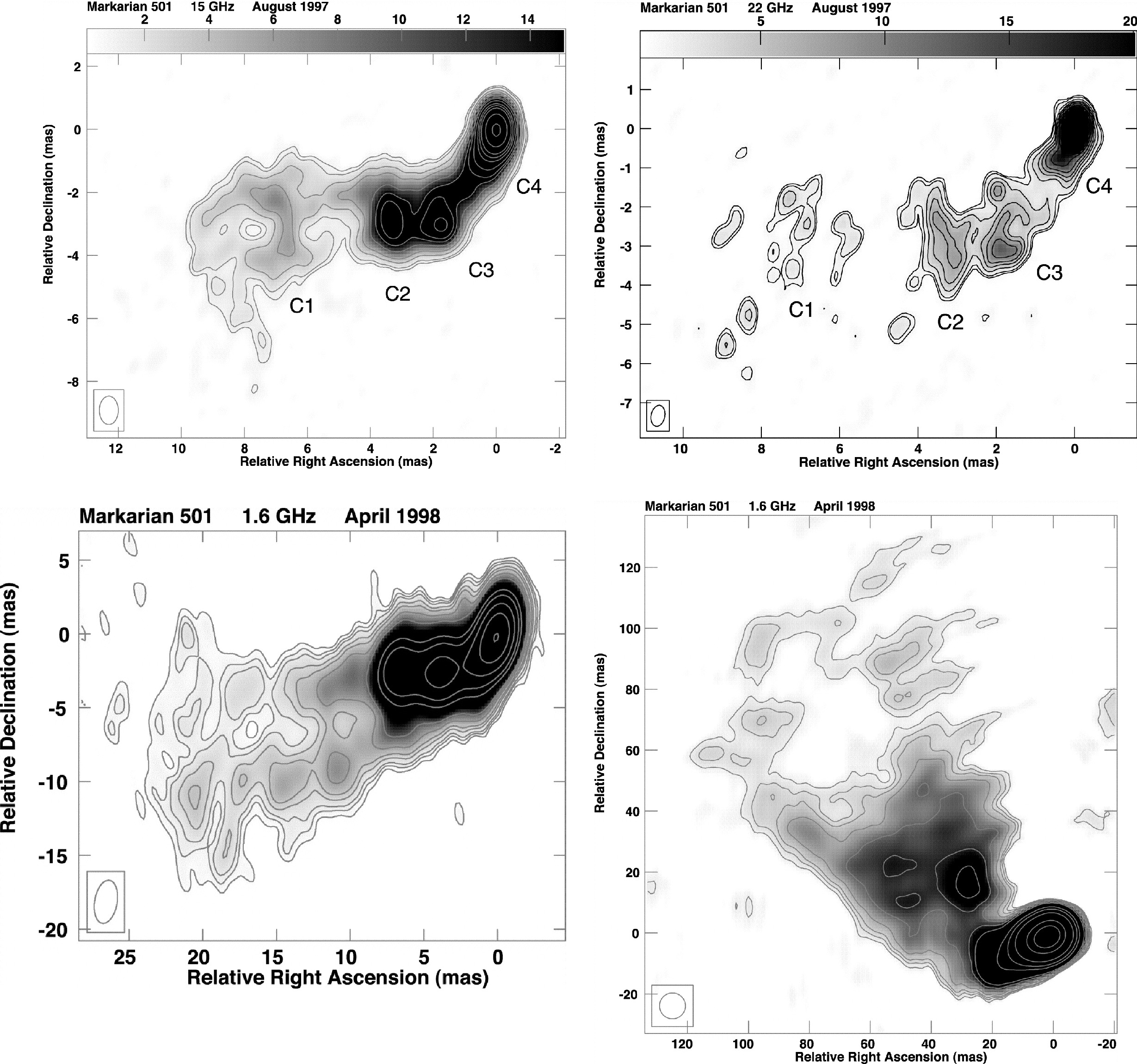} 
\caption[Radio images of MKN 501 jet]
{Grey-scale image of MKN 501 jet in radio at 15 GHz (top left), 22 GHz (top right)
and 1.6 GHz (bottom left and bottom right). Figures reproduced from 
Giroletti et al. \cite{2004v600p127}.}
\label{fig5:mkn501}
\end{center}
\end{figure}

MKN 501 is a nearby BL Lac object (z = 0.034) and also the second extra galactic source detected 
in TeV photon energies by ground-based Cherenkov Telescopes \cite{1996v456p83}. It was later 
detected in MeV photon energies by the satellite-based experiment EGRET \cite{1999v514p138}.
Radio images of MKN 501 show a jet emerging from a bright nucleus 
\cite{2000v52p1015,1999v159p439, 1999v159p427, 2004v600p127}. 
The high-resolution (milliarcsecond) radio images show a transverse jet structure with the 
edges being brighter than the central spine commonly referred as ``limb-brightened'' structure
\cite{2000v52p1015, 1999v159p439, 2004v600p127}  (Figure \ref{fig5:mkn501}). This feature is usually 
explained by the ``spine-sheath'' model where the velocity at the jet spine is larger compared 
to the velocity at the boundary. Such a radial stratification of velocity across the jet arises 
when jet moves through the ambient medium and the viscosity involved will cause a shear at the 
boundary. Three-dimensional hydrodynamic simulations of relativistic jets \cite{2000v528p85}
and two-dimensional simulations of relativistic magnetized jets \cite{2005v436p503} also support 
the presence of jet velocity stratification due to its interaction with the ambient medium. The 
existence of velocity shear at the jet boundary was first suggested by Owen, Hardee \& 
Cornwell \cite{1989v340p698} to explain the morphology of M87 jet. 
Perlman et al. \cite{1999v117p2185} later confirmed it through the polarization studies 
of M87 jet. Limb-brightening can occur in a misaligned jet with velocity stratification. 
For a proper combination of flow velocities, one will see a 
Doppler-boosted image of the boundary compared to the less boosted spine due to relativistic 
effects \cite{1990v16p284,1996v100p241}.
A possible consequence 
of the velocity shear is the alignment of the magnetic field at the boundary parallel to the flow 
velocity due to stretching of the frozen-in field lines of the plasma \cite{1983v202p553}. The 
polarization angle observed at the jet boundary of MKN 501 is perpendicular to the jet axis 
indicating a parallel magnetic field \cite{2005v356p859, 1999v159p427}. However, it should 
be noted here that the polarization angle at the jet spine indicates a perpendicular magnetic 
field, and this along with the parallel magnetic field at the jet boundary can be an outcome 
of a dynamically dominant toroidal magnetic field structure \cite{2005v356p859,1999v43p691,2005v14p363}.
The radial velocity stratification of the jet can introduce Kelvin-Helmholtz instability, and the 
stability of jets against this instability was studied by various authors 
\cite{1976v176p421,1976v176p443,1978v64p43,1979v234p47,1991v252p505}.

Giroletti et al. \cite{2004v600p127} have studied the limb-brightened structure 
of MKN 501 jet considering the differential Doppler boosting at the jet spine and the 
boundary \cite{1996v100p241,1990v16p284}, and concluded that the viewing angle (angle 
between the jet and the line of sight of the observer) of the radio jet should be more 
than $15^\circ$. However, high-energy studies of MKN 501 demand that the viewing angle of 
the jet should be $\approx 5^\circ$ in order to explain the observed rapid variability and 
the high-energy emission \cite{2001v367p809,2001v554p725}. Considering the fact that the 
$\gamma$-ray emission is originated from the inner part of the jet close to nucleus, Giroletti 
et al. \cite{2004v600p127} suggested that a bending of the jet may happen immediately 
after the $\gamma$-ray zone to explain the required large viewing angle of the radio jet. 
However, the mechanism required to bend the jets is still not well understood (jets deflected 
due to the pressure gradient in external medium are studied by Canto \& Raga 
\cite{1996v280p559}; Raga \& Canto \cite{1996v280p567}; Mendoza \& Longair 
\cite{2001v324p149})), and moreover the observed large bending of the jet in the radio maps can 
be the apparent one because of projection effects. This projection effects are even amplified
 when the jet is close to the line of sight. However it needs to be noted here that jets with 
large bending angle are indeed observed \cite{2006v647p172}.

The limb-brightened structure can also be explained if we consider the synchrotron emission 
from the particles accelerated at the boundary, and this inference does not require large 
viewing angle. Eilek \cite{1979v230p373,1982v254p472} considered the acceleration 
of particles due to turbulence initiated by Kelvin-Helmholtz and Rayleigh-Taylor instabilities 
at the jet boundary. Particles at the boundary can also be accelerated via shear acceleration 
\cite{1981v33p399,1981v7p352}, and this case is considered in the present work. The acceleration 
of particles in a shear flow or by turbulence is well studied by various authors for both 
relativistic and non-relativistic cases \cite{1988v331p91, 1989v340p1112,1990v238p435, 2002v578p763, 
2006v652p1044,2008v681p1725,2005v621p313}.

\section{Shear acceleration at MKN 501 jet boundary}{\label{sec5:shear}}
The particle acceleration process at the jet boundary can be described by the diffusion 
equation in momentum space. The evolution of an isotropic phase space distribution is 
given by \cite{1968v2p171}
\begin{align}
\label{eq:evol}
\frac{\partial f(p)}{\partial t}=\frac{1}{p^2}\;\frac{\partial}{\partial p} 
\left(p^2 D(p)\frac{\partial f(p)}{\partial p}\right)
\end{align}
where $D(p)$ is the momentum diffusion coefficient. The characteristic acceleration time-scale 
can be written as
\begin{align}
t_{acc}=p^3\left[\frac{\partial}{\partial p}\left( p^2 D(p)\right)\right]^{-1}
\end{align}
If we consider a sheared flow, the electrons are scattered across different velocity layers by 
turbulent structures which are embedded in the shear flow. Berezhko \cite{1981v33p399} 
showed in such case that there will be a net gain of energy in the electrons getting scattered, 
and this process is referred to as shear acceleration (\S \ref{sec2:shear_accln}). The momentum 
diffusion coefficient in case of a shear flow can be written as \cite{2006v652p1044,2007v309p119}
\begin{align}
\label{eq:shdiffcoeff}
D_s(p) = \chi\; p^2\, \tau
\end{align}
where $\tau$ is the mean scattering time given by $\tau \simeq \lambda/c$ with $\lambda$ 
being the mean free
path and $\chi$ is the shear coefficient given for a relativistic flow as 
\cite{2004v617p155}
\begin{align}
\chi = \frac{c^2}{15\,(\Gamma(r)^2-1)}\left(\frac{\partial \Gamma}{\partial r}\right)^2
\end{align}
where $\Gamma(r)$ is the bulk Lorentz factor of the flow and $r$ is the radial coordinate of the jet
cross section. Using equation (\ref{eq:shdiffcoeff}), the shear acceleration time-scale ($t_{acc,s}$) 
for $\tau = \tau_o p^\xi$ will be
\begin{align}
\label{eq:shacc}
t_{acc,s}= \frac{1}{(4+\xi)\chi\, \tau}
\end{align}
In case of turbulent acceleration(stochastic), the particles are scattered off by randomly moving 
scattering centres and gets energized by a second-order Fermi acceleration. The momentum diffusion 
coefficient in this case can be approximated as \cite{2007v309p119}
\begin{align}
\label{eq:turbdiffcoeff}
D_t(p)\simeq \frac{p^2}{3\,\tau}\left(\frac{V_A}{c}\right)^2 
\end{align}
where the Alfv$\acute{\textrm{e}}$n velocity($V_A$) is given by
\begin{align} 
V_A=\frac{B}{\sqrt{4\pi \rho}}
\end{align}
Here $B$ is the magnetic field and $\rho$ the mass density of the jet. 
Hence, the turbulent acceleration time-scale($t_{acc,t}$) 
will be
\begin{align}
t_{acc,t}=\frac{3\,\tau}{(4-\xi)}\left(\frac{c}{V_A}\right)^2
\end{align}
For shear acceleration to be dominant over turbulent acceleration $t_{acc,s}<t_{acc,t}$.
If we consider Bohm diffusion ($\xi = 1$) then the mean free path of the electron aligned 
to the magnetic field 
($\lambda_\parallel$) scales as the gyro radius ($r_g$) \cite{1994v285p687}, 
$\lambda_\parallel \simeq \eta \,\frac{\gamma\, m_e\, c^2}{e\,B}$, where $\eta$ is a numerical factor
($\eta>1$ for magnetized particles) and
$\gamma(\gg 1)$
is the Lorentz factor of the scattered electron. Since the magnetic field at the jet boundary of 
MKN 501 is parallel to the jet axis (or toroidal)
\cite{1999v159p427,2005v356p859,1999v43p691}, we
consider $\tau\simeq\lambda_\parallel/c$. Also if we consider
\begin{align}
\frac{\partial \Gamma}{\partial r}\simeq \frac{\Delta \Gamma}{\Delta r}
\end{align}
where $\Delta \Gamma$ is the difference between the bulk Lorentz factor at the jet spine and the 
jet boundary and $\Delta r$ is the thickness of the shear layer, then the condition for shear 
acceleration to be dominant over turbulent acceleration will be
\begin{align}
\label{eq:shthick1}
\Delta r < \frac{\eta\, \gamma\, m_e\,c^3\, (\Delta\Gamma)}{e\,B^2}\left[\frac{4\pi\, \rho}
{3\,(\Gamma(r)^2-1)}\right]^
{\frac{1}{2}}
\end{align} 
If we consider the mass density of the jet is dominated by cold protons
and if the number of protons is equal to the number of non-thermal electrons, then the 
jet mass density can be written in terms of equipartition magnetic field($B_{eq}$) as
\begin{align}
\rho \simeq \frac{m_p\,B_{eq}^2\,(2\,\alpha-1)}{16 \pi\, m_e\,c^2\, \alpha \,\gamma_{min}}
\end{align}
and equation (\ref{eq:shthick1}) will be 
\begin{align}
\label{eq:shthick2}
\Delta r < 0.29\,\frac{\eta\, \gamma\, c^2 (\Delta\Gamma)}{e\,B_{eq}}\left[\frac{m_e\,m_p\,(2\,\alpha-1)}
{\alpha\,\gamma_{min}\,(\Gamma(r)^2-1)}\right]^{\frac{1}{2}}
\end{align} 
where $\alpha$ is the observed photon spectral index, $m_p$ is the proton mass and $\gamma_{min}$ is
the Lorentz factor of electron responsible for the minimum observed 
photon frequency $\nu_{min}$. 
The equipartition magnetic field can be expressed in terms of observed 
quantities as 
\begin{align}
B_{eq}\simeq 9.62\,\frac{1}{\Gamma(r)}\,(m_e\,c\,e\,\nu_{min})^{\frac{1}{7}}\left[\frac{d_L^2\,F(\nu_{min})}
{V \sigma_T\, (2\,\alpha-1)}\right]^{\frac{2}{7}} G
\end{align}
where $F(\nu_{min})$ is the flux at the minimum observed frequency $\nu_{min}$, $d_L$ is the
luminosity distance, $V$ is the volume of the emission region and $\sigma_T$ is Thomson cross section. 
Hence, for $\Gamma(r)^2\gg1$ and $\alpha\simeq 0.7$, shear acceleration will dominate the particle
spectrum at the jet boundary of MKN 501 if the thickness of the shear layer
\begin{align}
\label{eq:shthick3}
\Delta r &< 7.22\times10^{-9}\times\left(\frac{\eta}{10}\right)
\left(\frac{\Delta \Gamma}{10}\right)
\left(\frac{\nu_{obs}}{1.6\;\textrm{GHz}}\right)^{\frac{1}{2}} 
\left(\frac{\nu_{min}}{10\;\textrm{MHz}}\right)^{-\frac{6}{14}} \times \nonumber \\
&\times
\left(\frac{F(10\;\textrm{MHz})}{910\;\textrm{mJy}}\right)^{\frac{-5}{14}}
\left(\frac{R}{1.5\;\textrm{pc}}\right)^{\frac{15}{14}}
\textrm{pc}
\end{align}
where $R$ is the radius of the spherical region considered. (We assume $10$ MHz as minimum 
observed frequency, and the flux at $10$ MHz is obtained from the 
flux at $1.6$ GHz considering 
the same spectral index.  The flux at $1.6$ GHz and $R$ in equation (\ref{eq:shthick3}) 
are obtained from a region around R.A $10$ mas and 
declination $-10$ mas from Figure \ref{fig5:mkn501} (bottom left)). 
The corresponding equipartition 
magnetic field $B_{eq}$ for $\Gamma=5$ is $1.2\times 10^{-3}$ G.

The electrons accelerated by shear acceleration cool via synchrotron radiation. The 
cooling time for synchrotron loss is given by
\begin{align}
\label{eq:syncool}
t_{cool}=\frac{6\pi\, m_e\,c}{\gamma\, \sigma_T\, B_{eq}^2}
\end{align}
Using equations (\ref{eq:shacc}) and (\ref{eq:syncool}), we find
\begin{align}
\label{eq:acccool}
\frac{t_{acc,s}}{t_{cool}}&\simeq 1.5 \times 10^{-12}\left(\frac{B}{1.2\times 10^{-3}\;\textrm{G}}\right)^3
\left(\frac{\eta}{10}\right)^{-1}\left(\frac{\Gamma(r)}{5}\right)^2 \times \nonumber \\
&\times \left(\frac{\Delta r}{10^{-9}\;\textrm{pc}}\right)^2\left(\frac{\Delta \Gamma}{10}\right)^{-2}
\end{align}
and since $t_{acc,s}\ll t_{cool}$, shear acceleration dominates over 
synchrotron cooling. 
It can be noted that equation
(\ref{eq:acccool}) is independent of the electron energy and hence the maximum energy of the electron
will be decided by the loss processes other than synchrotron loss 
(which are not considered in 
this simplistic treatment).

If we maintain the general form of mean scattering time $\tau=\tau_0p^\xi$, then for shear 
acceleration to dominate over turbulent acceleration the thickness of the shear layer 
($\Delta r$) should be
\begin{align}
\label{eq:shthickgen}
\Delta r < 1.7\times 10^6 \;\frac{\tau_0\, p^\xi\, (\Delta \Gamma)}{\Gamma(r)}
\,\left[\frac{(4+\xi)(2\,\alpha -1)}{\alpha\, (4-\xi)\, \gamma_{min}}\right]^{\frac{1}{2}}
\textrm{cm}
\end{align}
It can be noted that equations (\ref{eq:shthick1}) and (\ref{eq:shthickgen}) are equal, if we set in the latter
$\xi=1$ and $\tau_0 \;p^\xi = \eta \;r_g/c$.  

\section{Particle Diffusion at the jet boundary and Limb-brightening}

Particles accelerated at the shear layer of the jet boundary diffuse into the jet medium before
getting cooled off via synchrotron radiation. As the magnetic field at the jet boundary is 
parallel to the jet axis (or toroidal) \cite{1999v159p427,2005v356p859,1999v43p691}, the radial diffusion 
of the electron into the jet medium is determined by cross-field diffusion.  
The cross-field diffusion coefficient can be approximated as \cite{1965v13p115, 
1987v313p842,1994v285p687}  
\begin{align}
\label{eq:crossdiff}
\kappa_\perp \approx \frac{1}{3\,\eta}\, r_g\, c
\end{align}
where $\eta(>1)$ is the scaling factor determining the field-aligned mean free path 
(see \S\ref{sec5:shear}).

The radial distance $R_{diff}$ that the electron diffuse before getting cooled can then be 
approximated as 
\begin{align}
\label{eq:rdiff0}
R_{diff} \approx \sqrt{\kappa_\perp t_{cool}}
\end{align}
Using equations (\ref{eq:syncool}) and (\ref{eq:crossdiff}) and considering the equipartition magnetic 
field, we get
\begin{align}
\label{eq:rdiff}
R_{diff} \simeq 2.9 \times 10^{-4} \left(\frac{\eta}{10}\right)^{-\frac{1}{2}} 
\left(\frac{B}{1.2\times 10^{-3}\;\textrm{G}}\right)^{-\frac{3}{2}}\;\;\; \textrm{pc}
\end{align}
Since the thickness of the shear layer $\Delta r\ll R_{diff}$ (refer equations (\ref{eq:shthick3}) and 
(\ref{eq:rdiff})), the thickness of the limb-brightened 
structure will be $\approx R_{diff}$. 
This corresponds to an angular distance of $4.7\times10^{-4}$ mas which is 
beyond the resolution of present-day telescopes.

For $\tau=\tau_0\;p^\xi$, the cross-field diffusion coefficient will be 
\begin{align}
\kappa_\perp\simeq\frac{1}{3\,\tau_0}\,r_g^2\, p^{-\xi}
\end{align}
Using equations (\ref{eq:syncool}) and (\ref{eq:rdiff0}) we get 
\begin{align}
R_{diff} \simeq 5.2 \times 10^{15}\, B^{-2}\,\tau_0^{-\frac{1}{2}}\, p^{\frac{1-\xi}{2}}\;\;\;\textrm{cm}
\end{align}
and hence the thickness of the limb brightened structure will be energy dependent for $\xi \ne 1$.

\section{Spectral index}
\begin{figure}[tp]
\begin{center}
\includegraphics[width=90.mm,bb=0 0 1462 1213]{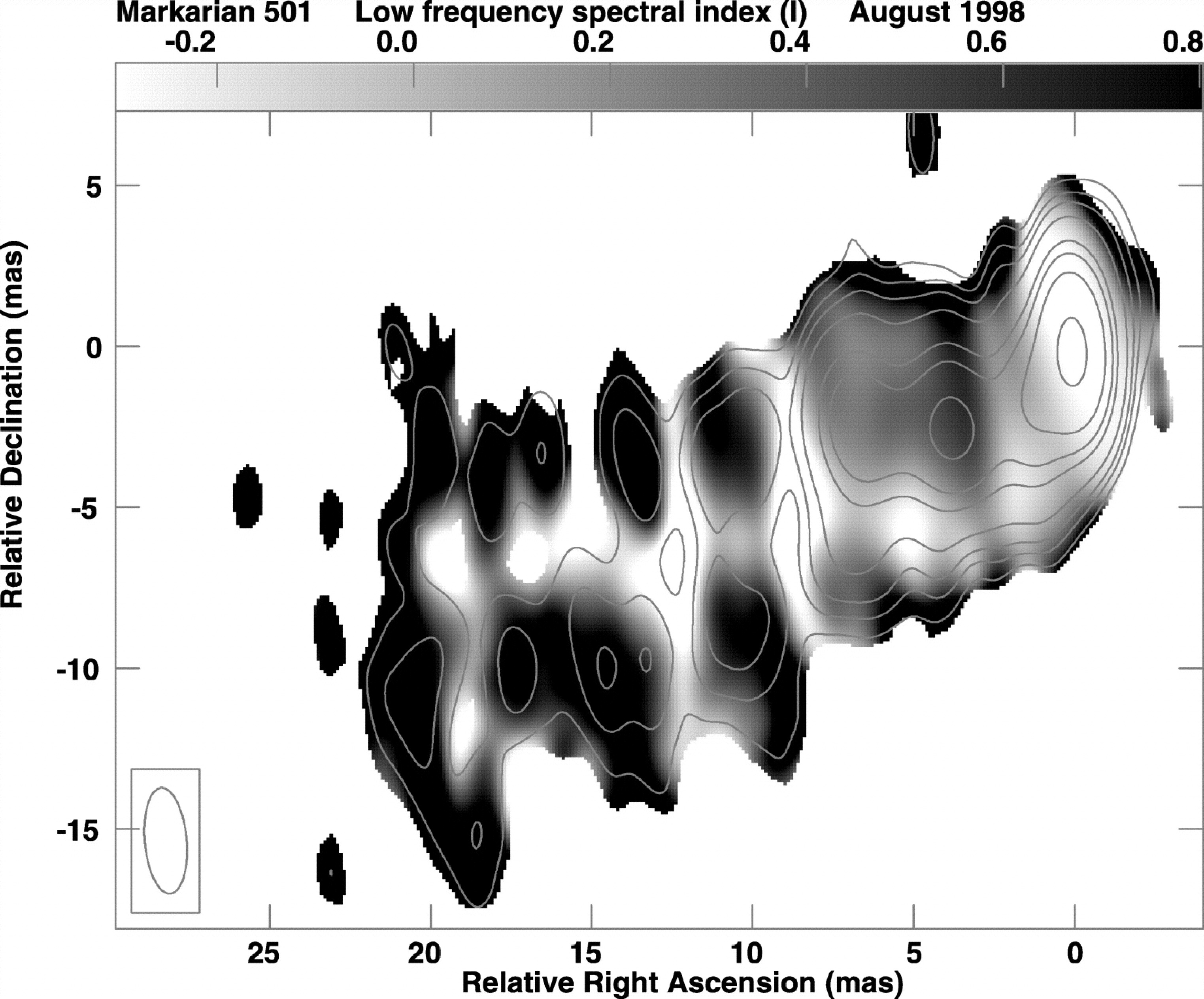} 
\caption[Spectral index map of MKN 501 jet]
{Low resolution 1.6 GHz-4.8 GHz spectral index map of MKN 501 jet. Figure reproduced from 
Giroletti et al. \cite{2004v600p127}.}
\label{fig5:mkn501_spec}
\end{center}
\end{figure}

{\label{sec5:spectral}}
If we add mono-energetic particle injection term ($\delta(p-p_o)$) and particle 
escape term ($-1/t_{esc}$)
in equation (\ref{eq:evol}), then the steady state equation in case of shear acceleration 
for $p>p_o$ and  $\xi=1$ can be written as 
\begin{align}
\label{eq:shearsteady1}
p^3\,\frac{d^2f_s}{dp^2}+5\,p^2\,\frac{df_s}{dp}-\frac{f_s}{\chi\,\tau_o \,t_{esc}} = 0
\end{align}
and in case of turbulent acceleration it will be  
\begin{align}
\label{eq:turbsteady}
p\,\frac{d^2f_t}{dp^2}+3\,\frac{df_t}{dp}-\frac{f_t}{\psi\, t_{esc}} = 0
\end{align}
where $\psi = \frac{V_A^2}{3\,c^2\,\tau_o}$. If we substitute $p = 1/x$ in equation (\ref{eq:shearsteady1})
we get 
\begin{align}
\label{eq:shearsteady}
x\,\frac{d^2f_s}{dx^2}-3\,\frac{df_s}{dx}-\frac{f_s}{\chi\, \tau_o\, t_{esc}} = 0
\end{align}
Equations (\ref{eq:turbsteady}) and (\ref{eq:shearsteady}) can be solved analytically 
\cite{1905v61p397}. The solutions are complex and are given by
\begin{align}
\label{eq:shearsol}
f_s&=\left(\frac{1}{\chi\,\tau_o\, p\, t_{esc}}\right)^2 \times \nonumber \\
	 &\times \left[a_s\, J_4\left(2i\sqrt{\frac{1}
	{\chi\,\tau_o\, p\, t_{esc}}}\right)+b_s\, Y_4\left(2i\,\sqrt{\frac{1}{\chi\,\tau_o\, p\, t_{esc}}}\right)\right]
\end{align}
and
\begin{align}
\label{eq:turbsol}
f_t=\left(\frac{\psi\, t_{esc}}{p}\right)\left[a_t\, J_2\left(2i\,\sqrt{\frac{p}{\psi\, t_{esc}}}
\right)+b_t\, Y_2\left(2i\,\sqrt{\frac{p}{\psi\, t_{esc}}}\right)\right]
\end{align}
where $J_n(z)$ and $Y_n(z)$ are the Bessel functions of first and second kind and $a_s$,
$b_s$, $a_t$ and $b_t$ are constants. For negligible escape ($t_{esc} \to \infty$), using the limiting
forms of Bessel functions \cite{book:abramowitz}, the solutions, equations (\ref{eq:shearsol}) 
and (\ref{eq:turbsol}),
approach a power-law $f_s \propto p^{-4}$ and $f_t \propto p^{-2}$. The shear-accelerated 
particle number density will then be $n_s(p)\propto p^{-2}$ and the corresponding synchrotron photon
flux will be $S_{\nu,shear} \propto \nu ^{-1/2}$. For turbulent acceleration, the number density will be 
independent of $p$ ($n_t(p)\propto p^0$) and hence the observed synchrotron photon flux will be 
a flat one $S_{\nu,turb} \propto \nu ^{1/3}$ \cite{book:pacholczyk}.
The spectral index map of MKN 501 jet indicates a steep photon spectrum at the boundary 
and a flat spectrum at the spine \cite{2004v600p127}(Figure \ref{fig5:mkn501_spec}). 
Hence, it can be argued that the shear acceleration may be dominant
at the jet boundary of MKN 501 and turbulent acceleration at the jet spine. However, $\xi$ is
usually related to the turbulent spectral index \cite{1987v322p643} which may be different at the
jet boundary and jet spine.

\section{Discussion}

As the AGN jet moves through the ambient medium, the viscosity involved will cause a shear at 
the jet boundary, and hence acceleration of particles in these shear layers is unavoidable. If 
the shear gradient $\partial \Gamma/\partial r$ is very steep or if the shear layer is very 
thin (equation (\ref{eq:shthick3})), then shear acceleration can dominate over the turbulent 
acceleration initiated by the instabilities at the jet boundary \cite{1982v254p472}. Turbulent 
acceleration may play an important role at the interior regions of the jet \cite{2005v621p313} 
and can provide an alternative to explain the emission from the inter knot regions of AGN jets
\cite{1996v175p195, 2001v373p447}. The observed hard spectrum at the jet spine \cite{2004v600p127} 
also support this inference since turbulent acceleration can produce a hard particle spectra 
\cite{2005v621p313} (also shown in \S \ref{sec5:spectral}). The electrons accelerated by the 
turbulence can be reaccelerated by shocks and can form a broken power-law electron spectrum. 
This can possibly explain the break in the radio-to-X-ray spectrum of the knots of FRI jets 
(\S \ref{sec4:m87}).

Giroletti et al. \cite{2004v600p127} calculated the jet viewing angle ($\theta$) using 
the correlation between the core power and the total power \cite{2001v552p508}. They estimated 
the jet viewing angle to be within $10^\circ<\theta<27^\circ$ by comparing the observed core 
radio power and the expected intrinsic core power derived from the correlation. However, this 
estimation may vary if the core flux density variability is more than a factor of $2$. Also 
considering the variation of the parameter values in the correlation with increased number of 
samples, this may not provide a strong constrain on the jet viewing angle. The estimate of 
$\theta$ based on the adiabatically expanding relativistic jet model \cite{1997v483p178} may 
not be a strong constraint as it considers a simplified situation. Also, the constrain is less 
severe in case of perpendicular magnetic fields, and observed polarization studies have indicated 
the presence of perpendicular magnetic fields at jet spine \cite{2005v356p859,1999v159p427}. 
Stawarz \& Ostrowski \cite{2002v578p763} proposed a model similar to the present one; however, 
their aim was to show the observational implications of the two-component particle spectrum 
(power-law distribution with high-energy pile-up) formed at the boundary shear layer and the 
complex beaming pattern.


\chapter{A Two zone Model for Blazar Emission Mechanism}
\label{chap:blazar}

The spectral energy distribution(SED) of blazars are characterized by a typical 
double-hump feature. The first component peaks at IR/optical energies for low energy 
peaked blazars and at UV/soft X-ray energies for high energy peaked blazars. 
Whereas, the second component peaks at hard X-ray/$\gamma$-ray region. The first 
component is generally modelled as the synchrotron emission due to the cooling of 
relativistic electrons in a magnetic field 
while the second component is considered
to be produced due to the inverse Compton scattering of soft photons by the
relativistic electrons.
The target photons for the inverse Compton process can be either synchrotron 
photons themselves
(SSC) \cite{1996v470p89,1998v509p608,1999v306p551,2000v536p299} 
or external radiation (EC) (\S  \ref{sec2:leptmod})
\cite{1993v416p458,1994v421p153,2000v545p107}.

Besides their non-thermal continuum, blazars also exhibit
rapid and strong variability 
(for example MKN 421 
\cite{2000v541p153, 2000v541p166,2000v542p105},
MKN 501 \cite{2000v353p97,1998v492p17}, 
PKS 2155-304 \cite{1999v521p552,2000v528p243,2001v554p274}, etc.). 
In some cases the variability detected at different frequencies are
non simultaneous and the flaring activity is referred as \emph{hard lag} or 
\emph{soft lag} depending upon the observed flare pattern. In a hard lag, the 
low frequency flare leads the high frequency one and for the soft lag it is the 
other way.
Takahashi et al. \cite{1996v470p89} reported a soft lag 
in ASCA observation of MKN 421, while Fossati et al. 
\cite{2000v541p153, 2000v541p166} reported a hard lag 
in the {\it BeppoSAX} observation of MKN 421
in 1997-1998. Similarly for PKS 2155-304, a soft lag was reported by 
Chiappetti et al. \cite{1999v521p552} and Kataoka et al. 
\cite{2000v528p243} in {\it BeppoSAX} observation of 
1997 and ASCA observation of 1994 respectively. However, 
Takahashi et al. \cite{2000v542p105} contradicted the conclusions of 
Fossati et al. \cite{2000v541p153, 2000v541p166} by reporting no spectral lag 
for the same data set. Also, Edelson et al. \cite{2001v554p274} reported 
no spectral lag for PKS 2155-304 from {\it XMM-Newton} observation in 2000. 
Nevertheless, the study of short time-scale variability at different
wavelength region and their interrelations is very important to understand 
the geometrical and causal connections between emission regions and emission 
processes.

The flux variability observed in blazars have been studied by several authors
using one zone and two zone models
\cite{1996v463p555,1998v333p452,1999v306p551,2000v536p299}. 
Kirk et al. \cite{1998v333p452} and Kusunose et al. \cite{2000v536p299} 
assumed a two zone model where particles are accelerated in a region presumably by a 
shock and escape into a cooling region where they lose their energy by radiative 
processes. On the contrary, one zone models assume instantaneous particle
acceleration and do not consider the particle acceleration process 
separately. Chiaberge \& Ghisellini \cite{1999v306p551} explained the short 
time variability observed in MKN 421 by dividing the emission region into thin slices. 
Moderski et al. \cite{2003v406p855} modelled the variability detected in 3C279 
considering SSC and EC processes.  

We consider a two zone model, namely acceleration region and  cooling region, 
to study the spectral behaviour of blazars. This model is similar to the one 
discussed in \S \ref{sec4:m87} used to model the knots of M87 jet.
However here we assume mono energetic particles are injected into the acceleration
region which are then accelerated. We studied the effect of particle 
acceleration time-scale on the variability pattern as well as spectral evolution
under two different scenarios: (a) acceleration time-scale is independent
of particle energy and (b) acceleration time-scale depends on particle energy.

\section[The Model]{Model}

We model the broadband emission of blazar from a spherical blob (cooling 
region) of radius $R$
permeated by a tangled magnetic field $B$, moving down the jet with a bulk 
Lorentz factor $\Gamma$. The  Doppler factor of the blob is given by
$\delta = [\Gamma(1-\beta\cos\theta)]^{-1}$, where $\theta$ is the
angle made by the jet with the line of sight.

Relativistic electrons are injected into the cooling region from an 
acceleration
region, around a shock front moving with velocity $v_s$. The acceleration 
region
is continuously fed with  mono-energetic electrons with Lorentz factor 
$\gamma_0$ at a rate
$Q_0$ electrons per unit volume. It is assumed that electrons are accelerated
by diffusive shock acceleration process \cite{1983v46p973} and 
their evolution $Q(\gamma,t)$ is described by,
\begin{align}
\label{eq6:kinetic}
\frac{\partial Q(\gamma,t)}{\partial t}+\frac{\partial}{\partial \gamma}
\left(\left[\left(\frac{d\gamma}{dt}\right)_{acc}-
\left(\frac{d\gamma}{dt}\right)_{syn}\right]Q(\gamma,t)\right)+\frac{Q(\gamma,t)}{t_{esc}}
=Q_0 \;\delta(\gamma-\gamma_0)
\end{align}
where
\begin{align}
\left(\frac{d\gamma}{dt}\right)_{acc}= \frac{\gamma}{t_{acc}}
\end{align}
and
\begin{align}
\left(\frac{d\gamma}{dt}\right)_{syn}= \frac{4}{3}\; 
\frac{\sigma_{T}}{m_e\,c}\;\gamma^2\;
                                    \frac{B_{acc}^2}{8 \pi}
\end{align}
are the particle acceleration rate and synchrotron loss rate of the particles
in the acceleration region.
Here $t_{acc}$ and $t_{esc}$ are the 
acceleration and escape time-scale of the electrons respectively 
and $B_{acc}$ is the magnetic field 
in the acceleration region.
In the frame work of diffusive shock acceleration, acceleration time-scale 
can be written as
\begin{align}
\label{eq6:tacc}
t_{acc}&=\left(\frac{20\,c}{3\,v_{s}^2}\right)
\left(\frac{m_e\,c^2}{e\,B_{acc}}\right)\xi\, \gamma \nonumber \\
&= t_{a,0}\, \gamma
\end{align}
where 
\begin{align} 
t_{a,0} = \left(\frac{20\,c}{3\,v_{s}^2}\right)\left(\frac{m_e\,c^2}{e\,B_{acc}}\right)\xi
\end{align}
The escape time-scale $t_{esc}$ is assumed to be $\alpha 
t_{acc}$ where $\alpha$
is a constant factor.

For energy independent scenario, $\gamma$ in equation (\ref{eq6:tacc}) 
is replaced by a constant value, $\gamma_{eff}$ \cite{2000v536p299} 
and for energy dependent scenario, equation (\ref{eq6:tacc}) is maintained as such. 
The dependence of $t_{esc}$ on  $t_{acc}$ is kept unchanged in both the cases.
The analytical solutions of equation (\ref{eq6:kinetic}), under these two 
physical conditions, have already been discussed by \cite{1998v333p452}.

Accelerated particles are injected into the cooling region from the acceleration 
region at a rate $Q(\gamma,t)/t_{esc}$ where they their lose energy via synchrotron and 
SSC processes. The evolution of the particle spectrum $N(\gamma,t)$
in the cooling region is governed by
\begin{align}
\label{eq6:crkin}
\frac{\partial N(\gamma,t)}{\partial t}-\frac{\partial}{\partial \gamma}
\left[
\left(\frac{d\gamma}{dt}\right)_{loss}N(\gamma,t)\right]+\frac{N(\gamma,t)}{\tau}
= \frac{Q(\gamma,t)}{t_{esc}}
\end{align}
where total energy loss rate of an electron,
\begin{align}
\left(\frac{d\gamma}{dt}\right)_{loss} = 
\left(\frac{d\gamma}{dt}\right)_{syn}
+\left(\frac{d\gamma}{dt}\right)_{SSC}
\end{align}
The synchrotron loss rate is given by
\begin{align}
\left(\frac{d\gamma}{dt}\right)_{syn}= \frac{4}{3}\; 
\frac{\sigma_{T}}{m_ec}\,\gamma^2\,
                                        \frac{B_{0}^2}{8 \pi}
\end{align}
and the energy loss rate due to SSC process is given by
\begin{align}
\left(\frac{d\gamma}{dt}\right)_{SSC}= \frac{4}{3}\; 
\frac{\sigma_{T}}{m_e\,c}\,
                                             \gamma^2\, U_{ph}
\end{align}
$U_{ph}$ is the energy density of synchrotron photons in the cooling region and can be 
calculated from the synchrotron specific intensity. Here $B_0$ is the magnetic field 
in cooling region and
$\sigma_T$ is the Thomson cross-section.

The equation (\ref{eq6:crkin}) is solved numerically using finite difference scheme  
\cite{1970v6p1,1999v306p551}. The specific 
intensity of synchrotron radiation at frequency $\nu$ is given by
\begin{align}
I_{s}(\nu,t)=\frac{j_\nu(t)}{\kappa_\nu(t)}
                       (1-e^{-\kappa_\nu(t)R})
\end{align}
where, synchrotron emissivity $j_\nu$ and synchrotron self-absorption
coefficient $\kappa_\nu$ are given by
(equations (\ref{eq3:synemiss}) and (\ref{eq3:ssacoeff}))
\begin{align}
j_\nu(t)=\frac{1}{4\pi} \int\limits_{\gamma_{min}}^{\gamma_{max}} 
                   N(\gamma,t)P(\gamma,\nu)\;d\gamma
\end{align}
and 
\begin{align}
\kappa_\nu(t)=-\frac{1}{8\pi\, m_e\,\nu^2}\int\limits_{\gamma_{min}}^{\gamma_{max}}
               \frac{N(\gamma,t)}{\gamma\, (\gamma^2-1)^{\frac{1}{2}}}
               \;\frac{d}{dt}[\gamma\, (\gamma^2-1)^{\frac{1}{2}} P(\gamma,\nu)]
\end{align}
respectively.
Here, $P(\gamma,\nu)$ is single particle emissivity (equation (\ref{eq3:synspe})).
The synchrotron photon energy density is then calculated using
\begin{align}
U_{ph}(\gamma,t)=\frac{4\pi}{c}\int\limits_{\nu_{s,min}}^{\nu_{max}(\gamma)}I_{s}(\nu,t)\;d\nu
\end{align}
where \cite{1986v305p45}
\begin{align}
\nu_{max}(\gamma)= min[\nu_{s,max}, 3\,m_e\,c^2/4h\gamma]
\end{align}
Here $\nu_{s,min}$ and $\nu_{s,max}$ are the minimum and maximum frequency of the 
synchrotron photons.
The specific intensity of inverse Compton scattering in the Thomson regime 
is calculated by using
the standard formulation \cite{book:rybicki}.

Different possible mechanisms for the generation of flare have been discussed 
in the literature. For example, Kusunose et al. 
\cite{2000v536p299} simulated flares by changing relation 
between acceleration
time-scale($t_{acc}$) and escape time-scale($t_{esc}$) in the acceleration 
region for certain
duration. Mastichiadis \& Kirk \cite{1997v320p19} discussed flares 
due to sudden changes 
in, (i) electron injection rate $Q_0$; (ii) maximum attainable energy of 
electrons $\gamma_{max}$ and (iii) cooling region magnetic field $B$.
We simulated flares by increasing the
injection into the acceleration region for a very short duration over and 
above the steady state injection.
Finally, the observed flux is computed taking into account the Doppler 
boosting \cite{1984v56p255} and cosmological effects.

\section{Results and Discussion}

\subsection{Spectral evolution}
We considered the quiescent spectrum of MKN 421 as a test case and deduce 
the best fit parameter
set for both energy dependent $t_{acc}$ and energy independent 
$t_{acc}$.
Values of the parameters are tabulated in Table \ref{ch6:table1} which are taken as the
standard values and all the results presented in this work are 
based on these values.
With these set of parameters $t_{acc}$ turned out to be  $1.4 
\times 10^5$ s (in source frame) for energy
independent case, while  $0.5 \gamma$ s (in source frame) for energy dependent case. 
Hence the particle acceleration rate ($\gamma/t_{acc}$) becomes 
$0.70 \times 10^{-5} \gamma$ s$^{-1}$ and $2$ s$^{-1}$ 
for energy independent and energy dependent cases respectively. 

\begin{table}[tp]\footnotesize
\caption[Two Zone Model Parameters used for fitting MKN 421]
{Parameters used for fitting MKN 421}
\label{ch6:table1}
\begin{tabular}{l l l l }
\toprule
Region & Parameters & Energy independent $t_{acc}$ & Energy 
dependent $t_{acc}$ \\
\midrule
Acceleration region& $\gamma_{o}$ & 1.00 & 1.00 \\
& $Q_0$ ($10^{-5}$ \# cm$^{-3}$ s$^{-1})$ & 2.1 & 2.1 \\
& $B_{acc}$ (G) & $0.137$ & $0.132$ \\
& $v_{s}$ (in units of c) & $1$ & $0.17$ \\
& $\gamma_{eff}$ & $10^7$ & $\ldots$ \\
& $\xi$ & $5 \times 10^3$ & $5 \times 10^3$ \\
& $\alpha$ & $1.70$ & $1.70$ \\
	\midrule
Cooling region& $B_{o}$ (G) & $0.18$ & $0.18$ \\
 & $R$ (cm) & $2.4 \times 10^{16}$ & $2.4 \times 10^{16}$ \\
& $\tau$ (s) & $\frac{2.5R}{c}$ & $\frac{2.5R}{c}$ \\
& $\delta$ & $12$ & $12$ \\
\bottomrule
\end{tabular}
\vskip .5cm
\end{table}

\begin{figure}[tp]
\begin{center}
\includegraphics[width=150.mm,bb=12 13 601 779]{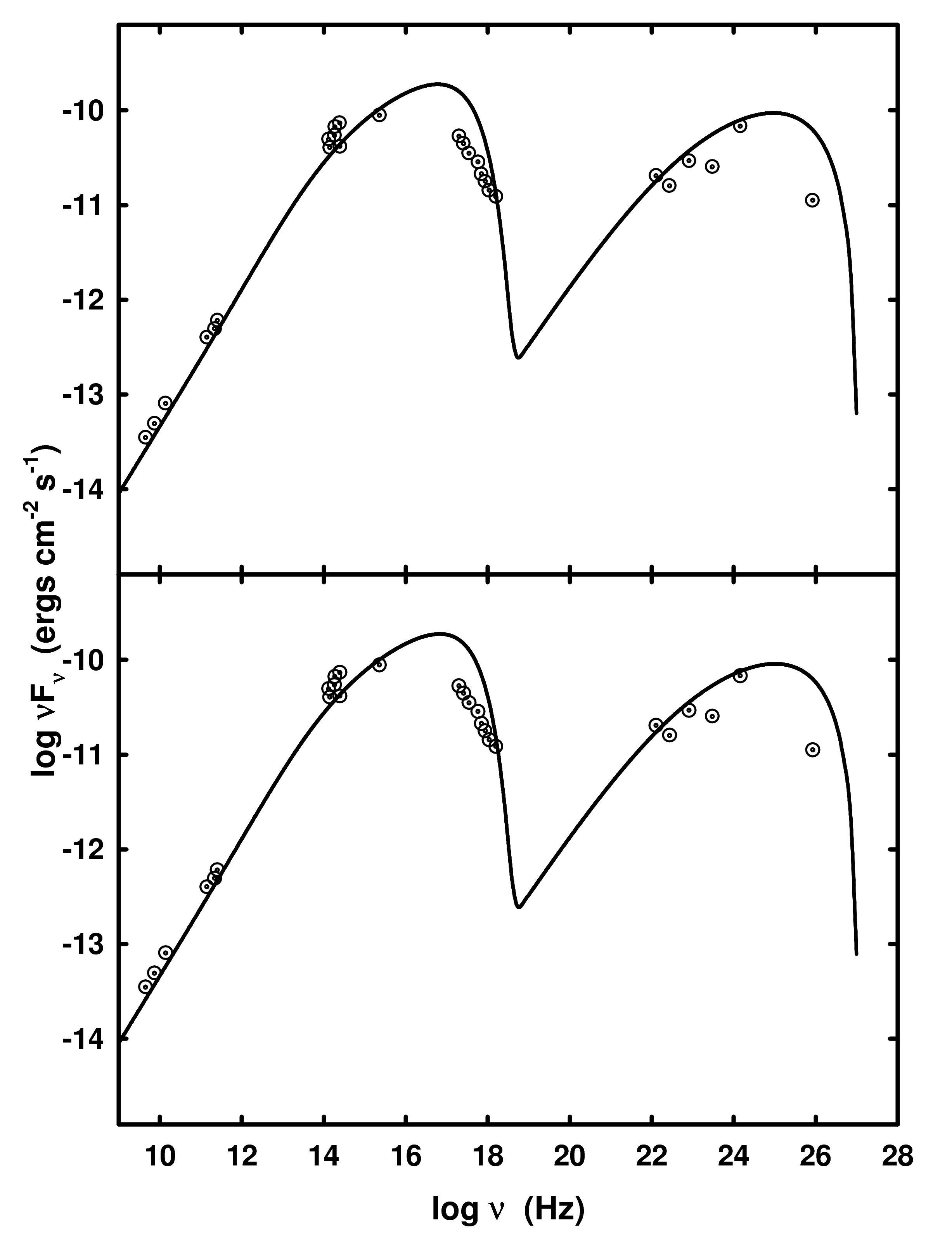} 
\caption[Spectral fits of MKN 421 using two zone model]
{Two zone model fitting of MKN 421 archival data considering 
energy independent acceleration time-scale (top) and energy dependent 
acceleration time-scale (bottom)
using the parameters given in Table \ref{ch6:table1}. The observed data are taken from
Kino et al. \cite{2002v564p97}.}
\label{fig6:fit}
\end{center}
\end{figure}

The fitted spectra of MKN 421 for both the cases of 
$t_{acc}$ are shown
in Figure \ref{fig6:fit}. The slope of the spectrum before synchrotron peak 
is decided by
$\alpha$, which relates the escape and the acceleration time-scale of electrons 
in the acceleration region. The escape
of the particles from the cooling region produces a break at 
$\nu\approx10^{14}$ Hz. As the second hump of the spectrum is produced due to the 
self-synchrotron Compton process, the break in the synchrotron component of the 
spectrum is also reflected there.  

\begin{figure}[tp]
\begin{center}
\includegraphics[width=150.mm,bb=11 4 592 786]{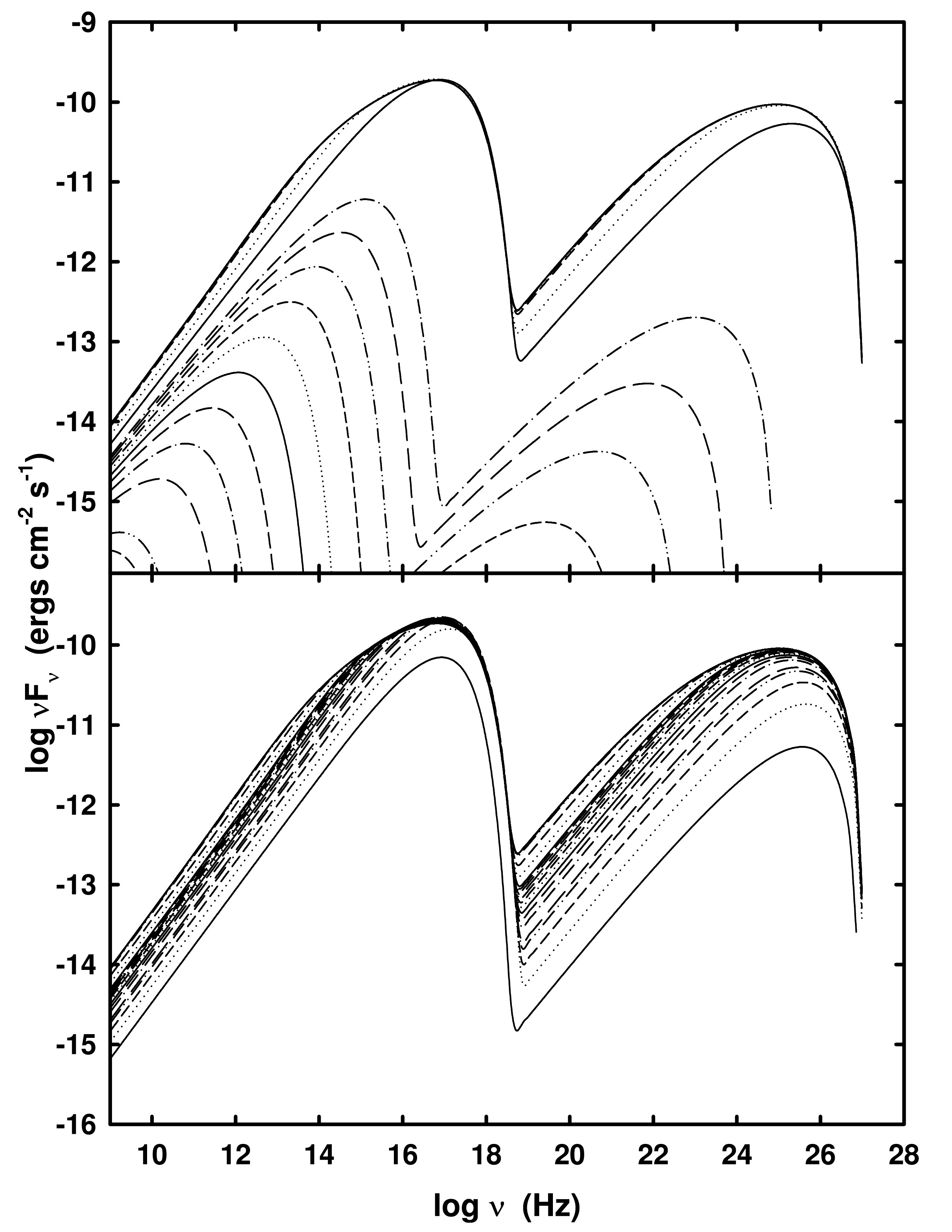} 
\caption[Evolution of the model spectra]{Evolution of the model spectra in case of 
energy independent acceleration time-scale (top) and energy dependent 
acceleration time-scale (bottom). The spectra are plotted for every 0.1 days
up to 1.5 days and there after at 2nd, 3rd, 6th and 8th days (in observer's frame).
The steady sate is arrived after 8 days in both the cases.} 
\label{fig6:evol}
\end{center}
\end{figure}

As shown in Figure \ref{fig6:fit}, the steady state spectrum of MKN 421 can be 
reproduced successfully by the two acceleration scenarios though rate of 
acceleration of particles in these scenarios are different. 
Hence the steady state spectrum of blazar cannot uniquely reflect the possible underlying 
electron acceleration mechanism and its dependencies on the electron energy. 
In case of energy independent $t_{acc}$, the rate of electron acceleration depends 
on the electron energy and is much less than the electron acceleration rate for 
energy dependent $t_{acc}$ (which is a constant in this case) for a wide range of $\gamma$.
Hence the evolution of the spectrum for energy independent $t_{acc}$ will be 
much slower compared to the energy dependent $t_{acc}$ case as shown in 
Figure \ref{fig6:evol}.
This feature in the spectral evolution will also be reflected in the flare phenomena, 
if the flare is produced due to a sudden increase 
in the rate of injection of particles $Q_0$ in the acceleration region for a small
duration. 


For energy independent case the system attains the steady state on the 8th day 
(in the observer's frame) of evolution. Although, in the energy dependent scenario,
the spectrum attains the steady state faster but to make a comparative study of flares, we considered
the steady state spectrum in both the scenarios only after the 8th day of evolution.
We discuss below the characteristics of the flare phenomena under the 
two different scenarios of particle acceleration process discussed above.

\subsection{Flare}
As mentioned above, we simulate a flare by increasing the electron 
injection rate into the acceleration region by a factor of 50 
for a duration
of 0.2 days (in the observer's frame) after the system reaches the 
steady state .

\subsubsection{Energy independent $t_{acc}$}
\begin{figure}[tp]
\begin{center}
\includegraphics[width=150.mm,bb=12 13 601 779]{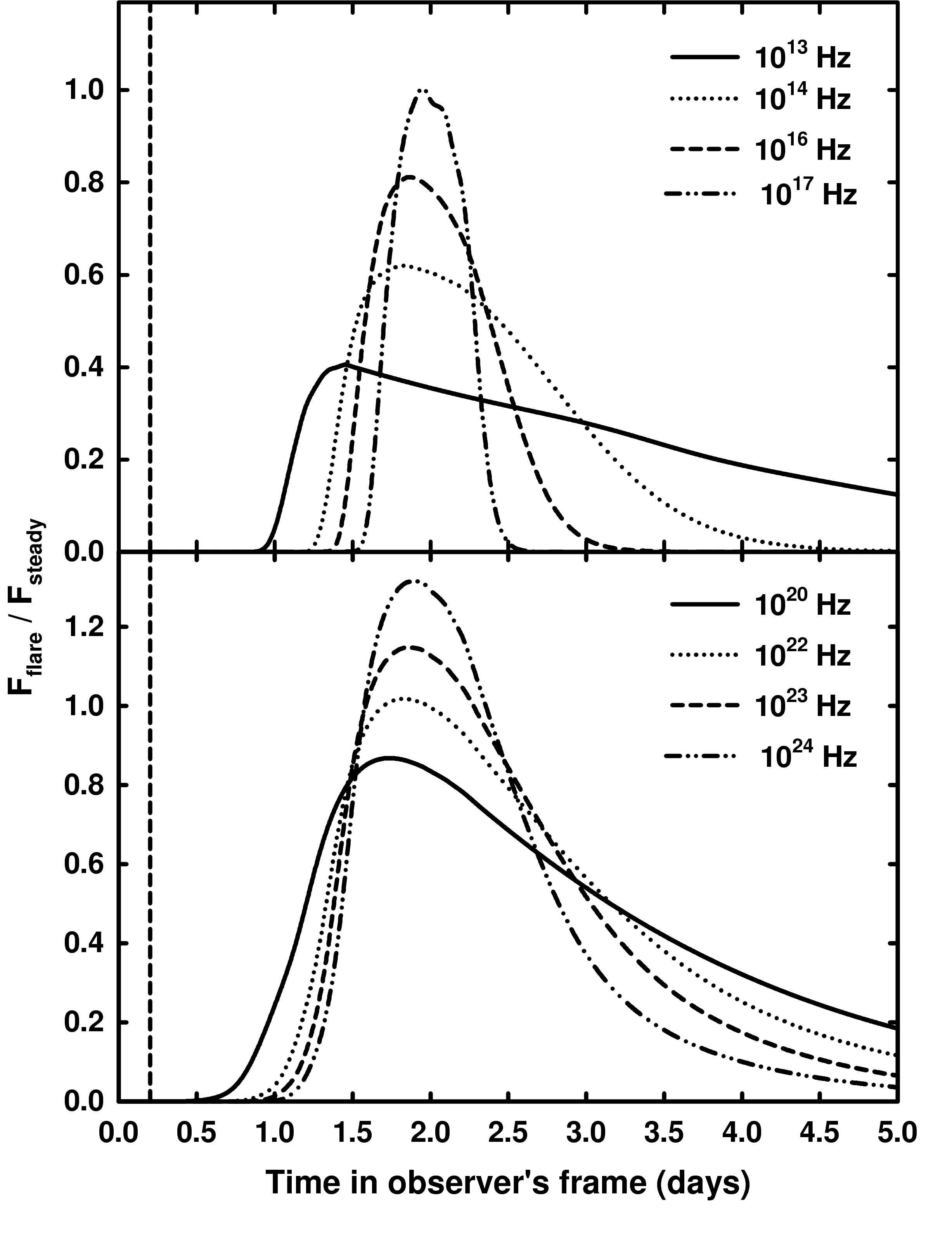} 
\caption[Synchrotron and SSC light curves (energy independent acceleration time-scale)]
{Light curves at different frequencies for 
energy independent acceleration time-scale during a simulated flare
created by the enhanced injection into the acceleration region (see text). 
The vertical dashed line represents the duration of the enhanced particle 
injection in the acceleration region. The top panel corresponds to 
the emission due to synchrotron process and bottom panel corresponds to
the SSC process.}
\label{fig6:lc1}
\end{center}
\end{figure}

The simulated light curves for energy-independent $t_{acc}$ are 
shown in Figure \ref{fig6:lc1}. It is evident from top panel of 
Figure \ref{fig6:lc1} that the high energy flares lag the low energy ones 
in time ($\sim$ kilo s) and it is also seen that the light curves are more
asymmetric at lower energies. The lag in the light curves arise since the 
rate of acceleration of electrons in the acceleration region is proportional 
to energy of the electrons and the high energy electrons which are 
responsible for the high energy synchrotron emission, take longer time
to attain the required energy. The asymmetry in the low energy light curves 
can be attributed to the longer synchrotron cooling time-scales of low energy 
electrons compared to the high energy ones. It can also be seen that the
variability amplitude increases with frequency of emission. These features 
are in qualitative agreement with the {\it BeppoSAX} observations of MKN 421 
\cite{2000v541p153,2000v541p166}. The lower panel of Figure \ref{fig6:lc1} 
describes the simulated flare patterns for SSC component of emission. Since, 
the synchrotron photons are Compton boosted by the same population of 
electrons, the variability features in SSC component are qualitatively same 
as in the synchrotron component.

\begin{figure}[tp]
\begin{center}
\includegraphics[width=150.mm,bb=12 13 601 779]{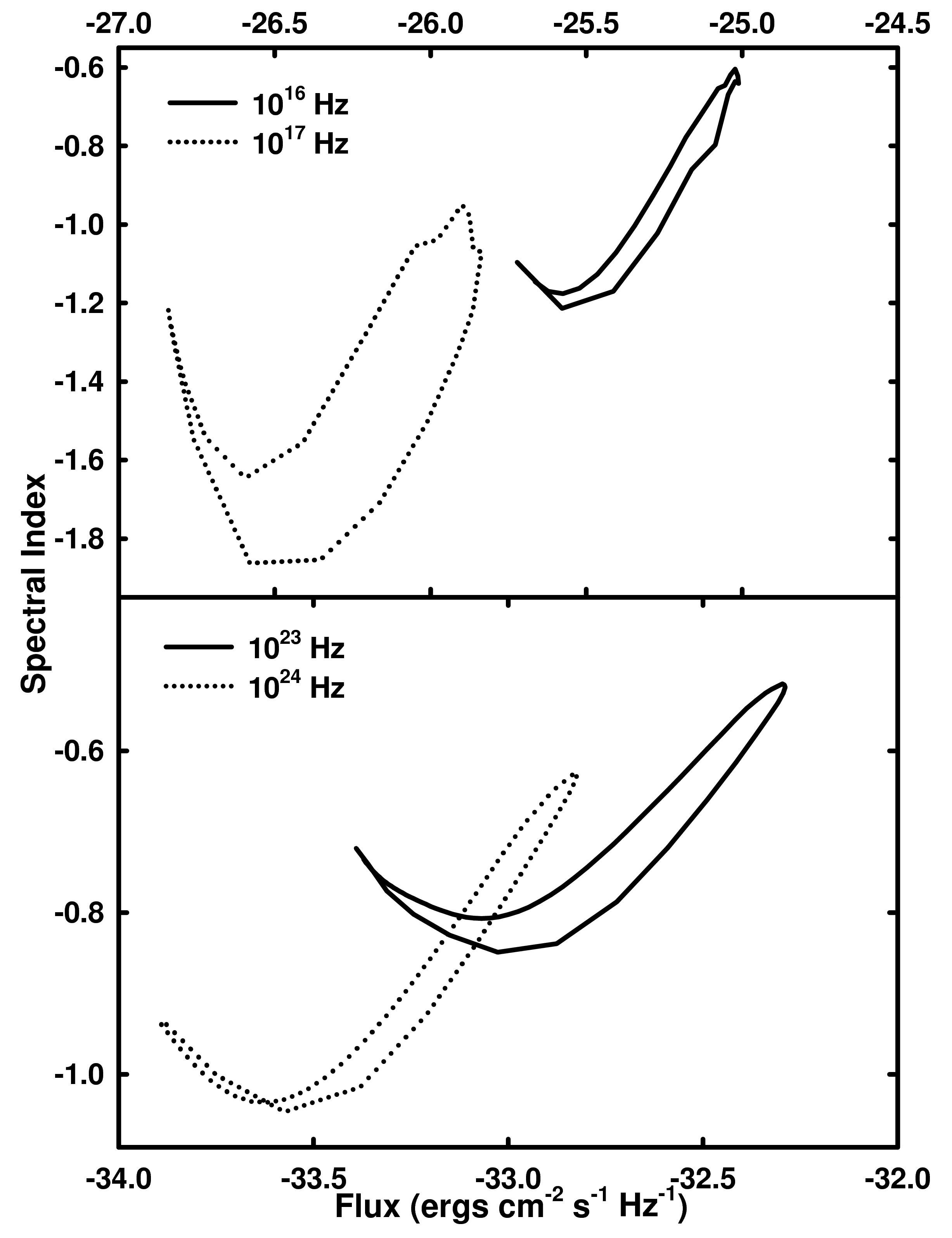} 
\caption[Hysteresis curves during a flare (energy independent acceleration time-scale)]
{The hysteresis curves showing the variation of spectral index with
respect to flux during a simulated flare for energy independent acceleration 
time-scale. The top panel corresponds to 
the emission due to synchrotron process and bottom panel corresponds to
the SSC process. The sense of rotation of the loops is anti clockwise.}
\label{fig6:hys1}
\end{center}
\end{figure}

The variation of the spectral index with respect to the flux ({\it hysteresis 
loops}) have also been studied at different energies and are shown in Figure 
\ref{fig6:hys1}. The sense of the loops is anti-clockwise representing the hard 
lag in the system.

\subsubsection{Energy dependent $t_{acc}$}
\begin{figure}[tp]
\begin{center}
\includegraphics[width=150.mm,bb=12 13 601 779]{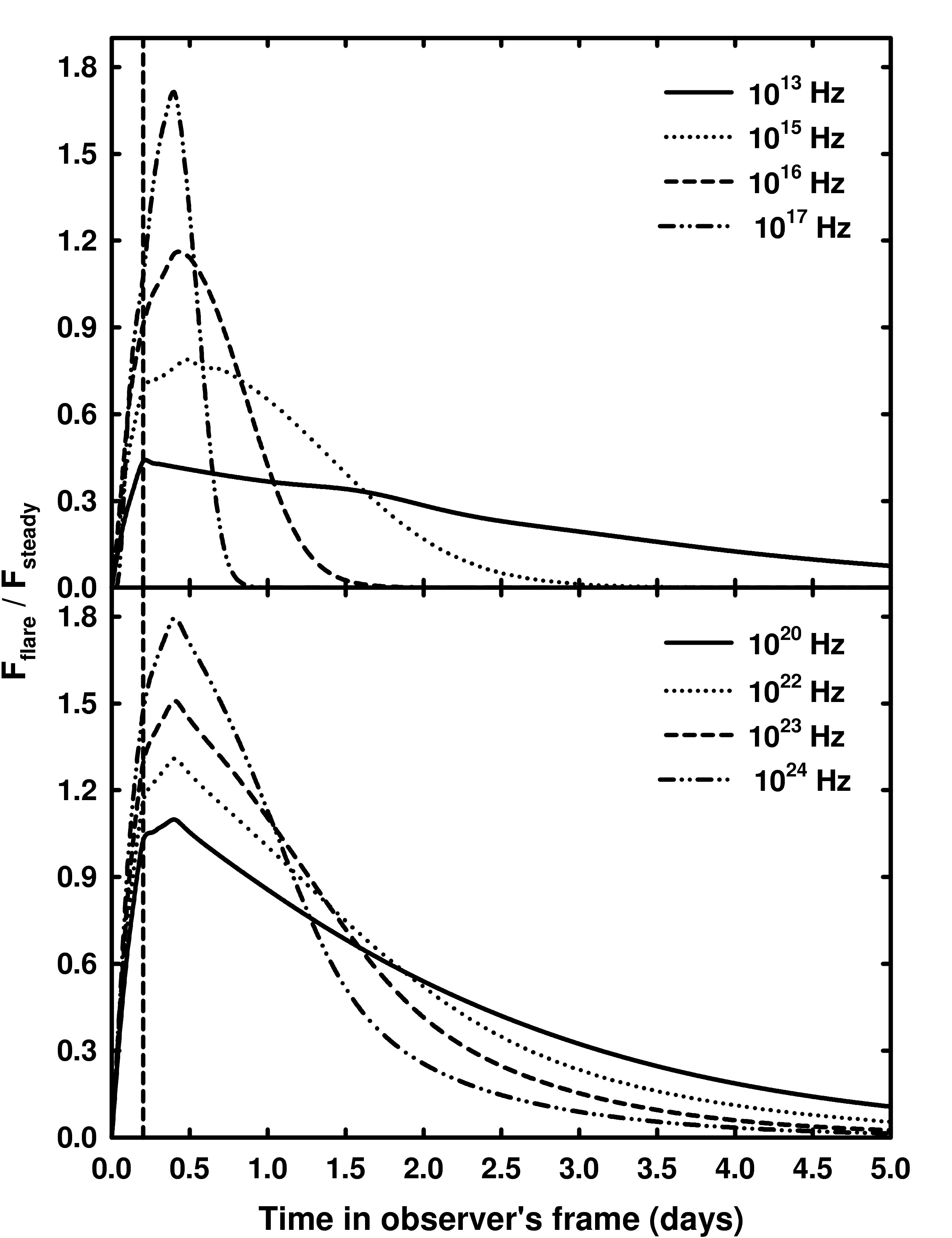} 
\caption[Synchrotron and SSC light curves (energy dependent acceleration time-scale)]
{Light curves at different frequencies for 
energy dependent acceleration time-scale during a simulated flare
created by the enhanced injection into the acceleration region (see text). 
The vertical dashed line represents the duration of the enhanced particle 
injection in the acceleration region. The top panel corresponds to 
the emission due to synchrotron process and bottom panel corresponds to
the SSC process.}
\label{fig6:lc2}
\end{center}
\end{figure}

\begin{figure}[tp]
\begin{center}
\includegraphics[width=150.mm,bb=12 13 601 779]{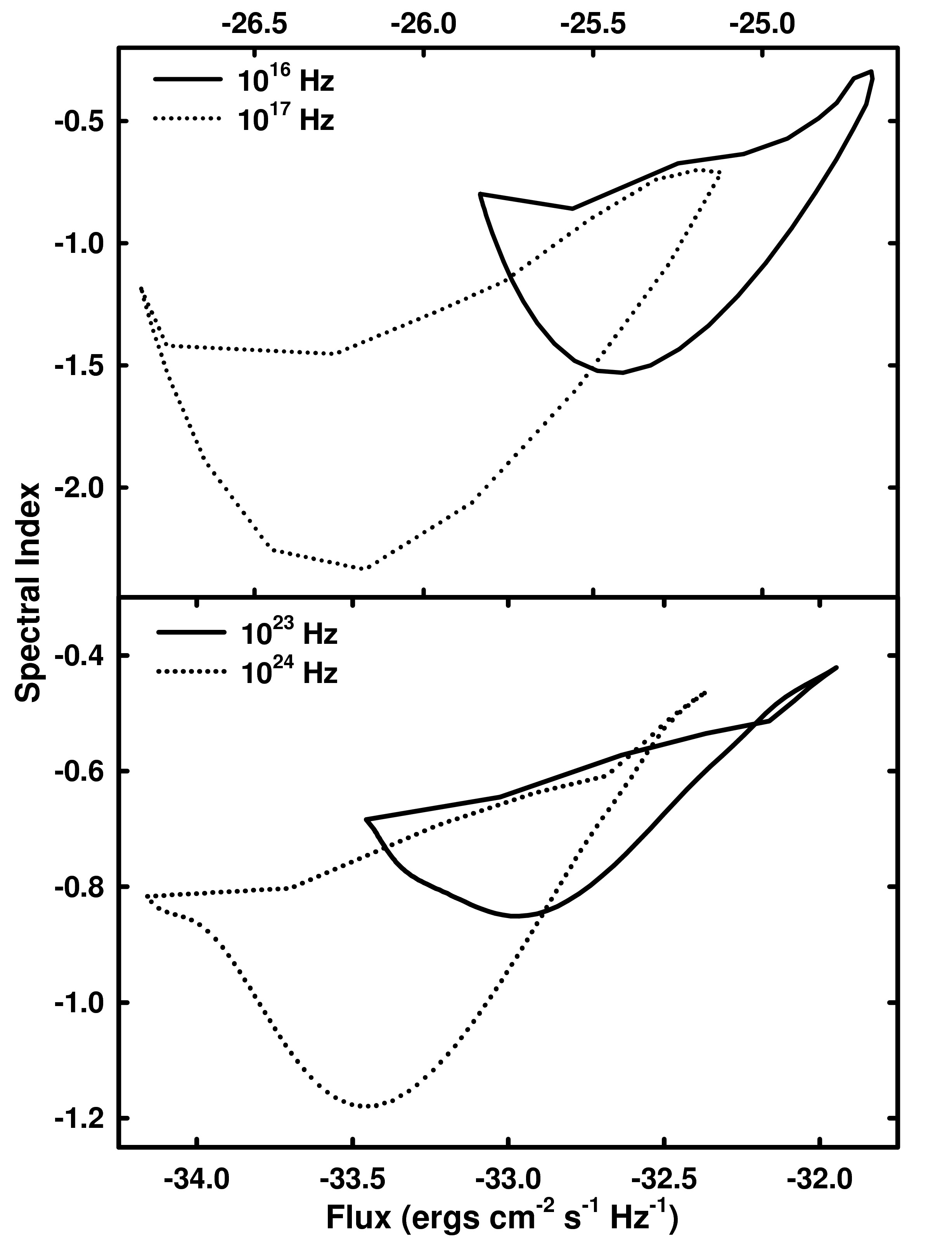} 
\caption[Hysteresis curves during a flare (energy dependent acceleration time-scale)]
{The hysteresis curves showing the variation of spectral index with
respect to flux during a simulated flare for energy dependent acceleration 
time-scale. The top panel corresponds to 
the emission due to synchrotron process and bottom panel corresponds to
the SSC process. The sense of rotation of the loops is clockwise.}
\label{fig6:hys2}
\end{center}
\end{figure}

In this case the acceleration time-scale is proportional to the energy of 
the electrons and hence the rate of acceleration is independent 
of electron energies ($1/t_{a,0}$). For the parameters given in table 
\ref{ch6:table1}, we find $t_{a,0} = 0.5$ s (in the source frame).  
As the rate of particle acceleration is high and same for all electron energies, 
the flares are initiated almost immediately after the extra injection into the 
acceleration region occurs. This fact is evident from near-simultaneous flare patterns 
shown in Figure \ref{fig6:lc2}. The light curves at optical/UV frequencies in 
the synchrotron component show prominent breaks in their rising part. Since, the 
rate of acceleration is high 
and same for all energies, electrons in acceleration region attain higher energy in 
shorter time-scale compared to the escape time-scale, ($t_{esc}$).
These high energy particles enter cooling region and get cooled giving rise 
to high energy emission. Then they join the freshly injected lower energy electrons 
and contribute to the low energy emission as well. The difference in the rates of these
contributions give rise to  breaks in the
light curves at low energy emissions. This phenomena gives rise to a tendency 
that the light curves at
optical and UV bands peak later than the light curves in the soft and hard 
X-ray bands. Although, this is not very prominent in the simulated light curves, 
it is reflected in the clockwise sense of the hysteresis loop shown in 
Figure \ref{fig6:hys2}. The absence of such breaks at very low 
energy emissions (namely far-IR band and below) can be justified by the fact 
that the escape time-scale from the cooling region
is smaller than the time required for the high energy electrons to cool to 
these energies.


Flares follow the same features of asymmetry as in the case of energy independent 
$t_{acc}$. Similar features are reflected in the flares of SSC component 
also.

Such variations in the photon spectral index during blazar flare (clockwise and
anti-clockwise) are observed for many sources 
\cite{1996v470p89,1999v521p552,2000v541p153,2000v515p19}.

\begin{figure}
\begin{center}
\includegraphics[width=115.mm,bb=12 13 779 601]{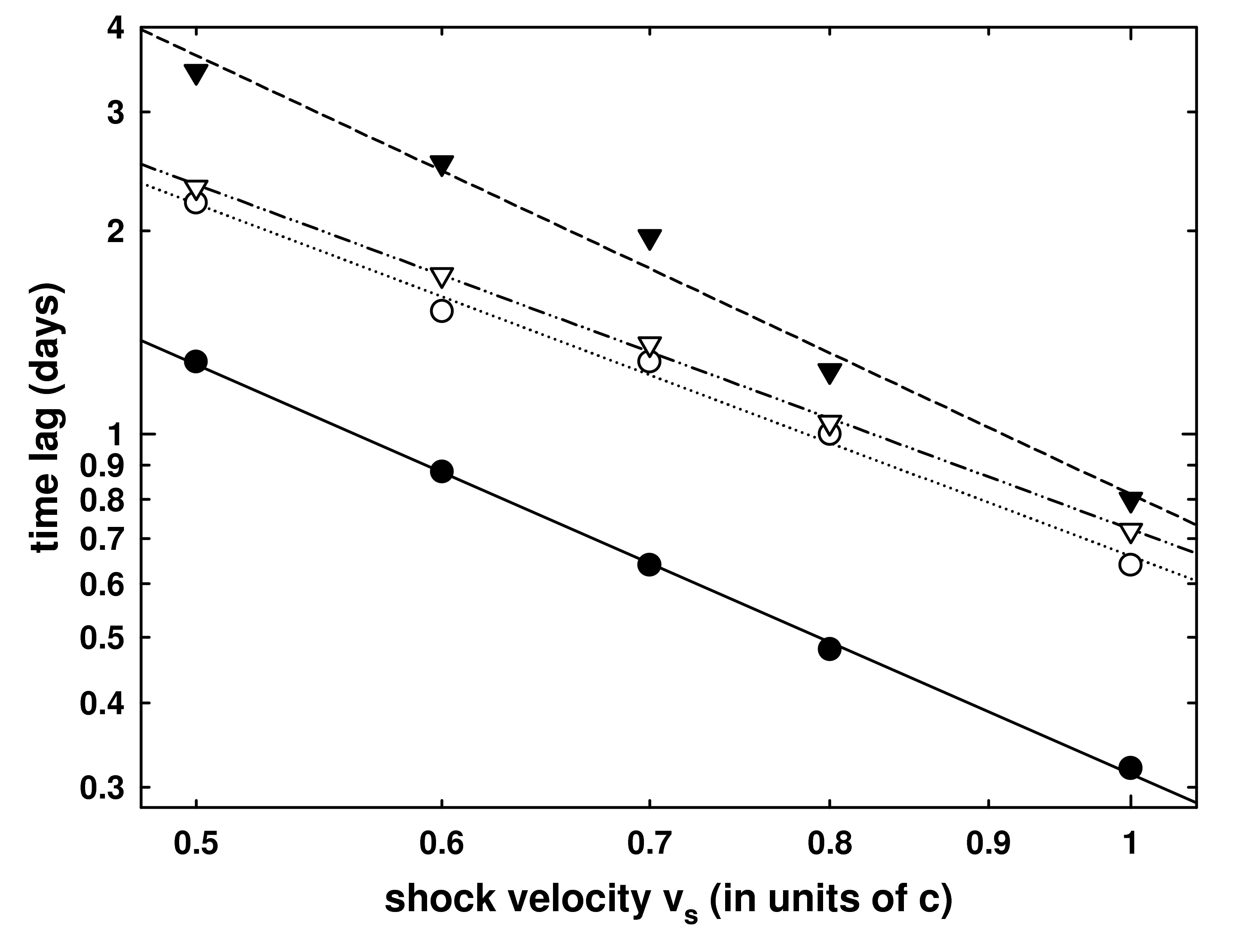} 
\caption[Variation of lag time-scale]
{Variation of lag time-scale with shock velocity in case of energy
independent acceleration. Filled circles corresponds to $10^{13}$ Hz, open 
circles corresponds to $10^{15}$ Hz, filled triangles corresponds to 
$10^{17}$ Hz and open triangles corresponds to $10^{24}$ Hz.}
\label{fig6:tacc}
\end{center}
\end{figure}
\begin{table}[H]\footnotesize
\centering
\caption{Values of $\eta$ for different frequencies}
\label{ch6:table2}
\begin{tabular}{c c}
\toprule
Frequency & $\eta$ \\
{\scriptsize (Hz)} &  \\
\midrule
$10^{13}$  & $2.05 \pm 0.02$ \\
$10^{15}$  & $1.71 \pm 0.09$ \\
$10^{17}$  & $1.99 \pm 0.16$ \\
$10^{24}$  & $1.67 \pm 0.04$ \\
\bottomrule
\end{tabular}
\vskip 0.5cm
\end{table}
\subsubsection{Dependence of lag time-scale on $t_{acc}$}
As shown in the previous section, the energy independent acceleration
time-scale gives rise to a hard lag in the flare patterns for different radiation
frequencies. The variation of lag time-scale for different shock velocities $v_s$ 
is shown in Figure \ref{fig6:tacc}. The lag time-scales 
are calculated with respect to the flare at $10^{11}$ Hz.  It is found that
the lag time-scale varies with $v_s$ following a power-law $(\propto {v_s}^{-\eta})$. 
The fitted value of $\eta$ for different frequencies are given in Table 
\ref{ch6:table2}.
As $t_{acc}$ depends on the shock velocity $v_s$ ($t_{acc} \sim {v_s}^{-2}$), so 
the lag time-scale is proportional to ${t_{acc}}^{\eta/2}$.



\chapter{Summary and Conclusions} 
\label{chap:summary}

The complete and comprehensive understanding of the physics of AGN jet requires
the study of different aspects of jet in different wavelength ranges. Such studies 
may give rise to a global picture of jets describing the dynamics  and radiation 
emission processes self-consistently. Although this is a mammoth task and requires 
a long time, an attempt has been made in this thesis to study certain aspects of 
jet physics phenomenologically and to understand the possible inter-connections 
between them. Here, we studied the dynamics and emission processes from knots in jets 
observed in radio-optical-X-ray bands. Similarly, we considered blazars where we 
studied the limb brightening effect in Mkn 501 observed in radio wave band. This 
feature has intimate connection with physics of jet flow and the particle 
acceleration process at jet boundary. Our work has been further extended to 
include the study of flares in blazars in infra-red-to-TeV energy band.

To model the emission from the knots we used the archival radio, optical and X-ray data.
A continuous injection plasma model, where non-thermal relativistic electrons are 
injected into an expanding spherical region with a tangled magnetic field, is used to 
study the broadband emission from the knots. Injected relativistic electrons lose energy 
by synchrotron process in the tangled magnetic field of the knot and by up-scattering 
the cosmic microwave background photons. The expansion also introduces an adiabatic loss.
For a given observation time, the particle 
distribution in the emission region will be a broken power-law with a break at an energy 
where the cooling time-scale is equal to the observation time. The resultant spectrum 
is fitted to the observed data to yield the parameters of the model. The parameters 
obtained from the spectral fitting are physically reasonable and they are used to obtain 
the kinetic powers of the jets.

Above work has been further extended to include the dynamics of knots. Assuming that 
the knots are produced due to the collision of matter shells ejected randomly in time 
from the central engine, the complete kinematics of the shells are used to study 
their location of collision and the energetics. Non-thermal relativistic electrons 
are produced in the shock generated due the collision of shells. These electrons 
emit radiation by synchrotron and inverse Compton process as described above. Apart 
from fitting the observed spectrum, the other main conclusion of this work is that 
the location of the knots can be reproduced from the physically acceptable choice 
of parameters in the kinematics of shell collision. Therefore internal shocks can be 
considered as one of the viable mechanism of knot generation. It is to be noted that 
the timescales obtained from the kinematics of shells and internal shock, 
convincingly support the continuous injection scenario.

A two zone model was proposed to explain the emission from the 
knots of the M87 jet since simple models involving 
continuous injection/one-time injection of non-thermal particles failed
to reproduce the observed X-ray flux and the spectral index. In the 
proposed model, we consider the injection of a power-law distribution of particles into 
an acceleration region where they are accelerated further. The particles 
then escape from the acceleration region into a
cooling region where it lose energy mostly via synchrotron 
radiation. The particle distribution in the cooling region will be a double
broken power-law with one break at energy corresponding to the cutoff
energy of the initial injected power-law into the acceleration region
and the next break at energy for which the cooling time-scale
equals to the age of the knot. The observed 
radio-optical-X-ray spectrum from the knots in
M87 jet are reproduced by the resultant synchrotron emission. 
In its simplest form,
the model does not consider any specific acceleration process but
assumes an energy independent acceleration time-scale. The model can 
successfully reproduce the broadband spectrum from the knots/jets of other
FRI galaxies, namely 3C 66B, 3C 346 and 3C 296, which are not explained by
synchrotron emission from simple one zone models.

High-resolution radio maps of the jet of the BL Lac object MKN 501 shows a 
limb-brightened feature and an explanation of this feature based on the 
differential Doppler boosting of a stratified jet, requires large viewing
angle ($>15^\circ$). The viewing angle constraints inferred from the high-energy 
$\gamma$-ray studies of the source is very small ($\sim5^\circ$). Since the 
$\gamma$-ray emission originates from the inner region jet, close to the central 
engine, this model requires the jet to be bent to accommodate the viewing
angle conflict. However, the observed limb-brightened structure of the MKN 501 
jet can be explained if we consider the shear acceleration of particles at the 
boundary due to velocity stratification and their diffusion into the jet medium. 
This inference does not require a large viewing angle as demanded by the 
explanation based on differential Doppler boosting of the jet spine and boundary.
We have shown that shear acceleration dominates over turbulent acceleration
at the boundary if we consider thin shear layer or a sharp velocity gradient. 
Also for the estimated set of parameters, shear acceleration time-scale is much 
smaller than synchrotron cooling time-scale allowing acceleration of electrons to 
be possible. The thickness of the limb-brightened structure will be decided by 
the distance electrons have diffused into the jet medium before loosing its energy 
via synchrotron radiation. However the estimated thickness is beyond the resolution 
of present day telescopes. Simple analytical solution of the steady state diffusion 
equation considering mono-energetic injection and particle escape indicates a steep 
particle spectra for the electrons accelerated at the shear layer in comparison with 
turbulent acceleration. The radio spectral index map of MKN 501 jet is also observed 
to have steep spectrum at the boundary supporting the presence of shear acceleration. 

The temporal behaviour of the blazar emission is studied under the framework of a 
two zone model. The spectral evolution has been examined for two different physical 
conditions of diffusive shock acceleration mechanism, namely energy independent 
acceleration time-scale and energy dependent acceleration time-scale. 
The model is applied on the BL Lac object, MKN 421, to study the implications on 
the flare characteristics for the above mentioned conditions.
We found that in case of energy independent particle acceleration, the photon 
spectrum evolves at a slower rate compared to the energy dependent case though their 
steady state spectra are not differentiable. The flare patterns at different frequencies show 
a hard lag in the energy independent acceleration scenario while they are near 
simultaneous in the energy dependent scenario. Hence, the presence/absence 
of time lags in the flare pattern has direct bearing on the underlying particle 
acceleration mechanism in a blazar jet. Also the presence of a break in the 
rising part of 
high energy light curves in the case of near-simultaneous flares suggests that the 
acceleration time-scale may depend on particle energy. In the case of energy 
independent $t_{acc}$, it is also shown that the time-lag between two given frequencies
has a power-law dependence on the shock velocity.  Hence, a simultaneous multi 
wavelength 
study of blazar variability with good time resolution may be useful to constrain 
the physical parameters of the blazar jet and may also reveal the nature of 
underlying particle acceleration process which is crucial in understanding the 
dynamics of the jet.

The models described here to study the different aspects of jets from
different AGN can be improved further to determine and constrain the
model parameters unambiguously if better quality data are available. This
requires truly simultaneous long-term observation of AGN using both 
ground-based and space-based telescopes in different wavelength bands. 
Particularly, radio observations with high spatial resolution are necessary 
to study the structure of knots as well as the boundary layers of the 
jets. This will shed further light on the jet dynamics, instabilities and
the possible particle acceleration sites along the jet.

To study the flux variability of blazars it is important to have simultaneous
measurements of flaring events at different wavelengths. The high sensitivity
timing study is particularly important at GeV-TeV energies where blazars 
show the fastest flux variations. More observations of blazars with presently 
operating ground-based Cherenkov telescopes like MAGIC, HESS, VERITAS 
telescopes are extremely important in this respect. The upcoming Cherenkov
telescopes with higher sensitivity like MACE and CTA
will of course enhance the quality of data. This will in turn help us to
improve the theoretical models to have better understanding of AGN jets. 

In fact the present work of blazar can further be extended to study the 
effects of extragalactic background light (EBL) on the blazar spectrum and 
possible estimation of EBL. This is an important issue in blazar research
at high energies and this will be pursued in future.



\cleardoublepage
\addcontentsline{toc}{chapter}{Bibliography}
\bibliographystyle{test}
\bibliography{sunderthesis}


\end{document}